\documentclass[12pt,a4paper]{article} 
\usepackage{amsmath,amsfonts,amssymb,amsthm}
\usepackage{fullpage}

\title{Noncommutative Bundles and Instantons in Tehran}

\author{Giovanni Landi$^1$, Walter D. van Suijlekom$^2$ \\[10mm]
$^1$ Dipartimento di Matematica e Informatica, Universit\`a di Trieste\\
Via A. Valerio 12/1, I-34127 Trieste, Italy \\ 
\texttt{landi@univ.trieste.it} \\[5mm] 
$^2$ Max Planck Institute for Mathematics, \\
Vivatsgasse 7
D-53111 Bonn, Germany \\
\texttt{waltervs@mpim-bonn.mpg.de}
}
\date{27 February 2007} %

\renewcommand{\bar}[1]{\overline{#1}}
\renewcommand{\tilde}[1]{\widetilde{#1}}

\def\det{\mathrm{det} \mb}
\newcommand{\half}{{\mathchoice{\oh}{\oh}{\shalf}{\shalf}}} 
\newcommand{\oh}{{\tfrac{1}{2}}}    
\newcommand{\shalf}{{\scriptstyle\frac{1}{2}}} 
\def\II{\mathbb{I}}
\def\ii{\mathrm{i}} 
\def\into{\hookrightarrow}

\def\t{\mathrm{t}}

\def\isom{\simeq}

\def\C{\mathbb{C}} 
\def\bH{\mathbb{H}}
\def\I{\mathbb{I}} 
 
\def\bR{\mathbb{R}}

\def\bT{\mathbb{T}}
\def\Z{\mathbb{Z}} 

\def\A{\mathcal{A}} 

\def\cB{\mathcal{B}}
\def\B{\mathcal{B}}
\def\cD{\mathcal{D}}
\def\E{\mathcal{E}}

\def\cH{\mathcal{H}}

\def\cO{\mathcal{O}}

\def\cS{\mathcal{S}} 
\def\cU{\mathcal{U}}

\def\cZ{\mathcal{Z}}
\def\g{\mathfrak{g}}

\def\G{\mathcal{G}}

\def\SL{\mathrm{SL}}
\def\Sp{\mathrm{Sp}}

\newcommand{\ket}[1]{|#1\rangle}    
\newcommand{\bra}[1]{\langle#1|}    
\newcommand{\braket}[2]{\langle#1|#2\rangle}

\newcommand{\br}[1]{\langle #1 \rangle}

\def\chern{\mathrm{ch}}

\def\dix{\mathrm{Tr}_\omega}

\def\inv{\mathrm{Inv}}

\def\res{\mathrm{Res}}
\def\resz{\underset{z=0}{\res}}

\def\ad{\mathrm{ad}}
\def\Ad{\mathrm{Ad}}
\def\Cinf{C^\infty}
\def\class{{(0)}}

\def\M{M_\theta}
\def\n{{(n)}}
\def\P{P_\theta}
\def\R{\bR_\theta}
\def\S{S_\theta}
\def\Sk{S_{\theta'}}
\def\Sym{\mathrm{Sym}}

\def\Top{\mathrm{Top}}
\def\un{\mathrm{un}}
\def\YM{\mathrm{YM}}
 
\def\lt{\triangleright}    
\def\rt{\triangleleft}    

\def\bd{\begin{displaymath} }
\def\ed{\end{displaymath} }
\def\be{\begin{equation}}
\def\ee{\end{equation}}
\def\bea{\begin{eqnarray}}
\def\eea{\end{eqnarray}}
\def\bmult{\begin{multline}}
\def\emult{\end{multline}}
\def\nn{\nonumber}
\newtheorem{thm}{Theorem}
\newtheorem{corl}[thm]{Corollary}
\newtheorem{lma}[thm]{Lemma} 
\newtheorem{prop}[thm]{Proposition}
\newtheorem{defn}[thm]{Definition}
\newtheorem{ex}[thm]{Example}
\newtheorem{rem}[thm]{\bf Remark}
\def\mb{\mbox{ }}
\def\rightbox{\begin{flushright}$\Box$\end{flushright} }

\newcommand{\bean}{\begin{eqnarray*}}
\newcommand{\eean}{\end{eqnarray*}}
\newcommand{\cinf}{C^\infty}       
\numberwithin{equation}{section}
\newcommand{\ca}{{\mathcal A}}

\newcommand{\ce}{{\mathcal E}}
\newcommand{\cf}{{\mathcal F}}

\newcommand{\ch}{{\mathcal H}}

\newcommand{\cs}{{\mathcal S}}

\newcommand{\cu}{{\mathcal U}}

\newcommand{\IC}{{\mathbb C}}

\newcommand{\IR}{{\mathbb R}}

\newcommand{\IT}{{\mathbb T}}
\newcommand{\IZ}{{\mathbb Z}}

\def\lra{\longrightarrow}

\newcommand{\wh}{\widehat}
\newcommand{\wt}{\widetilde}
\def\bar#1{\overline{#1}}

\newcommand{\omca}{\Omega {\mathcal A}}

\newcommand{\oca}[1]{\Omega^{#1}{\mathcal A}}

\newcommand{\comca}{{\mathcal E}\otimes_{\mathcal A}\Omega {\mathcal A}}
\newcommand{\coca}[1]{{\mathcal E}\otimes_{\mathcal A}\Omega^{#1}{\mathcal A}}

\newcommand{\ota}{\otimes_{\mathcal A}}

\newcommand{\otc}{\otimes_{\IC}}
\newcommand{\ot}{\otimes}

\newcommand{\op}{\oplus}
\newcommand{\hs}[2]{\left\langle #1,#2\right\rangle}

\DeclareMathOperator{\id}{id}
\DeclareMathOperator{\Mat}{Mat}
\DeclareMathOperator{\Tr}{Tr}

\DeclareMathOperator{\tr}{tr}
\DeclareMathOperator{\SU}{SU}
\DeclareMathOperator{\GL}{GL}
\DeclareMathOperator{\U}{U} 
\DeclareMathOperator{\SO}{SO}
\DeclareMathOperator{\Spin}{Spin}
\DeclareMathOperator{\Aut}{Aut} 
\DeclareMathOperator{\Hom}{Hom}
\DeclareMathOperator{\End}{End}
\DeclareMathOperator{\ind}{index}
\newcommand{\dd}{{\rm d}} 
\newbox\ncintdbox \newbox\ncinttbox
\setbox0=\hbox{$-$} \setbox2=\hbox{$\displaystyle\int$}
\setbox\ncintdbox=\hbox{\rlap{\hbox to
\wd2{\hskip-.125em\box2\relax \hfil}}\box0\kern.1em}
\setbox0=\hbox{$\vcenter{\hrule width 4pt}$}
\setbox2=\hbox{$\textstyle\int$}
\setbox\ncinttbox=\hbox{\rlap{\hbox to
\wd2{\hskip-.175em\box2\relax \hfil}}\box0\kern.1em}
\newcommand{\ncint}{\mathop{\mathchoice{\copy\ncintdbox}%
{\copy\ncinttbox}{\copy\ncinttbox}{\copy\ncinttbox}}\nolimits}

\begin{document}

\maketitle

\begin{abstract}
We present an introduction to the use of noncommutative geometry for gauge theories with emphasis on a construction of instantons for a class of four dimensional  toric noncommutative manifolds. These instantons are solutions of self-duality equations and are critical points of an action functional. We explain the crucial role of twisted symmetries as well as methods from noncommutative index theorems.  
\end{abstract}

\vfill
\noindent 
Based on lectures delivered by GL at the \\
\textit{International Workshop on Noncommutative Geometry NCG2005}, \\
Institute for Studies in Theoretical Physics and Mathematics (IPM), \\
Tehran, Iran, September 2005.

\thispagestyle{empty}

\newpage
\tableofcontents

\newpage

\section{Introduction}\label{se:intro}

These notes are intended to be a pedagogical introduction to noncommutative gauge theories. They report on attempts to explore the `Noncommutative Pardis' \cite{pardis} seeking instantons. We started the journey with the  guiding image of an instanton as  `a rank two complex vector bundle on a four dimensional manifold endowed with a self-dual connection' , but 
we usually travel with a mind open to diversity. 
Descriptions of parts of the region were noncommutative instantons grow                                                                                                                                                                                                                                                                                                                                                                                           have already been reported \cite{NS98} and resulted in a host of interesting and current developing activities. 

We concentrate on gauge theories on  toric noncommutative manifolds \cite{CL01} and in fact, working out explicit examples, we give a detailed construction of a family of $\SU(2)$ gauge instantons on a four dimensional noncommutative sphere $\S^4$ as constructed in \cite{LS04,LS06}. 

Here is a brief synopsis of these notes. 
We start in Sect.~\ref{se:introgauge}
with some elements of classical gauge theory on principal and vector bundles and review the theory of connections on modules -- the algebraic substitute for bundles. 
Sect.~\ref{se:toric-ncm} is devoted to toric noncommutative manifolds $\M$ -- where $\theta$ is an antisymmetric matrix of deformation parameters -- which are deformations of Riemannian manifolds $M$ along torus actions. 
In Sect.~\ref{se:hopf-fibration}, we focus on two such manifolds, $\Sk^7$ and $\S^4$,  and  exhibit a one-parameter family of noncommutative $\SU(2)$ principal  fibrations $\Sk^7\to \S^4$, with $\theta'$ a simple function of $\theta$.

In Sect.~\ref{se:ym}, we first develop gauge theories  on $\S^4$ by defining a Yang-Mills action functional in terms of the curvature of a connection on a projective module over $\Cinf(\S^4)$. We derive the `absolute minima' of this functional as connections with (anti)self-dual curvature -- the instantons. These are characterized by a `topological charge' -- an integer which is the $K$-theory class of the projective module on which the instanton is given,  and which  is calculated by a noncommutative index theorem.
We also sketch gauge theories on any four dimensional toric noncommutative manifold.

The sphere $\S^4$ carries an action of twisted rotations. This is shown in Sect.~\ref{se:twist}, after a description of the general procedure to twist Hopf algebras and their actions. 
The twisted infinitesimal symmetries of $\S^4$ make up the Hopf algebra $\U_\theta(so(5))$, which also leave invariant a basic instanton. The latter is introduced in Sect.~\ref{se:constrinst} where we show how to obtain 
a family of gauge non-equivalent instantons  by acting on the basic instanton with twisted infinitesimal conformal transformations, encoded in the Hopf algebra $\U_\theta(so(5,1))$. There we also prove -- by using noncommutative index theoretical arguments --  that this collection is the complete set of (infinitesimal) charge 1 instantons. 
We finish in Sect.~\ref{se:fire}, with a mention of alternative $\SU_q(2)$ bundles over quantum four spheres.

\subsection*{Acknowledgements} GL thanks the organizers of the `International Workshop on Noncommutative Geometry NCG2005', for the kind invitation. The organizers, the participants and all people at IPM in Tehran are warmly thanked for the very nice time spent there. GL enjoyed immensely to travel in Iran with Jacopo who was a perfect companion.

\section{Elements of gauge theories}\label{se:introgauge}

Classically, Yang-Mills gauge theories are described by means of principal and vector bundles and connections on them. From a noncommutative point of view, there are suitable substitutes and an analogous theory can be developed.

\subsection{Connections on principal and vector bundles}

With $P$ and $X$ smooth manifolds and $G$ a Lie group, the surjection $\pi : P \to X$ is a principal $G$--bundle on $X$ if it is a fibre bundle with typical fibre $G$ and $G$ acts freely and transitively on $P$, so that $X = M / G$, the space of orbits. The group $G$ is called the structure (or sometime the gauge) group; we denote with $R_g$ the right action of the element $g\in G$ on $P$ and write also $R_g(p)=pg$.
If $\g$ is the Lie algebra of $G$, the fundamental vector field $\xi^\#$ associated to $\xi \in \g$ acts on a smooth function $f$ on $P$ by
$$
\xi^\# f(p) = \tfrac{d }{d  t}_{|_{t=0}} f(p \exp(t \xi)).
$$
A connection on a principal bundle is mostly easily given via a connection form.
\begin{defn}
 A {\rm connection one-form} on $P$ is a one-form $\omega$ taking values in $\g$ and satisfying the conditions: 
\begin{itemize}
\item[(i)] $\omega(\xi^\#)=\xi$,
\item[(ii)] $R_g^* \omega = \Ad_{g^{-1}} \omega$,
\end{itemize}
where $R_g^*$ is the induced action of $G$ on the one-forms $\Omega^1(P)$ on $P$ and 
$$
\left( \Ad_{g^{-1}} \omega\right)_p (Y_p) := g^{-1} \omega_p (Y_p) g, 
$$
with $Y_p \in T_pP$, the tangent bundle at the point $p\in P$.
\end{defn}
\noindent
Given a connection one-form $\omega$, the corresponding {\rm horizontal subspace} $HP$ is the kernel of $\omega$ in the tangent bundle $TP$,
$$
H_p P = \left\{Y_p \in T_p P : \omega(Y_p) =0 \right\}, 
$$
and $Y^H=Y-\omega(Y)$ will denote the projection of $Y \in \Gamma(TP)$ onto $\Gamma(HP)$. 

With $W$ a vector space (later to carry a representation of $G$), 
the covariant derivative of any vector valued form on $P$, $\varphi \in \Omega^r(P) \ot W$,  is then defined as, 
\begin{align*}
&\mathrm{D}: \Omega^r(P)\ot W \to \Omega^{r+1}(P) \ot W , \\
&\mathrm{D} \varphi( Y_1, \ldots, Y_{r+1} ) = \dd_P \varphi( Y_1^H, \ldots, Y_{r+1}^H ).
\end{align*}
where $\dd_P$ is the exterior derivative on $P$.
\begin{defn}
The {\rm curvature} $\Omega$ of $\omega$ is the covariant derivative of $\omega$, 
$$
\Omega=\mathrm{D} \omega 
$$
\end{defn}
\noindent
A more explicit form of the curvature is given in terms of Cartan's structure equation,
$$
\Omega(X,Y) = \dd_P \omega(X,Y) + [\omega(X),\omega(Y)] ,
$$
which is usually written as $\Omega = \dd_P \omega+ \omega \wedge\omega$. The Bianchi identity is the statement that the covariant derivative of the curvature vanishes automatically:
\bd
\mathrm{D}\Omega=0 .
\ed

\medskip

If $\rho$ is a (finite-dimensional) representation of $G$ on the vector space $W$, the associated bundle to $P$ by $W$ is defined to be the vector bundle $E :=P\times_G W$ having typical fibre $W$. It is a classical result in differential geometry that the space of sections $\Gamma(E)$ can be given as the collection of $G$-equivariant maps from $P$ to $W$: 
$$
C_G(P,W):=\left\{ \varphi \in C(P,W):= C(P) \ot W : \varphi( p\cdot g) = \rho_g(\varphi( p)) \right\}.
$$
This identifications being as (right) $C(X)$-modules: one multiplies sections (or equivariant maps with $C(X)$ realized as a subalgebra of $C(P)$) by functions pointwise.

With the module identification, $\Gamma(E)\simeq C_G(P,W)$, a {\it connection} or {\it covariant derivative} on $E$ is defined as the map, 
\bd
\nabla : \Gamma(E) \to  \Gamma(E)\ot_{C(X)} \Omega^1(X) , \qquad
\nabla (\varphi) := \dd_P \varphi + \omega \varphi ,
\ed
and is a particular case of the above definition of the covariant derivative of vector valued forms on $P$. Then, the  curvature of the connection is also the map $\nabla^2$ and  one finds that
\bd
\nabla^2(\varphi)\simeq\Omega(\varphi),
\ed
with both maps $\nabla^2$ and $\Omega$ being $C(X)$-linear.

There is also an equivalent description of connections in terms of local charts of $X$. Choose a local section of $P \to X$ and define $A$ to be the pull-back of $\omega$ under this section. Then $A$ is a (locally defined) one-form on $X$, taking values in $\g$, and is called {\it gauge potential}. The pull-back $F$ of the curvature (the {\it field strength}) is,  in terms of $A$,
\bd
F=\dd A + A \wedge A, 
\ed
with Bianchi identity $\dd F + A \wedge F= 0$. It turns out that $F$ is a two-form taking values in the adjoint bundle $\ad(P):=P \times_G \g$, where $G$ acts on $\g$ with the adjoint representation. 

We are at the crucial notion of a gauge transformation: it is just a section of the bundle of automorphisms of $E$. More precisely, the infinite dimensional group $\G$ of gauge transformations consists of sections of the bundle $P \times_G G$ where $G$ acts on itself by conjugation. A gauge transformation $f$ acts on a connection as 
\bd
\nabla \mapsto f^{-1} \nabla f,
\ed
inducing the familiar transformation rule for the connection one-form $A$,
\bd
A \mapsto f^{-1} A f + f^{-1} \dd f,
\ed
together with 
\bd
F \mapsto f^{-1} F f.
\ed
for its curvature. 
This transformation make evident the invariance under gauge transformations of the 
Yang-Mills action funcional defined (up to a constant factor) by  
\bd
S[A] =\| F\|^2 := - \int_X \tr (F \wedge \ast F), 
\ed
where $\ast$ is the Hodge star operator of a Riemannian metric on $X$ -- that, from now one we take to be four dimensional. The corresponding critical points are solutions of the equation
\bd
\mathrm{D} \ast F = 0 .
\ed
If we decompose $F=F_+ \oplus F_-$ into its self-dual and antiself-dual part, {i.e.} $\ast F_\pm=\pm F_\pm$, we can relate this action to the second Chern number
\bd
c_2 = \int_X \chern_2(\nabla) = \frac{1}{2} \Big(\frac{\ii}{2\pi}\Big)^2 \int \tr F \wedge F.
\ed
In fact, this is a topological quantity -- a topological charge in physicist parlance -- depending only on the bundle and not on the connection and taking integral values. If $c_2=k \in \Z$, one has that
\bd
S[A] = \|F_+ \|  + \| F_-\| , \qquad
8 \pi^2 k =  \|F_+ \|  - \| F_-\|
\ed
from which we deduce the lower bound $S \geq 8 \pi^2 |k|$. Equality holds if $\ast F=\pm F$. Connections with self-dual or antiself-dual  curvature are called instantons (or anti-instanton) and are absolute minima of the Yang-Mills action; for them the Bianchi identity automatically implies the field equations.

Instantons can be used to obtain an approximation for the path integral
\bd
Z(t)=\int \cD[A] e^{-t S[A]} ~,
\ed
with the (formal) integral taken over the space of all gauge potentials. One is really interested in integrating over the moduli space of gauge connections modulo gauge transformations.
As $t\to0$, the path integral is essentially modelled on the integral over the moduli space of instantons. 

Instantons are most elegantly described via the so-called ADHM construction \cite{ADHM78,Ati79}. 
These ideas have culminated in Donaldson's construction of invariants of smooth four-dimensional manifolds \cite{Don90,DK90}. In \cite{Wit88} it is shown how to recover the Donaldson invariants from the path integral $Z(t)$ as $t$ tends to 0.

The generalization in \cite{NS98} of the ADHM method on a noncommutative space $\IR^4$ has found several important applications notably in brane and superconformal theories.

\bigskip

Let us briefly illustrate the above structure with the crucial example of the principal Hopf fibration $S^7 \to S^4$ with $\SU(2)$  as structure group -- more on this bundle is in Sect.~\ref{se:modules} later on.  Let $E$ be the vector bundle associated to the fundamental representation: $E= S^7 \times_{\SU(2)} \C^2$. In this case, an instanton is a connection on this rank two complex vector bundle on $S^4$ -- which we can define by a gauge potential $A$ -- having self-dual curvature. The basic instanton constructed in  \cite{BPST75} can be given on the local chart $\bR^4$ -- with local coordinates 
$\{ \zeta_\mu, \zeta_{\mu}^* \}$ coming from stereographical projection -- by the connection, 
\begin{multline}
A:=\frac{1}{1+| \zeta |^2} \Big( (\zeta_1 \dd \zeta_1^* - \zeta_1^* \dd \zeta_1 - 
\zeta_2 \dd \zeta_2^* + \zeta_2^* \dd \zeta_2 ) \sigma_3 \nn \\
 \qquad  + 2 (\zeta_1 \dd \zeta_2^* - \bar{\lambda} \zeta_2^* \dd \zeta_1) \sigma_+ + 
2 (\zeta_2 \dd \zeta_1^* - \lambda \zeta_1^* \dd \zeta_2) \sigma_- \Big) \, ,
\end{multline}
with $\sigma_3, \sigma_\pm$ generators of the Lie algebra $su(2)$.
The corresponding self-dual curvature is 
\[
F = \frac{1}{(1+| \zeta |^2)^2} \Big( (\dd \zeta_1 \dd \zeta_1^* - \dd \zeta_2^* \dd \zeta_2) \sigma_3  
+ 2 (\dd \zeta_1 \dd \zeta_2^*) \sigma_+ + 2(\dd \zeta_2 \dd \zeta_1^* ) \sigma_- \Big) \, ,
\]
and the values of the topological number is computed to be $1$.

By acting with the conformal group $SL(2,\bH)$ of $S^4$ on the above gauge potential, one obtains different non gauge equivalent instantons \cite{Ati79}. The conformal group leaves both the (anti)self-dual equation $\ast F=\pm F$ and the Yang-Mills  action invariant, thus transforming instantons into instantons, with the subgroup $\Spin(5)\simeq \Sp(2,\bH) \subset SL(2,\bH)$ yielding gauge equivalent instantons (in fact leaving invariant the basic instanton). Thus,  there is a five-parameter family of instantons up to gauge transformations. On the local chart $\bR^4$ of $S^4$, these five parameters correspond to one scaling $\rho$ and four `translations' of the basic instanton. They form the five-dimensional moduli space of charge $1$ instantons.

\subsection{Connections on noncommutative vector bundles}
\label{se:connections}

We now review the notion of a (gauge) connection on a (finite projective) module $\ce$ over an algebra $\ca$ with respect to a given calculus; we take a right module structure. Also, we recall gauge transformations in this setting.

Let us suppose we have an algebra $\ca$ with a differential calculus $(\omca=\op_p \oca{p}, \dd)$. 
A  {\it connection} on the right $\ca$-module $\ce$ is a
$\IC$-linear map
\[
\nabla : \coca{p} \lra \coca{p+1} , 
\]
defined for any $p \geq 0$, and satisfying the Leibniz rule
\[
\nabla(\omega \rho) = (\nabla \omega) \rho + (-1)^{p} \omega \dd \rho  , 
~~\forall ~\omega \in \coca{p} , ~\rho \in \omca .  
\]
\noindent
A connection is completely determined by its restriction 
\be
\nabla : \ce  \to \coca{1}  ,
\ee
 which satisfies
\be\label{ulei}
\nabla (\eta a) = (\nabla \eta) a + \eta \ota \dd a  , 
~~\forall ~\eta \in \ce , ~a \in \ca ,  
\ee
and which is extended by the Leibniz rule. It is again the latter property that implies the $\omca$-linearity of  
the composition, 
\[
\nabla^2 = \nabla \circ \nabla : \coca{p} \lra \coca{p+2} .
\]
Indeed, for any $\omega \in \coca{p}, \rho \in \omca$ one has
$
\nabla^2 ({\omega \rho}) = \nabla \left( (\nabla \omega) \rho + (-1)^{p} \omega \dd \rho \right)   = (\nabla ^2 \omega) \rho + (-1)^{p+1} (\nabla \omega) \dd \rho + (-1)^{p} (\nabla \omega) \dd \rho 
+ \omega \dd^2 \rho  = (\nabla ^2 \omega) \rho 
$.
The restriction of $\nabla^2$ to $\ce$ is the 
{\it curvature} 
\be
F : \ce  \to \coca{2} ,
\ee 
of the connection. It is $\ca$-linear, $F(\eta a) =  F(\eta)a$ for any
$\eta\in\ce, a \in \ca$, and satisfies
\be
\nabla^2(\eta \ota \rho) =  F(\eta) \rho , ~~\forall ~\eta \in \ce , ~\rho \in \omca . 
\ee
Thus, $ F\in \Hom_{\ca}(\ce, \coca{2} )$, the latter being the collection of (right) $\A$-linear endomorphisms of $\E$, with values in the two-forms $\Omega^2\A$.

In order to have the notion of a Bianchi identity we need some generalization. Let 
$\End_{\omca}(\comca)$ be the collection of all $\omca$-linear endomorphisms of $\comca$. It is an  algebra under composition. The curvature $ F$ can be thought of as an element of $\End_{\omca}(\comca)$. There is then a well-defined map
\begin{align} \label{def:conn-end}
[\nabla, \cdot~ ] & ~:~ \End_{\omca}(\comca) \lra \End_{\omca}(\comca) \nn \\
[\nabla, T] &:= \nabla \circ T - (-1)^{|T|}~ T \circ \nabla  .
\end{align}
where $|T|$ denotes the degree of $T$ with respect to the $\IZ^2$-grading of $\Omega\A$. Indeed, for any $\omega \in \coca{p} , ~\rho \in \omca$, it follows that 
\begin{align*}
[\nabla, T] ({\omega \rho}) & = \nabla (T ( \omega \rho) ) -  (-1)^{|T|} ~ T ( \nabla ( \omega \rho)) \nn \\
& = \nabla \big( T ( \omega) \rho\big) - (-1)^{|T|} ~ T \big( (\nabla \omega) \rho + (-1)^{p} \omega \dd \rho \big) \nn \\
& = \big(\nabla (T(\omega))\big) \rho + (-1)^{p+|T|}~ T(\omega) \dd \rho 
- (-1)^{|T|}~ T (\nabla \omega)  \rho - (-1)^{p+|T|}~ T(\omega) \dd \rho \nn \\
& = \big(\nabla (T(\omega)) -  (-1)^{|T|} ~T (\nabla \omega) \big) \rho  
= \big([\nabla, T] ( \omega)  \big) \rho   , 
\end{align*}
proving $\omca$-linearity.
It is easily checked that $[\nabla, \cdot ~]$ is a graded derivation for the algebra $\End_{\omca}(\comca)$: 
\be
[\nabla,S \circ T]=[\nabla,S]\circ T + (-1)^{|S|} S \circ [\nabla,T].
\ee
\begin{prop}\label{ubianchi}
The curvature $ F$ satisfies the  {\rm Bianchi identity},
\be
[\nabla,  F ] = 0 .  \label{ubia}
\ee
\end{prop}
\begin{proof} 
Since $ F\in\End^0_{\omca}(\comca)$, the map $[\nabla,  F ]$ makes sense. Furthermore,
\[
[\nabla,  F ] = \nabla \circ \nabla^2 - \nabla^2 \circ \nabla = \nabla^3 -
\nabla^3 = 0 . 
\] 
\end{proof}
\noindent
In Sect.~II.2 of \cite{C85}, such a Bianchi identity was implicitly used in the construction of a so-called canonical cycle from a connection on a finite projective $\A$-module $\E$. 

Connections always exist on a projective module.
On the module $\ce = \IC^N \otc \ca \simeq \ca^N$, which is  free, a connection is given by the operator 
\[
\nabla_0 = \II \ot \dd : \IC^N \otc \oca{p} \lra \IC^N \otc \oca{p+1} .
\]
With the canonical identification $\IC^N \otc \omca = (\IC^N \otc \ca) \ota \omca ~\simeq~ (\omca)^N$, one thinks of $\nabla_0$ as acting on $(\omca)^N$ as the
operator $\nabla_0 = (\dd, \dd, \cdots, \dd)$  ($N$-times).  
Next, take a  finite projective module $\ce$ with inclusion map, $\lambda : \ce \to \ca^N$, which identifies $\ce$ as a direct summand of the free module $\ca^N$, and idempotent $p : \ca^N \to \ce$ which allows one to identify $\ce = p \ca^N$. Using these maps and their natural extensions to $\ce$-valued forms,  
a connection $\nabla_0$ on $\ce$ (called {\it Levi-Civita} or {\it Grassmann}) is the composition,
\[
\coca{p} ~\stackrel{\lambda}{\lra}~ \IC^N \otc \oca{p} 
~\stackrel{\II \ot  \dd}{\lra}~ \IC^N \otc \oca{p+1} 
~\stackrel{p}{\lra}~ \coca{p+1}, 
\]
that is  
\be\label{ugras}
\nabla_0 = p \circ (\II \ot  \dd ) \circ \lambda.    
\ee
One indicates it simply by $\nabla_0 = p \dd$.  \

\begin{rem}
For the universal calculus $(\omca, \dd)=(\omca_\un,\delta)$,
the existence of a connection on the module $\ce$ is
equivalent to it being projective 
\cite{CQ95a}.
\end{rem}

The space $C(\ce)$ of all connections on $\ce$ is an affine space modeled on $\Hom_{\ca}({\ce,\coca{1}})$.
Indeed, if $\nabla_1, \nabla_2$ are two connections on $\ce$, their difference is
$\ca$-linear,
\[
(\nabla_1 - \nabla_2)(\eta a) = ((\nabla_1 - \nabla_2)(\eta )) a , \quad \forall ~ \eta \in \ce , ~a \in \ca ,
\]
so that $\nabla_1 - \nabla_2 \in \Hom_{\ca}({\ce,\coca{1}})$. Thus, any connection can be written as 
\be
\nabla = p \dd + \alpha  , \label{uconn}
\ee
where $\alpha$ is any element in $\Hom_{\ca}({\ce,\coca{1}})$.  The `matrix of $1$-forms'
$\alpha$ as in \eqref{uconn} is called the {\it gauge potential}  of the connection $\nabla$. The
corresponding curvature $ F$ of $\nabla$ is
\be
 F = p \dd p \dd p + p \dd \alpha + \alpha^2  . \label{ucurv} 
\ee

\bigskip
Next, let the algebra $\ca$ have an involution $^*$ extended to the whole of $\omca$ by the requirement 
$(\dd a)^* = \dd a^*$ for any $a\in\ca$.
A {\it Hermitian structure} on the module $\ce$ is a map $\hs{ \cdot}{\cdot} : \ce \ot \ce \to \ca$ with the properties
\begin{align} \label{hest}
&\hs{\eta a }{\xi } = a^* \hs{\xi}{\eta}  , \qquad
\hs{\eta}{\xi }^* = \hs{\xi}{\eta}  , \nn \\
&\hs{\eta}{\eta} \geq 0 ,  \hs{\eta}{\eta} = 0 \iff \eta = 0 , 
\end{align}
for any $\eta,\xi\in\ce$ and $a\in\ca$ (an element $a\in\ca$ is positive if it is of the form $a=b^* b$ for some $b\in\ca$).
We shall also require the Hermitian 
structure to be {\it self-dual}, {i.e.} every right $\ca$-module homomorphism 
$\phi: \E \to \ca$ is represented by an element of $\eta \in \E$, by the 
assignment $\phi(\cdot) = \hs{\eta}{\cdot}$, the latter having the correct 
properties by the first of \eqref{hest}.
With a self-dual Hermitian structure, any $T \in 
\End_{\ca}(\ce)$ is adjointable, that is it admits an adjoint, an $\ca$-linear 
map $T^*:\ce\to\ce$ such that,
\[ 
\hs{T^* \eta}{\xi} = \hs{\eta}{T \xi} , \quad 
\forall ~\eta, \xi \in \ce . 
\]

The Hermitian structure is naturally extended to an $\omca$-valued linear map on the product $\comca \times \comca$ by
\be\label{uhext}
\hs{ \eta \ota \omega}{\xi \ota \rho }~ = (-1)^{|\eta| |\omega|} \omega^* \hs{\eta}{\xi}\rho , \quad \forall ~\eta, \xi \in \comca , ~\omega,\rho\in
\omca.
\ee
A connection $\nabla$ on $\ce$ and a Hermitian structure $\hs{ \cdot}{\cdot}$ on $\ce$ are said to be compatible if the following condition is satisfied \cite{C94},
\be\label{ucompat}
\hs{\nabla \eta}{\xi} + \hs{\eta}{\nabla \xi} = \dd \hs{\eta}{\xi}, \quad \forall ~\eta, \xi \in \ce .
\ee
It follows directly from the Leibniz rule and \eqref{uhext} that this extends to
\be\label{ucompat-ext}
\hs{\nabla \eta}{\xi} + (-1)^{|\eta|} \hs{\eta}{\nabla \xi} = \dd \hs{\eta}{\xi} , \quad \forall ~\eta, \xi \in \comca .
\ee

On the free 
module $\A^N$ there is a canonical Hermitian structure given by 
\be\label{hs-canonical} \hs{\eta}{\xi} = \sum_{j=1}^N \eta_j^* \xi_j , \ee with 
$\eta = (\eta_1, \cdots, \eta_N)$ and $\eta = (\eta_1, \cdots, \eta_N)$ any two 
elements of $\A^N$.
Under suitable regularity conditions on the algebra $\A$ all Hermitian structures 
on a given finite projective module $\E$ over $\A$ are isomorphic to each other 
and are obtained from the canonical structure \eqref{hs-canonical} on $\A^N$ by 
restriction \cite[II.1]{C94}. Moreover, if $\E=p \A^N$, then $p$ is self-adjoint: 
$p=p^*$, with $p^*$ obtained by the composition of the involution $^*$ in the 
algebra $\A$ with the usual matrix transposition. The Grassmann connection 
\eqref{ugras} is easily seen to be compatible with this Hermitian structure, 
\be 
\dd \hs{\eta}{\xi} = \hs{\nabla_0 \eta}{\xi} + \hs{\eta}{\nabla_0 \xi} . 
\ee
For a general connection \eqref{uconn}, the compatibility with the Hermitian
structure reduces to
\be
\hs{ \alpha \eta}{\xi } + \hs{\eta}{\alpha \xi }= 0 , \quad  \forall ~\eta, \xi \in \ce , 
\ee
which just says that the gauge potential is skew-hermitian,
\be
\alpha ^* = -\alpha . 
\ee
We still use the symbol $C(\ce)$ to denote the space of compatible 
connections on $\ce$.

Let $\End^s_{\omca}(\comca)$ denote the space of elements $T$ in $\End_{\omca}(\comca)$ which are skew-hermitian with respect to the Hermitian structure \eqref{uhext}, {i.e.}  satisfying
\be \label{def:skew}
\hs{T \eta}{\xi} + \hs{\eta}{T\xi}=0, \qquad \forall \eta,\xi\in \E.
\ee
\begin{prop} \label{prop:conn-skew}
The map $[\nabla, \cdot~]$  in \eqref{def:conn-end} restricts to $\End^s_{\omca}(\comca)$ as a derivation
\begin{align}
[\nabla, \cdot~ ] & ~:~ \End^s_{\omca}(\comca) \lra \End^s_{\omca}(\comca) , 
\end{align}
\end{prop}
\begin{proof}
Let $T\in\End^s_{\omca}(\comca)$ be of order $|T|$; it then satisfies 
\be\label{inter}
\hs{T \eta}{\xi} + (-1)^{|\eta||T|} \hs{\eta}{T\xi}=0,
\ee
for $\eta,\xi\in \comca$. 
Since $[\nabla,T]$ is $\omca$-linear, it is enough to show that
$$
\hs{[\nabla,T]\eta}{\xi}+ \hs{\eta}{[\nabla,T]\xi}=0, \qquad \forall \eta,\xi \in \E. 
$$
This follows from eqs. \eqref{inter} and \eqref{ucompat-ext}, 
\begin{align*}
\hs{[\nabla,T]\eta}{\xi}+ \hs{\eta}{[\nabla,T]\xi} &= \hs{\nabla T \eta}{\xi}- (-1)^{|T|} \hs{T \nabla \eta}{\xi}+\hs{\eta}{\nabla T \xi} - (-1)^{|T|} \hs{\eta}{T\nabla \xi} \\
&= \hs{\nabla T \eta}{\xi}- \hs{\nabla \eta}{T \xi}+\hs{\eta}{\nabla T \xi} - (-1)^{|T|} \hs{T\eta}{\nabla \xi} \\
&=\dd \big( \hs{T\eta}{\xi}+\hs{\eta}{T\xi} \big) = 0.
\end{align*}
\end{proof}

\subsection{Gauge transformations}

The group $\cu(\ce)$ of unitary endomorphisms of $\ce$ is given by 
\be\label{unit}
\cu(\ce) := \{ u \in \End_{\ca}(\ce) ~|~ u u^* = u^* u = \id_\ce  \} .
\ee
This group  plays the role of the {\it infinite dimensional group of gauge transformations}. 
It naturally acts on compatible connections by
\be\label{ugcon}
(u, \nabla) \mapsto \nabla^u := u^* \nabla u , \quad  \forall ~u\in \cu(\ce),  ~\nabla \in C(\ce) ,
\ee
where $u^*$ is really $u^* \ot \id_{\omca}$; this will always be understood in the following. Then the curvature transforms in a covariant way
\be
(u,  F) \mapsto  F^u = u^*  F u , \label{ugcur}
\ee
since, evidently, $ F^u = (\nabla^u)^2 = u^* \nabla u u^* \nabla u^* = u^* \nabla^2 u = u^*  F u$. \\
As for the gauge potential, one has the
usual affine transformation
\be
(u, \alpha) \mapsto \alpha^u := u^* p \dd u + u^* \alpha u . \label{ugpot}
\ee
Indeed,  
$\nabla^u (\eta) = u^*( p \dd + \alpha) u \eta = u^* p \dd (u \eta) + u^* \alpha u \eta  
= u^* p u \dd \eta + u^* p (\dd u) \eta  + u^* \alpha u \eta  =  p \dd \eta + (u^* p \dd u + u^* \alpha u) \eta$
for any $\eta\in\ce$, which yields \eqref{ugpot} for the transformed potential.

\bigskip

In order to proceed, we add the additional requirement that the 
algebra $\A$ is a Fr\'echet algebra and that $\E$ a right Fr\'echet module. 
That is, both $\A$ and $\E$ are complete in the topology defined by a family of 
seminorms $\| \cdot \|_i$ such that the following condition is satisfied: for all 
$j$ there exists a constant $c_j$ and an index $k$ such that 
\be \| \eta a \|_j 
\leq c_j \| \eta \|_k \|a\|_k. 
\ee 
These are also used to make $\End_{\ca}(\ce)$ a Fr\'echet algebra with corresponding family of seminorms: for $T \in \End_{\ca}(\ce)$ given by, 
\be 
\| T \|_i = \sup_\eta \left\{ \| T 
\eta \|_i : \| \eta\|_i \leq 1 \right\}. 
\ee

The `tangent vectors' to the gauge group $\cU(\E)$ constitute the 
vector space of infinitesimal gauge transformations. Suppose $\{u_t\}_{t \in 
\bR}$ is a differentiable family of elements in $\End_\A(\E)$ (in the topology 
defined by the above sup-norms) and define $X:=(\partial u_t /\partial t)_{t=0}$. 
Unitarity of $u_t$ then induces that $X=-X^*$. In other words, for $u_t$ to be a 
gauge transformation, $X$ should be a skew-hermitian endomorphisms of $\E$. In 
this way, we understand $\End^s_\A(\E)$ as the collection of {\it infinitesimal 
gauge transformations}.  It is a real vector space whose complexification 
$\End^s_\A(\E)\ot_\bR \C$ can be identified with $\End_\A(\E)$.

Infinitesimal gauge transformations act on a connection in a natural way. Let the gauge transformation $u_t$, with $X=(\partial u_t /\partial t)_{t=0}$, act on 
$\nabla$ as in \eqref{ugcon}. From the fact that 
$(\partial (u_t \nabla u_t^*)/\partial t)_{t=0} = [\nabla, X]$, we conclude that an element $X \in 
\End^s_\A(\E)$ acts infinitesimally on a connection $\nabla$ by the addition of 
$[\nabla,X]$, 
\be\label{uXcon} 
(X, \nabla) \mapsto \nabla^X = \nabla + t 
[\nabla,X] + \cO(t^2) , \quad \forall ~X \in \End^s_\A(\E), ~\nabla \in C(\ce). 
\ee 
As a consequence, for the  transformed curvature one has 
\be\label{uXcur}
(X,  F) \mapsto F^X = F + t [F,X] + \cO(t^2), 
\ee
since $F^X = (\nabla + t [\nabla,X])\circ (\nabla + t [\nabla,X]) = \nabla^2 + t [\nabla^2, X] + \cO(t^2)$.

\subsection{Noncommutative principal bundles}\label{se:hg-general}

The datum of a Hopf-Galois extension is an efficient encoding of the notion of a noncommutative principle bundle (see e.g. \cite{BM93}, \cite{HM99}).
Let us recall some relevant definitions \cite{KT81,Mon93}. 
Recall that we work over the field $k=\IC$.

\begin{defn}\label{defn:hopfgalois}
Let $H$ be a Hopf algebra and $P$ a right $H$-comodule
algebra with
multiplication $m: P\ot P \to P$ and  coaction $\Delta_R : P \to P 
\ot H$. Let
$B\subseteq P$ be the subalgebra of coinvariants, i.e.
$B:=\{p\in P ~|~ \Delta_R(p) = p \ot 1\}$. The extension $B\hookrightarrow P$ is called an
$H$ Hopf-Galois extension if the canonical map
\begin{eqnarray}\label{can} && \chi : P \otimes_B P \lra P \ot H \; , 
\nn \\ && \chi
:= (m \ot id) \circ (id \otimes_B \Delta_R) \; , \quad p' \ot_B p
\mapsto
\chi(p' \otimes_B p) = p' p_{(0)} \ot p_{(1)} \; ,
\end{eqnarray} is bijective.
\end{defn}
\noindent We use Sweedler-like notation $\Delta_R p=p_{(0)} \ot p_{(1)}$. 
By construction, the 
canonical map
is left $P$-linear and right
$H$-colinear and is a morphism (an isomorphism  for Hopf-Galois 
extensions) of left
$P$-modules and right $H$-comodules. It is also clear that $P$ is 
both a left and a
right $B$-module.  

Classically, the above requirement corresponds to freeness and transitivity of the action of a Lie group $G$ on a manifold $X$.
The injectivity of the canonical map dualizes the condition of a group  action
$X
\times G
\to X$ to be free: if $\alpha$ is the map $\alpha: X \times G \to X
\times_M X, ~(x,g) \mapsto (x,x\cdot g)$ then $\alpha^*=\chi$ with
$P,H$ the algebras of functions on $X,G$ respectively and the action 
is free if and
only if $\alpha$ is injective. Here $M:=X/G$ is the space of orbits 
with projection
map $\pi: X \to M, ~\pi(x\cdot g)=\pi(x)$, for all
$x\in X, g\in G$. Moreover, $\alpha$ is surjective if and only if for all
$x\in X$, the fibre $\pi^{-1}(\pi(x))$ of $\pi(x)$ is equal to the 
residue class $x
\cdot G$, that is, if and only if $G$ acts transitively on the fibres of $\pi$.

In differential geometry a principle bundle is more than just a free and effective action
of a Lie group. However, for a structure Hopf algebra $H$ which is cosemisimple and has bijective antipode -- as is clearly the case for the example $H=\A(\SU(2))$ of the present paper -- from Th.~I of \cite{Sch90} further nice
properties can be established. 
In particular the surjectivity of the canonical map implies bijectivity and faithfully flatness
of the extension.

An additional useful result \cite{ScP00} is that the map $\chi$ is
surjective whenever, for any generator $h$ of $H$, the element $1
\ot h$ is in its image. This follows from the  left $P$-linearity and right
$H$-colinearity of the map $\chi$. Indeed, let $h,~k$ be two elements
of $H$ and
$\sum p_i' \ot p_i,~\sum q_j' \ot q_j \in P \ot P$ be such that
$\chi(\sum p_i' \ot_B p_i)=1 \ot h,~\chi(\sum q_j' \ot_B q_j)=1 \ot k$. Then
$\chi(\sum p_i'q_j' \ot_B q_j p_i)=1 \ot kh$, that is $1 \ot kh$ is 
in the image of
$\chi$. Surjectivity of the map $\chi$ then follows from its left  $P$-linearity.
It is also easy to write down an explicit expression for the inverse of the canonical map: in the above notation one has  $\chi^{-1} (1 \otimes k h) = \sum p'_i q'_j \otimes_B q_i' p_j'$  and its general form follows again from left $P$-linearity. 

\bigskip
We are only interested in $H$-Hopf-Galois extension $B \into P$, for a cosemisimple Hopf algebra $H$ with bijective antipode. 

An important consequence of being a faithfully flat Hopf-Galois extension is the existence of a so-called \textit{strong connection}. Constructing a strong connection is an alternative way to
prove that one has a Hopf Galois extension \cite{Haj96,DGH01}.
If $H$ is cosemisimple and has a bijective antipode, then a $H$-Hopf-Galois extension $B \into P$ is \textit{equivariantly projective}, that is, there exists a left $B$-linear right $H$-colinear splitting $s:P\to B\otimes P$ of the multiplication map $m:B \otimes P \to P$, $m\circ s=\id_P$. Such a map characterizes a strong connection. 
If $H$ has an invertible antipode, a possible description
of a strong connection can be given in terms of a map
$\ell: H \to P \ot P$ satisfying a list of conditions \cite{MA99, BH04}
\bea \label{ell}
\ell(1) &=& 1\otimes 1 , \nn \\ 
\bar\chi(\ell(h))&=&1\otimes h ,  \nn \\
(\ell \otimes \id)\circ \Delta &=& (\id \otimes \Delta_R) \circ \ell , \nn \\ 
(\id \otimes \ell)\circ \Delta &=& (\Delta_L \otimes \id) \circ \ell ,
\eea
Here the map $\bar\chi: P\otimes P \to P \otimes H$ is the lift of the canonical map as $\bar\chi(p'\otimes p)=p'p_{(0)}\otimes p_{(1)}$; $\Delta$ is the comultiplication on $H$ and  $\Delta_L:P \to H \otimes P$, $p \mapsto S^{-1} p_{(1)} \otimes p_{(0)}$. Then, one defines 
a  `strong connection one-form' with respect to the universal calculus, $\omega: H \to \Omega^1_\un P$ by
\be\label{strongc1f}
\omega:h \mapsto \ell(h) - \epsilon(h) 1\otimes 1.\nn
\ee
Indeed, if one writes $\ell(h)=h^{\br{1}} \otimes h^{\br{2}}$ (a summation is understood) and applies $\id \otimes \epsilon$ to the second formula in \eqref{ell}, one has $h^{\br{1}}h^{\br{2}}=\epsilon(h)$. Therefore, 
\be\label{strongomega}
\omega( h)= h^{\br{1}} \delta  h^{\br{2}}\nn
\ee
where $\delta: P \to \Omega^1_\un P, ~p \mapsto 1\otimes p -p \otimes 1$ is the universal differential. 
Equivariant projectivity of $B\into P$ follows by taking as splitting of the multiplication the map
$s:P \to B\otimes P,  ~p \to p_{(0)} \ell(p_{(1)})$. 
The form \eqref{strongomega} enjoys a list of properties which could be given as its definition, 
\begin{enumerate}
\item $\bar\chi \circ \omega = 1 \otimes (\id -\epsilon)$, \qquad  (\textit{fundamental vector field condition})
\item $\Delta_{\Omega^1_\un (P)} \circ \omega = (\omega \otimes \id) \circ \Ad_R$, \qquad 
(\textit{right adjoint colinearity})
\item $\delta p - p_{(0)} \omega(p_{(1)}) \in ( \Omega^1_\un B) P$, \, $\forall  p \in P$, 
\qquad (\textit{strongness condition}).
\end{enumerate}
Here $\Omega_\un(B) \subset \Omega_\un(P)$ are the universal calculi on $B$ and $P$ with differential $\delta$; 
the comultiplication $\Delta_R : P \to P \otimes H$, is extended to  $\Delta_{\Omega^1_\un (P)}$ on $\Omega^1_\un P \subset P \otimes P$ in a natural way by 
$ \Delta_{\Omega^1_\un (P)} (p'\otimes p) \mapsto p'_{(0)} \otimes p_{(0)} \otimes p'_{(1)} p_{(1)}$; and $\Ad_R(h) = h_{(2)}\otimes S(h_{(1)}) h_{(3)}$ is the right adjoint coaction of $H$ on itself. 

\bigskip

A strong connection on the extension $B \into P$ induces connections -- in the sense of Sect.~\ref{se:connections} and with the universal calculi -- on the associated modules \cite{HM99}. 
\begin{defn}
Let $\rho: W \to H \otimes W$ be an left $H$-comodule and denote $\rho(v)=v_{(0)} \otimes v_{(1)}$. The right $B$-module $\Hom^\rho(W,P)$ associated to $B \into P$ by $(\rho, W)$ consists of $H$-coequivariant maps $\varphi: W \to P$, {i.e.} they satisfy
$$
\varphi( v_{(1)}) \otimes S v_{(0)}=\Delta_R \varphi( v), \qquad v \in W.
$$
\end{defn}
For $\varphi \in \Hom^{\rho} (W, P)$, we set 
\be
\nabla_\omega (\varphi)(v) \mapsto \delta \varphi( v) + \omega(v_{(0)}) 
\varphi( v_{(1)}), \nn
\ee
were $\omega$ is the form defined in \eqref{strongc1f}.
Using the right adjoint colinearity of $\omega$ and a little algebra one shows that $\nabla_\omega (\varphi)$ satisfies the following coequivariance condition
$$
\nabla_\omega (\varphi)(v_{(1)}) \otimes S v_{(0)} = \Delta_{\Omega^1_\un(P)} \big( \nabla_\omega (\varphi)(v) \big)
$$
so that 
$$
\nabla_\omega: \Hom^{\rho} (W, P) \to \Hom^{\rho} (W, \Omega_\un^1(P)).
$$ 
In fact, from properties of the form $\omega$, one establishes that $\nabla_\omega$ is a map 
$$
\nabla_\omega: \Hom^{\rho} (W, P) \to \Hom^{\rho} (W, P) \otimes_{B} \Omega^1_\un(B).
$$

Classically, a notion that is close -- although not equivalent -- to triviality of a principal bundle is the one of being cleft. In general one says that a Hopf-Galois extension is {\rm cleft} if there exists a (unital) convolution-invertible colinear map $\phi: H \to P$, called a {\rm cleaving map}. 
Furthermore, if a Hopf-Galois extension is cleft, its associated modules are trivial, {i.e.} isomorphic to the free module $B^N$ for some $N$ \cite{BM93, HM99}.

\section{Toric noncommutative manifolds}\label{se:toric-ncm}

We start by recalling the general construction of toric noncommutative manifolds given in \cite{CL01} where they were called isospectral deformations. These are deformations of a classical
Riemannian manifold and  satisfy all the properties  of noncommutative spin geometry~\cite{C96}. They are the content of the following result taken from~\cite{CL01}, 

\begin{thm}\label{Theorem6}
Let $M$ be a compact spin Riemannian manifold whose isometry group 
has rank $r \geq 2$. Then $M$ admits a natural 
one parameter isospectral deformation to noncommutative geometries $M_{\theta}$.
\end{thm}

The idea of the construction is to deform the standard spectral triple describing the Riemannian geometry of $M$ along a torus embedded in the isometry group, thus obtaining a family of spectral
triples describing noncommutative geometries.

\subsection{Deforming along a torus}
\label{subse:def-torus}
Let $M$ be an $m$ dimensional compact Riemannian manifold equipped with an isometric smooth action $\sigma$ of an $n$-torus $\bT^n$,  $n \geq 2$.  We denote by $\sigma$ also the corresponding action of $\bT^n$ by automorphisms on the algebra $\Cinf(M)$ of smooth functions on $M$, obtained by pull-back. The algebra $\cinf(M)$ may be decomposed
into spectral subspaces which are indexed by the dual group $\IZ^n = \wh\IT^n$. Now, with $s=(s_1, \cdots s_n)\in\bT^n$, 
each $r \in \IZ^n$ labels a character $e^{2\pi i s} \mapsto e^{2\pi i r\cdot s}$ of $\IT^n$, with the scalar product
$r\cdot s := r_1 s_1 +\cdots+ r_n s_n$. The $r$-th spectral
subspace for the action $\sigma$ of $\IT^n$ on $\cinf(M)$ consists of those smooth
functions $f_r$ for which
\be\label{actor}
\sigma_s (f_r) = e^{2\pi \ii r\cdot s} \,f_r ,
\ee
and each $f \in \cinf(M)$ is the sum of a unique (rapidly convergent) series $f = \sum_{r\in\IZ^n} f_r$.
Let now $\theta = (\theta_{j k} = - \theta_{k j})$ be a real antisymmetric $n\times n$ matrix. 
The $\theta$-deformation of $\cinf(M)$ may be defined by replacing the ordinary product by a
deformed product, given on spectral subspaces by
\be
\label{eq:star-product}
f_r \times_\theta g_{r'} := f_r ~ \sigma_{\half r \cdot \theta}( g_{r'}) = e^{ \pi \ii r \cdot \theta \cdot r' } f_r g_{r'} ,
\ee
where $r\cdot \theta =(r_j \theta_{j 1}, \ldots, r_j \theta_{j n} )\in \IR^n$. This product is then extended linearly to all functions in $\Cinf(M)$. We denote $\cinf(M_\theta) := (\cinf(M),\times_\theta)$ and note that the action $\sigma$ of $\IT^n$ on $\cinf(M)$ extends to an action on $\cinf(M_\theta)$ given again by \eqref{actor} on the homogeneous elements.

\bigskip
Next, let us take $M$ to be a spin manifold with $\cH:=L^2(M,\cS)$ the Hilbert space of spinors and $D$ the usual Dirac operator of the metric of $M$. Smooth functions act on spinors by pointwise multiplication thus giving a representation $\pi : \Cinf(M) \to \B(\cH)$, the latter being the algebra of bounded operators on $\ch$. 

There is a double cover $c: \widetilde{\bT}^n \to \bT^n$ and a representation of $\tilde \bT^n$ on $\cH$ by unitary operators $U(s), s \in \tilde\bT^n$, so that 
\be
U(s) D U(s)^{-1} = D,
\ee
since the torus action is assumed to be isometric, and such that for all $f \in \Cinf(M)$, 
\be
U(s) \pi(f) U(s)^{-1} = \pi(\sigma_{c(s)}(f)).
\ee
Recall that an element $T\in \B(\cH)$ is called smooth for the action of $\tilde \bT^n$ if the map 
\bd
\tilde \bT^n \ni s \mapsto \alpha_s(T) := U(s) T U(s)^{-1},
\ed
is smooth for the norm topology. From its very definition, $\alpha_s$ coincides on $\pi(C^\infty(M)) \subset \B(\cH)$ with the automorphism $\sigma_{c(s)}$. Much as it was done before for the smooth functions, we shall use the torus action to give a spectral decomposition of smooth elements of $\B(\cH)$.  Any such a smooth element $T$ is written as a (rapidly convergent) series $T =\sum T_{r}$ with $r\in\IZ$ and each $T_{r}$ is homogeneous of degree $r$ under the action of $\tilde \bT^n$, {i.e.}
\be\label{homocomp}
\alpha_s(T_{r}) =e^{2 \pi \ii r \cdot s } T_{r} ,  \quad \forall \quad s  \in \tilde \bT^n .
\ee
Let ($P_1, P_2,\ldots, P_n$) be the infinitesimal generators of the action of $\tilde \bT^n$ so that we can write $U(s)=\exp{2 \pi \ii s \cdot p}$.  
Now, with $\theta$ a real $n\times n$ anti-symmetric matrix as above,  one defines a twisted representation 
of the smooth elements $\B(\cH)$ on $\cH$ by
\be\label{twist}
L_\theta(T):=\sum_r T_r U( \half r\cdot \theta) 
= \sum_r T_r \exp \big\{ \pi \ii \,  r_j \theta_{jk} P_k \big \}, 
\ee
The twist $L_\theta$ commutes with the action $\alpha_s$ of $\tilde \bT^n$ 
and preserves the spectral components of smooth operators: 
\be\label{spectwist} 
\alpha_s(L_\theta(T_r))  = U(s) ~T U(\half r\cdot \theta)~ U(s)^{-1}
=U(s) T U(s)^{-1} U(\half r\cdot \theta) = e^{2 \pi i r \cdot s } L_\theta (T_r).
\ee
 
In particular, taking smooth functions on $M$ as elements of $\B(\cH)$, via the representation $\pi$, the previous 
definition gives an algebra $L_\theta(\Cinf(M))$ which we may think of as a representation (as bounded operators on $\ch$) of the algebra $\Cinf(\M)$. Indeed, by the very definition of the product $\times_\theta$ in \eqref{eq:star-product} one establishes  that 
\be
L_\theta(f\times_\theta g)= L_\theta(f) L_\theta(g),
\ee
proving that the algebra $\Cinf(M)$ equipped with the product $\times_\theta$ is isomorphic to the algebra $L_\theta(\Cinf(M))$. 
It is shown in \cite{Rie93a} that there is a natural completion of the algebra $\cinf(M_\theta)$ to a $C^*$-algebra $C(M_\theta)$ whose smooth subalgebra -- under the extended action of $\IT^n$ -- is
precisely $\cinf(M_\theta)$. Thus, we can understand $L_\theta$ as a {\it quantization map} 
\be
L_\theta : \Cinf(M) \to \Cinf(\M), 
\ee
which provides a strict deformation quantization in the sense of Rieffel.
More generally, in \cite{Rie93a} one considers a (not necessarily commutative) $C^*$-algebra $A$ carrying an action of $\bR^n$. For an anti-symmetric $n \times n$ matrix $\theta$, one defines a star product $\times_\theta$ between elements in $A$ much as we did before. The algebra $A$ equipped with the product $\times_\theta$ gives rise to a $C^*$-algebra denoted by $A_\theta$. Then the collection $\{A_{\hbar\theta} \}_{\hbar\in [0,1]}$ is a continuous family of $C^*$-algebras providing a strict deformation quantization in the direction of the Poisson structure on $A$ defined by the matrix $\theta$. 

Our cases correspond to the choice $A=C(M)$ with an action of $\bR^n$ that is periodic or, in other words, an action of $\bT^n$. The smooth elements in  the deformed algebra make up
the algebra $\Cinf(\M)$.
The quantization map plays a key role in the following, allowing us to extend differential geometric techniques from $M$ to the noncommutative space $\M$.

\bigskip

It was shown in \cite{CL01} that $(L_\theta(C^\infty(M)), \cH, D)$ satisfies  all axioms of a noncommutative spin geometry  \cite{C96, GVF01}; there is also a grading $\gamma$ (for the even case) and a real structure $J$. 
In particular, boundedness of the commutators $[D,L_\theta(f)]$ for $f \in C^\infty(M)$ follows from 
$[D,L_\theta(f)]=L_\theta([D,f])$, $D$ being of degree $0$ (since $\bT^n$ acts by isometries, each $P_k$ commutes with $D$). This noncommutative geometry is an isospectral deformation of the classical Riemannian geometry of $M$, in that the spectrum of the operator $D$ coincides with that of the Dirac operator $D$ on $M$. 
Thus all spectral properties are unchanged. In particular, the triples are $m^+$-summable and there is a noncommutative integral as a Dixmier trace \cite{Dix66},
\be\label{dix}
\ncint L_\theta (f) := \dix \big( f |D|^{-m}  \big), 
\ee
with $f \in \Cinf(\M)$ understood in its representation on $\cH$. A drastic simplification of this noncommutative integral is given by the Lemma~\cite{GIV05}. 
\begin{lma} \label{lma:dix}
If $f \in \Cinf(M)$ then
$$\ncint L_\theta (f) = \int_{M} L_\theta (f) \dd \nu.$$
\end{lma}
\begin{proof}
Any element $f \in  \Cinf(M)$ is given as an infinite sum of functions that are homogeneous under the action of $\bT^n$. Let us therefore assume that $f$ is homogeneous of degree $k$ so that $\sigma_s(L_\theta(f))=L_\theta(\sigma_s(f))= e^{2\pi i k \cdot s} L_\theta (f)$. From the tracial property of the noncommutative integral and the invariance of $D$ under the action of $\bT^n$, we see that 
$$
\dix\big(\sigma_s(L_\theta(f))|D|^{-m}\big) = \dix\big(U(s) L_\theta(f) U(s)^{-1} |D|^{-m}\big) = \dix (L_\theta(f) |D|^{-m}).
$$
In other words, $e^{2\pi i k\cdot s} \dix(L_\theta(f) |D|^{-m} )=\dix(L_\theta(f) |D|^{-m})$ from which we infer that this trace vanishes if $k\neq 0$. If $k=0$, then $L_\theta (f) = f$, leading to the desired result. 
\end{proof}

\subsection{Examples: planes and spheres}
\label{se:nc-planes-spheres}
For $\lambda_{\mu\nu}=e^{2\pi \ii \theta_{\mu\nu}}$, where $\theta_{\mu\nu}$ is an anti-symmetric real-valued matrix,
the algebra $\A(\R^{2n})$ of polynomial functions on the noncommutative $2n$-plane is defined to be the unital $*$-algebra generated by $2n$ elements $\{ z_\mu, z_\mu^*, \mu=1,\ldots, n \}$ with relations
\be
z_\mu z_\nu = \lambda_{\mu\nu} z_\nu z_\mu,\quad z_\mu^* z_\nu = \lambda_{\nu\mu} z_\nu z_\mu^*,\quad z_\mu^* z_\nu^* = \lambda_{\mu\nu} z_\nu^* z_\mu^* . \nn
\ee
The involution $*$ is defined by putting $(z_{\mu})^* = z_\mu^*$. For $\theta=0$ one recovers the commutative $*$-algebra of complex polynomial functions on $\bR^{2n}$. 

For any value of the index $\mu$, the element $z_\mu^* z_\mu = z_\mu z_\mu^*$  is central.  Let $\A(\S^{2n-1})$ be the $*$-quotient of $\A(\R^{2n})$ by the two-sided ideal generated by the central element 
$\sum_\mu z_\mu^* z_\mu - 1$. We will denote the images of $z_\mu$ under the quotient map again by $z_\mu$.

The abelian group $\bT^n$ acts on $\A(\bR^{2n}_\theta)$ by 
$*$-automorphisms. For $s =(s_\mu) \in \bT^n$,  
$\sigma_s$ is defined on the generators by $\sigma_s(z_\mu) = e^{2\pi \ii s_\mu} z_\mu$. Clearly, $s \mapsto \sigma_s$ is a group-homomorphism from $\bT^n \to \Aut(\A(\R^{2n}))$. At the special case, $\theta=0$, the map $\sigma$ is induced by a smooth action of $\bT^n$ on the manifold $\bR^{2n}$. 
Since the ideal generating the algebra $\A(\S^{2n-1})$ is invariant under the action, $\sigma$ induces a group-homomorphism from $\bT^n$ into the group of automorphisms on the quotient $\A(\S^{2n-1})$ as well. 

We continue by defining the unital $*$-algebra $\A(\R^{2n+1})$ of polynomial functions on the noncommutative $(2n+1)$-plane which is given by adjoining a central self-adjoint generator $z_0$ to the algebra $\A(\R^{2n})$, {i.e.} $z_0^*=z_0$ and $z_0 z_\mu = z_\mu z_0$, for $\mu=1,\ldots, n$. The action of $\bT^n$ is extended trivially by $\sigma_s(z_0) = z_0$. Let $\A(\S^{2n})$ be the $*$-quotient of $\A(\R^{2n+1})$ by the ideal generated by the central element $\sum z_\mu^* z_\mu+z_0^2 -1$. As before, we will denote the canonical images of $z_\mu$ and $z_0$ again by $z_\mu$ and $z_0$, respectively. Since $\bT^n$ leaves this ideal invariant, it induces an action by $*$-automorphisms on the quotient $\A(\S^{2n})$.

Next, we construct a differential calculus on $\R^m$. For $m=2n$, the complex unital associative graded $*$-algebra $\Omega(\R^{2n})$ of forms is generated by $2n$ elements $z_\mu, z_\mu^*$ of degree $0$ and $2n$ elements $dz_\mu, dz_\mu^*$ of degree $1$ with relations:
\bea\nn  \label{rel:diff-Rm}
&dz_\mu dz_\nu+ \lambda_{\mu\nu} dz_\nu dz_\mu =0 ,\quad dz_\mu^* dz_\nu+ \lambda_{\nu\mu} dz_\nu dz_\mu^* =0, \quad dz_\mu^* dz_\nu^*+ \lambda_{\mu\nu} dz_\nu^* dz_\mu^* =0,& \\
&z_\mu dz_\nu = \lambda_{\mu\nu} dz_\nu z_\mu,\quad z_\mu^* dz_\nu = \lambda_{\nu\mu} dz_\nu z_\mu^*, \quad z_\mu^* dz_\nu^* = \lambda_{\mu\nu} dz_\nu^* z_\mu^*.&
\eea
There is a unique differential $\dd$ on $\Omega(\R^{2n})$ such that $\dd: z_\mu\mapsto dz_\mu$. The involution $\omega \mapsto \omega^*$ for $\omega \in \Omega(\R^{2n})$ is the graded extension of $z_\mu \mapsto z_\mu^*$, {i.e.} it is such that $(\dd \omega)^*=\dd \omega^*$ and $(\omega_1\omega_2)^* = (-1)^{p_1 p_2}\omega_2^* \omega_1^*$ for $\omega_i\in \Omega^{p_i}(\R^{2n})$.
For $m=2n+1$, we adjoin to $\Omega(\R^{2n})$ one generator $z_0$ of degree $0$ and one generator $dz_0$ of degree $1$ with commutations
\be
z_0 dz_0 = dz_0 z_0, \quad z_0 \omega = \omega z_0, \quad dz_0 
\omega= (-1)^{|\omega|} \omega dz_0.\nn
\ee
We extend the differential $\dd$ and the graded involution $\omega \mapsto \omega^*$ of $\Omega(\R^{2n})$ to $\Omega(\R^{2n+1})$ by setting $z_0^*=z_0$ and $(\dd z_0)^*=\dd z_0$, so that $(d z_0)^*= d z_0$. 

The differential calculi $\Omega(\S^m)$ on the noncommutative spheres $\S^m$ are defined to be the quotients of $\Omega(\R^{m+1})$ by the differential ideals generated by the central elements $\sum_\mu z_\mu^* z_\mu - 1$ and $\sum z_\mu^* z_\mu+z_0^2 -1$, for $m=2n-1$ and $m=2n$ respectively. 

The action of $\bT^n$ by $*$-automorphisms on $\A(\M)$ is extended to the differential calculi $\Omega(\M)$, for $\M=\R^m$ and $M=\S^m$, by imposing
that $\sigma_s \circ \dd = \dd \circ \sigma_s$.  

\subsection{Vector bundles on $\M$}
\label{se:nc-vb}
 We will presently give vector bundles on $\M$  in terms of a deformed or $\ast$-product.

Let $E$ be a {\it $\sigma$-equivariant} vector bundle  $M$, that is a bundle which carries an action $V$ of $\bT^n$ by automorphisms, covering the action {\it $\sigma$} of $\bT^n$ on $M$,
\be
\label{eq:sigma-equivariant}
V_s (f \psi) = \sigma_s(f) V_s(\psi), \quad \forall \, f \in \Cinf(M) , \, \psi \in \Gamma^\infty(M,E) .
\ee
The $\Cinf(\M)$-bimodule $\Gamma^\infty(\M, E)$ is defined as the vector space $\Gamma^\infty(M,E)$ but with the bimodule structure given by 
\be\label{eq:left-right-action}
f \lt_\theta \psi = \sum_k f_k V_{\half k\cdot \theta} (\psi) , \qquad
\psi \rt_\theta f = \sum_k V_{-\half k\cdot \theta} (\psi) f_k ,
\ee
where $f=\sum_k f_k$, with $f_k \in \Cinf(M)$ homogeneous of degree $k$ under the action of $\bT^n$ -- as in \eqref{homocomp} -- and $\psi$ is a smooth section of $E$. By using the explicit expression \eqref{eq:star-product} for the star product  and eq.~\eqref{eq:sigma-equivariant}, one checks that these are indeed actions of $\Cinf(\M)$. 

The $\Cinf(\M)$-bimodule $\Gamma^\infty(\M,E)$ is finite projective \cite{CD02} and still carries an action of $\bT^n$ by $V$ with equivariance as in \eqref{eq:sigma-equivariant} for both the left and right action of $\Cinf(\M)$. Indeed, the group $\bT^n$ being abelian, one establishes that 
\be
V_s (f \lt_\theta \psi) = \sigma_s(f) \lt_\theta V_s(\psi), \quad \forall \, f \in \Cinf(\M) , \, \psi \in \Gamma^\infty(\M,E),
\ee
and a similar property for the right structure $\rt_\theta$.
Indeed, due to the fact that the category of $\sigma$-equivariant finite projective module over $\Cinf(M)$ is equivalent to the category of $\sigma$-equivariant finite projective modules over $\Cinf(\M)$ \cite{Jul81}, all equivariant finite projective modules on $\Cinf(\M)$ are of the above type. 
This also reflects the result in \cite{Rie93b} that the K-groups of a $C^*$-algebra deformed by an action of $\bR^n$ are isomorphic to the K-groups of the original $C^*$-algebra: as mentioned above, the noncommutative manifolds $\M$ are a special case -- in which the starting algebra is commutative and the action periodic -- of the general formulation in \cite{Rie93a} of deformations of $C^*$-algebras under an action of $\bR^n$.

From the very definition of $\Gamma^\infty(\M,E)$ the following lemma is true.
\begin{lma} \label{lma:modulesA}
If $E$ $F$ are isomorphic as $\sigma$-equivariant vector bundles, then $\Gamma^\infty(\M,E)$ and $ \Gamma^\infty(\M,F)$ are isomorphic as $\Cinf(\M)$-bimodules.
\end{lma}

Although we defined the above left and right actions on the sections with respect to an action of $\bT^n$ on $E$, the same construction can be done for vector bundles carrying instead an action of the double cover $\tilde \bT^n$. We have already seen an example of this in the case of the spinor bundle, where we defined a left action of $\Cinf(\M)$ in terms of \eqref{twist}. 

\subsection{Differential calculus on $\M$}
\label{se:diff-calc}

It is straightforward to construct a differential calculus on $\M$. This can be done in two equivalent manners, either by extending to forms the quantization maps, or by using the general construction in \cite{C94} by means of the Dirac operator. 

Firstly, let $(\Omega(M),\dd)$ be the usual differential calculus on $M$, with $\dd$ the exterior derivative. The quantization map $L_\theta: \Cinf(M) \to \Cinf(\M)$ is extended to $\Omega(M)$ by imposing that it commutes with $\dd$. The image $L_\theta(\Omega(M))$ will be denoted $\Omega(\M)$. 
Equivalently, $\Omega(\M)$ could be defined to be $\Omega(M)$ as a vector space but equipped with an `exterior deformed product' which is the extension of the product  \eqref{eq:star-product}  to $\Omega(M)$ by the requirement  that it commutes with $\dd$. 
Indeed, since the action of $\bT^n$ commutes with $\dd$, an element in $\Omega(M)$ can be decomposed  into a sum of homogeneous elements for the action of $\bT^n$ -- as was done for $\Cinf(M)$. Then one defines a star product $\times_\theta$ on homogeneous elements in $\Omega(M)$ as in \eqref{eq:star-product} and denotes $\Omega(\M)=(\Omega(M), \times_\theta)$. This construction is in concordance with the previous section, when $\Omega(M)$ is considered as the $\Cinf(M)$-bimodule of sections of the cotangent bundle. The extended action  
of $\bT^n$  from $\Cinf(M)$ to $\Omega(M)$ is used to endow the space $\Omega(\M)$ with the structure of a $\Cinf(\M)$-bimodule with the left and right action given in \eqref{eq:left-right-action}.

As mentioned, a differential calculus $\Omega_D(\Cinf(\M))$ on $\Cinf(\M)$ can also by obtained from a general procedure \cite{C94} by means of the isospectral Dirac operator $D$ on $\M$.
The $\Cinf(\M)$-bimodule $\Omega^p_D(\Cinf(\M))$ of $p$-forms is made of classes of operators of the form
\be\label{formdirac}
\omega = \sum_j a_0^j [D, a_1^j]\cdots [D, a_p^j], \quad a_i^j \in \Cinf(\M), 
\ee
modulo the sub-bimodule of operators (the so-called `junk forms')
\be\label{formjunk}
\Big\{ \sum_j [D,b_0^j] [D, b_1^j]\cdots [D, b_{p-1}^j]:~ b_i^j \in \Cinf(\M),~ b_0^j [D, b_1^j]\cdots [D, b_{p-1}^j]=0 \Big\} .
\ee
With brackets $[ ~\cdot~ ]$ denoting the corresponding equivalence classes, the exterior differential $\dd_D$ is  
\be
\dd_D\bigg[\sum_j a_0^j [D, a_1^j]\cdots [D, a_p^j]\bigg] := \bigg[\sum_j [D,a_0^j] [D, a_1^j]\cdots [D, a_p^j]\bigg],
\ee
and satisfies $\dd_D^2=0$. One introduces an inner product on forms by declaring that forms of different degree are orthogonal, while  the inner product of two $p$-forms $\omega_1,\omega_2$ is defined to be
\be \label{def:innerprod-connes}
( \omega_1,\omega_2 )_D:=\ncint \omega_1^* \omega_2.
\ee
Here the noncommutative integral is the natural extension of the one in \eqref{dix},
\be\label{dix2}
\ncint T  := \dix \big( T |D|^{-m}  \big), 
\ee
with $T$ an element in a suitable class of operators.
Not surprisingly, these two construction of forms agree \cite{CD02}, that is, the differential calculi $\Omega(\M)$ and $\Omega_D(\Cinf(\M))$ are isomorphic. This allows us in particular to integrate forms of top dimension, by defining
\be
\int_{\M} \omega := \ncint \omega_D, \qquad \mathrm{for} \quad 
\omega \in \Omega^m(\M),
\ee
where $\omega_D$ denotes the element in $\Omega_D^m(\Cinf(\M))$ corresponding to $\omega$ (replacing every $\dd$ in $\omega$ by $\dd_D$). We have the following noncommutative Stokes theorem. 
\begin{lma}
\label{lma:stokes}
If $\omega \in \Omega^{m-1} (\M)$ then
$$
\int_{\M} \dd \omega = 0 .
$$
\end{lma}
\begin{proof}
From the definition of the noncommutative integral, 
$$
\int_{\M} \dd \omega = \ncint \dd_D\omega_D = \ncint \dd_D L_\theta(\omega_D^\class) ,
$$
with $\omega_D^\class$ the classical analogue of $\omega$, {i.e.} $\omega=L_\theta (\omega_D^\class)$. At this point one remembers that $D$ commutes with $L_\theta$ (see Sect.~\ref{subse:def-torus}), and realizes that there is an analogue of  Lemma~\ref{lma:dix} for forms, i.e. $\ncint L_\theta (T) = \int_{M} T$. 
One concludes that the above integral vanishes since it vanishes in the classical case.
\end{proof}

\bigskip

The next ingredient is a Hodge star operator on $\Omega(\M)$. Classically, the Hodge star operator is a map $\ast: \Omega^p(M) \to \Omega^{m-p}(M)$ depending only on the conformal class of the metric on $M$. On the one end, since $\bT^n$ acts by isometries, it leaves the conformal structure invariant and therefore, it commutes with $\ast$. On the other hand, with the isospectral deformation one does not change the metric.  
Thus it makes sense to define a map $\ast_\theta : \Omega^p(\M)\to \Omega^{m-p}(\M)$ by 
\be
\ast_\theta L_\theta(\omega) = L_\theta(\ast \omega), \quad \mathrm{for} \quad \omega \in \Omega(\M) .
\ee

With this Hodge operator, there is an alternative definition of an inner product on $\Omega(\M)$. 
Given that $\ast_\theta$ maps $\Omega^p(\M)$ to $\Omega^{m-p}(\M)$, we can define for $\alpha,\beta \in \Omega^p(\M)$
\be
\label{eq:inner-product-forms}
( \alpha,\beta )_2 = \ncint \ast_\theta(\alpha^* \ast_\theta \beta),
\ee
since $\ast_\theta(\alpha^* \ast_\theta \beta)$ is an element in $\Cinf(\M)$. 
\begin{lma} \label{lma:innerprod-forms}
Under the isomorphism $\Omega_D(\Cinf(\M)) \simeq \Omega(\M)$, the inner product $(\cdot,\cdot)_2$ coincides with $(\cdot,\cdot)_D$.
\end{lma}
\begin{proof}
Let $\omega_1, \omega_2$ be two forms in $\Omega_D(\Cinf(M))$, so that $L_\theta(\omega_i)$ are two generic forms in $\Omega_D(\Cinf(\M)) \isom L_\theta (\Omega_D(\Cinf(M))$. Then, using Lemma~\ref{lma:dix}
it follows that 
\be
\ncint L_\theta(\omega_1)^* L_\theta(\omega_2) = \ncint L_\theta(\omega_1^* \times_\theta \omega_2) = \ncint \omega_1^* \times_\theta \omega_2,
\ee
 Now, the inner product $(~,~)_D$ coincides with $(~,~)_2$ as defined by \eqref{eq:inner-product-forms} in the classical case -- under the above isomorphism $\Omega_D(\Cinf(M)) \isom \Omega(M)$; see for example \cite[VI.1]{C94}.  It follows that the above expression equals
\be
\ncint \ast(\omega_1^* \times_\theta (\ast \omega_2) ) = \ncint \ast_\theta(L_\theta(\omega_1)^*  (\ast_\theta L_\theta(\omega_2) ) ),
\ee
using Lemma~\ref{lma:dix} for forms once more, together with the defining property of $\ast_\theta$. 
\end{proof}

\begin{lma}
\label{lma:d-dstar}
The formal adjoint $\dd^*$ of $\dd$ with respect to the inner product $(\cdot,\cdot)_2$ -- {i.e.} so that $(\dd^* \alpha, \beta)_2=(\alpha,\dd \beta)_2$ -- is given on $\Omega^p(\M)$ by 
$$
\dd^* = (-1)^{m(p+1)+1} \ast_\theta \dd \ast_\theta
$$
\end{lma}
\begin{proof}
Just as in the classical case, this follows from Lemma~\ref{lma:stokes}, together with 
$$
\int_{\M} \omega = \ncint *_\theta \omega, \qquad \omega \in \Omega^m(\M),
$$
a result again established from the classical case using the mentioned analogue of Lemma~\ref{lma:dix} for forms. 
\end{proof}

\subsection{Local index formula on toric noncommutative manifolds}\label{se:liftoric}

For the toric noncommutative manifolds the local index formula of Connes and Moscovici \cite{CM95} -- that we shall use later on -- simplifies drastically. 

We recall the general form of the local index formula; and we limit ourself to the `even' case, the only relevant one for the present paper.
Suppose in general that $(\A,\cH,D,\gamma)$ is an even $p$-summable spectral triple with discrete and simple dimension spectrum.  For a projection $e \in M_N(A)$, the operator 
\[
D_e = e(D \ot \II_N)e
\] 
is a Fredholm operator, thought of as the Dirac operator with coefficient in the module determined by $e$. 
Then the local index formula of Connes and Moscovici \cite{CM95} provides a method to compute its index via the pairing of suitable cyclic cycles and cocycles. 

Let $C_*(\A)$ be the chain complex over the algebra $\A$; 
in degree $n$, $C_n(\A):=\A^{\otimes(n+1)}$. On this complex there are defined the Hochschild operator $b:C_n(\A)\to C_{n-1}(\A)$ and the boundary operator $B:C_n(\A)\to C_{n+1}(\A)$, satisfying $b^2=0, B^2=0, bB+Bb=0$; thus $(b+B)^2=0$. From general homological theory, one defines a bicomplex $CC_*(\A)$ by $CC_{(n,m)}(\A):= CC_{n-m}(\A)$ in bi-degree $(n,m)$. Dually, one defines $CC^\ast(\A)$ as functionals on $CC_*(\A)$, equipped with the dual Hochschild operator $b$ and coboundary operator $B$ (we refer to \cite{C94} and \cite{Lod92} for more details on this). 
\begin{thm}[Connes-Moscovici \cite{CM95}]\label{thm:cm} ~~
\begin{itemize}
\item[(a)] An even cocycle $\phi^*=\sum_{k\geq0} \phi^{2k}$ in $CC^*(\A)$, $(b+B)\phi^*=0$, defined by the following formul{\ae}. For $k=0$,
\bd
\phi^0(a):=\resz  ~z^{-1} \tr( \gamma a |D|^{-2z} ) ;
\ed
whereas for $k > 0$
\bd
\phi^{k}(a^0, \ldots, a^{2k}) := \sum_{\alpha} c_{k,\alpha} \resz ~ \tr \big( \gamma a^0 [D,a^1]^{(\alpha_1)} \cdots [D,a^{2k}]^{(\alpha_{2k})} |D|^{-2(|\alpha| +k + z)} \big) ,
\ed
with 
\bd
c_{k,\alpha}=(-1)^{|\alpha|} \Gamma(k+|\alpha|) \big( \alpha! (\alpha_1+1) (\alpha_1+\alpha_2+2) \cdots (\alpha_1+ \cdots +\alpha_{2k} + 2k)\big)^{-1} ,
\ed 
and $T^{(j)}$ denotes the j'th iteration of the derivation $T \mapsto [D^2,T]$. 
\item[(b)]
For $e \in K_0(\A)$, the Chern character $\chern_*(e) = \sum_{k\geq 0} \chern_k(e)$ is the even cycle in $CC_*(\A)$, $(b+B)\chern_*(e)=0$, defined by the following formul{\ae}. For $k=0$,
\bd
\chern_0(e):= \tr(e),
\ed
whereas for $k > 0$
\bd
\chern_k(e):=(-1)^k \frac{(2k)!}{k!}  \sum (e_{i_0 i_1}-\frac{1}{2}\delta_{i_0 i_1}) \otimes {e_{i_1 i_2} 
\otimes e_{i_2 i_3} \otimes \cdots \otimes e_{i_{2k} i_0} }.
\ed
\item[(c)] The index is given by the natural pairing between cycles and cocycles
\be
\ind(D_{e})=\langle \phi^\ast,\chern_\ast(e) \rangle .
\ee
\end{itemize}
\end{thm}

For toric noncommutative manifolds, the above local index formula simplifies drastically \cite{LS04}.
\begin{thm}
\label{prop:lif-theta}
For a projection $p \in M_N(\Cinf(\M))$, we 
have 
$$ \ind D_p = \resz z^{-1} \tr \bigg(\gamma p |D|^{-2z} \bigg)+
\sum_{k \geq 1} \frac{(-1)^k}{k} \resz \tr \bigg( \gamma \big(p-\frac{1}{2} \big) [D,p]^{2k} |D|^{-2(k+z)}
\bigg), 
$$ 
and now the trace $\tr$ comprises a matrix trace as well. 
\end{thm}
\begin{proof}
Recall that the quantization map $L_\theta$ preserves the spectral decomposition, for the toric action of $\tilde \bT^n$, of smooth operators (see eq.~\eqref{spectwist}). 
Then, once extended the deformed $\times_\theta$-product to $\Cinf(\M) \bigcup [D,\Cinf(\M)]$ -- unambiguously since $D$ is of degree 0 --  we write the local cocycles $\phi^{k}$ in Thm~\ref{thm:cm} in terms of the quantization map $L_\theta$:
\begin{multline}
\phi^{k}\big( L_\theta(f^0),L_\theta(f^1),\ldots, L_\theta(f^{2k}) \big) \\ 
 =  \resz \tr \Big( \gamma L_\theta \big( f^0 \times_\theta [D,f^1]^{(\alpha_1)} \times_\theta \cdots \times_\theta [D,f^{2k}]^{(\alpha_{2k})} \big) 
|D|^{-2(|\alpha| +k + z)}\Big) .
\end{multline}
Suppose now that $f^0,\ldots, f^{2k}\in C^\infty(M)$ are homogeneous of degree $r^0, \ldots, r^{2k}$ respectively, so that the operator 
$f^0 \times_\theta [D,f^1]^{(\alpha_1)} \times_\theta \cdots \times_\theta [D,f^{2k}]$ is a homogeneous element of degree $r=\sum_{j=0}^{2k} r^j$.  
It is in fact a multiple of $f^0 [D,f^1] \cdots  [D,f^{2k}]$ as it can be established by working out the $\times_\theta$-products.
Forgetting about this factor -- which is a power of the deformation parameter $\lambda$ -- we obtain from \eqref{twist} that
\be
L_\theta(f^0 \times_\theta [D,f^1]^{(\alpha_1)} \times_\theta \cdots \times_\theta [D,f^{2k}]) = f^0 [D,f^1] \cdots [D,f^{2k}] U(\half r \cdot \theta) .
\ee
Each term in the local index formula for $(C^\infty(\M),\cH,D)$ then takes the form
\bd
\resz  \tr \big( \gamma f^0 [D,f^1]^{(\alpha_1)} \cdots [D,f^{2k}]^{(\alpha_{2k})} |D|^{-2(|\alpha| +k + z)} U(s) \big)
\ed
for $s=\half r \cdot \theta \in \tilde \bT^n$. The appearance of the operator $U(s)$ here is a consequence of the close relation with the index formula for 
$\bT^n$-equivariant Dirac spectral triples. In \cite{CH97} 
an even dimensional compact spin manifold $M$ on which a (connected compact) Lie group $G$ acts by isometries was studied. The equivariant Chern character was defined as an equivariant version of the JLO-cocycle, the latter being an element in equivariant entire cyclic cohomology. The essential point is that  an explicit formula for the above residues was obtained. In the case of a $\bT^n$-action on $M$ this is 
\begin{multline}
\resz  \tr \big( \gamma f^0 [D,f^1]^{(\alpha_1)} \cdots [D,f^{2k}]^{(\alpha_{2k})} |D|^{-2(|\alpha| +k + z)} U(s) \big) \\
=\Gamma^\infty(|\alpha|+k) \lim_{t \to 0} t^{|\alpha|+k}  \tr \big( \gamma f^0 [D,f^1]^{(\alpha_1)} \cdots [D,f^{2k}]^{(\alpha_{2k})} e^{-tD^2} U(s) \big) ,
\end{multline}
for every $s \in \tilde \bT^n$. Moreover, from Thm 2 in \cite{CH97} the limit vanishes when $|\alpha|\neq0$. This finishes the proof of our theorem.
\end{proof}

By inserting the symbol $\pi$ for the algebra representation, 
the components of the Chern character are represented as operators on  the Hilbert space $\ch$ by explicit formul{\ae}, 
\be\label{pid}
\pi_D(\chern_k(e)) := (-1)^k \frac{(2k)!}{k!}  \sum (\pi(e_{i_0 i_1})-\frac{1}{2}\delta_{i_0 i_1}) [D, {\pi(e_{i_1 i_2})]
[D, \pi(e_{i_1 i_2})] \cdots [D,\pi(e_{i_{2k} i_0})] } ,
\ee
for $k>0$, while $\pi_D(\chern_0(e)) = \sum \pi(e_{i_0 i_0})$.

\section{The Hopf fibration on $\S^4$}
\label{se:hopf-fibration}
We now focus on the two noncommutative spheres $\S^4$ and $\Sk^7$ starting from the algebras $\A(\S^4)$ and $\A(\Sk^7)$ of polynomial functions on them. The latter algebra carries an action of the (classical) group $\SU(2)$ by automorphisms in such a way that its invariant elements are exactly the polynomials on $\S^4$. The anti-symmetric $2\times 2$ matrix $\theta$ is given by a single real number also denoted by $\theta$.  
On the other hand, the requirements that $\SU(2)$ acts by automorphisms and that  $\S^4$ makes the algebra of invariant functions, give the matrix $\theta'$ in terms of $\theta$ as well. This yields a one-parameter family of noncommutative Hopf fibrations $\Sk^7 \to \S^4$. Moreover, there is an inclusion of the differential calculi $\Omega(\S^4) \subset \Omega(\Sk^7)$, defined in Sect.~\ref{se:diff-calc}.

For each irreducible representation $W^\n:=\Sym^n(\C^2)$ of $\SU(2)$ we construct the noncommutative vector bundles $E^\n$ associated to the fibration $\Sk^7 \to \S^4$. These bundles are described by the $\Cinf(\S^4)$-bimodules of `equivariant maps from $\Sk^7$ to $W^\n$'. As expected, these modules are finite projective and we construct explicitly the projections $p_\n \in M_{4^n}(\A(\S^4))$ such that these modules are isomorphic to the image of $p_\n$ in $\A(\S^4)^{4^n}$. In the special case of the defining representation, we recover the basic instanton projection on the sphere $\S^4$ constructed in \cite{CL01}. 
Then, one defines connections $\nabla=p_\n \dd$ as maps from $\Gamma^\infty(\S^4,E^\n)$ to $\Gamma^\infty(\S^4,E^\n) \otimes_{\A(\S^4)} \Omega^1(\S^4)$. The corresponding connection one-form $A$ turns out to be valued in a representation of the Lie algebra $su(2)$. 
By using the projection $p_\n$, the Dirac operator with coefficients in the noncommutative vector bundle $E^\n$ is given by $D_{p_\n}:= p_\n D p_\n$. We compute its index by using the very simple form of the local index theorem of Connes and Moscovici \cite{CM95} in the case of toric noncommutative manifolds as obtained in Thm \ref{prop:lif-theta}.

Finally, we show that the fibration $\Sk^7 \to \S^4$ is a `not-trivial principal bundle with structure group $\SU(2)$'. This means that the inclusion $\A(\S^4) \into \A(\Sk^7)$ is a not-cleft Hopf-Galois extension \cite{KT81,Mon93}; in fact, it is a principal extension \cite{BH04}. We find an explicit form of the so-called strong connection \cite{Haj96} which induces connections on the associated bundles $E^\n$ as maps from $\Gamma^\infty(\S^4,E^\n)$ to $\Gamma^\infty(\S^4,E^\n) \otimes_{\A(\S^4)} \Omega_\un^1(\A(\S^4))$, where $\Omega_\un^*(\A(\S^4))$ is the universal differential calculus on $\A(\S^4)$. We show that these connections coincide with the Grassmann connections $\nabla =p_\n \dd$ on $\Omega(\S^4)$.

\subsection{Let us roll noncommutative spheres}\label{se:nc-spheres}
With $\theta$ a real parameter, the algebra $\A(\S^4)$ of polynomial functions on the sphere $\S^4$ is 
 generated by elements  $z_0=z_0^*, z_j, z_j^*$, $j=1,2$, subject to relations
\be\label{s4t}
z_\mu z_\nu = \lambda_{\mu\nu} z_\nu z_\mu, \quad  z_\mu z_\nu^* = \lambda_{\nu\mu} z_\nu z_\mu^*,
\quad z_\mu^* z_\nu^* = \lambda_{\mu\nu} z_\nu^* z_\mu^*, \quad \mu,\nu = 0,1,2 ,
\ee
together with the spherical relation $\sum_\mu z_\mu^* z_\mu=1$. The deformation parameters are given by $\lambda_{\mu\mu}=1$ and
\be
\lambda_{1 2} = \bar{\lambda}_{2 1} =: \lambda=e^{2\pi \ii \theta}, 
\quad \lambda_{j 0} = \lambda_{0 j } = 1, \quad j=1,2 ,
\ee
At $\theta=0$ one recovers 
the $*$-algebra of complex polynomial functions on the usual $S^4$.

The differential calculus $\Omega(\S^4)$ is generated as a graded differential $*$-algebra by the elements $z_\mu, z_\mu^*$ of degree 0 and elements $d z_\mu, d z_\mu^*$ of degree 1 satisfying the relations:
\be\label{rel:diff-S4theta}
\begin{aligned}
&dz_\mu dz_\nu+ \lambda_{\mu\nu} dz_\nu dz_\mu =0, \\
&z_\mu dz_\nu = \lambda_{\mu\nu} dz_\nu z_\mu,
\end{aligned}
\qquad
\begin{aligned}
&dz_\mu^* dz_\nu+ \lambda_{\nu\mu} dz_\nu dz_\mu^* =0 ,\\
&z_\mu^* dz_\nu = \lambda_{\nu\mu} dz_\nu z_\mu^* .
\end{aligned}
\ee
with $\lambda_{\mu\nu}$ as before. There is a unique differential $\dd$ on $\Omega(\S^4)$ such that  $\dd:z_\mu \mapsto d z_\mu$ and the involution on $\Omega(\S^4)$ is the graded extension of $z_\mu \mapsto z_\mu^*$.

\bigskip

With $\lambda'_{a b} = e^{2 \pi \ii \theta'_{ab}}$ and $(\theta'_{ab}=-\theta'_{ba})$ a real antisymmetric matrix, 
 the algebra $\A(\Sk^7)$ of polynomial functions on the sphere $\Sk^7$ is generated by elements  
$\psi_a, \psi_a^*$, $a=1,\dots,4$, subject to relations
\be\label{s7t}
\psi_a \psi_b = \lambda'_{a b} \psi_b \psi_a, \quad  \psi_a \psi_b^* = \lambda'_{b a} \psi_b^* \psi_a,
\quad \psi_a^*\psi_b^* = \lambda'_{a b} \psi_b^* \psi_a^* ,
\ee
and with the spherical relation $\sum_a \psi_a^* \psi_a=1$. 
The above is a deformation of the $*$-algebra of 
 complex polynomial functions on the sphere $S^7$. As before, a differential calculus $\Omega(\Sk^7)$ can be defined to be generated by the elements $\psi_a, \psi_a^*$ of degree 0 and elements $d \psi_a, d \psi_a^*$ of degree 1 satisfying relations similar to the ones in \eqref{rel:diff-S4theta}.

\subsection{The $\SU(2)$ principal fibration on $S^4$}
\label{se:modules}
We review the classical construction of the instanton bundle on $S^4$ taking the approach of \cite{Lnd00}. We generalize slightly and construct complex vector bundles on $S^4$ associated to all finite-dimensional irreducible representations of $\SU(2)$. Let 
\begin{align*}
S^7&:=\{\psi=(\psi_1,\psi_2, \psi_3,\psi_4)\in \C^4:|\psi_1|^2+|\psi_2|^2+|\psi_3|^2+|\psi_4|^2=1\}, \nn \\
S^4&:=\{z=(z_1,z_2,z_0)\in \C^2 \op \bR:z_1^* z_1 + z_2^* z_2 +z_0^2=1\}, \nn \\
~\nn \\
\SU(2)&:=\{ w \in GL(2,\C) : w^* w=w w^* =1, \det w=1\} \\ 
&= \left\{ w=\begin{pmatrix} w^1 & w^2 \\-\bar{w}^2 & \bar{w}^1\end{pmatrix} :
w^1 \bar{w}^1+w^2 \bar{w}^2=1 \right\}.
\end{align*}
The space $S^7$ carries a right $\SU(2)$-action, 
$S^7 \times \SU(2) \to S^7$, given on generators by
\bd
\big((\psi_1, -\psi_2^*, \psi_3, -\psi_4^*),w\big) \mapsto (\psi_1, -\psi_2^*, \psi_3, -\psi_4^*) \begin{pmatrix} w  & 0 \\ 0 &w \end{pmatrix}.
\ed
It might seem unnatural to define an action that mixes the $\psi$'s and $\psi^*$'s. However, this is only a more convenient labelling for the left action of $Spin(5)$ on $S^7$ as we will see later on (cf. eq.~\eqref{act7}).
The Hopf projection map is defined as a map $\pi(\psi)\mapsto(z)$ with
\begin{align*}
& z_0= \psi_1^* \psi_1+\psi_2^* \psi_2-\psi^*_3 \psi_3-\psi_4^* \psi_4 ,\\
& z_1 = 2(\psi_1 \psi^*_3+\psi_2^* \psi_4), \quad
z_2 = 2(-\psi_1^* \psi_4 +\psi_2 \psi_3^*),\nn
\end{align*}
and one computes that $z_1^*z_1 + z_2^*z_2 +z_0^2=(\sum_a \psi_a^* \psi_a)^2=1$. 

The finite-dimensional irreducible representations of $\SU(2)$ are labeled by a positive integer $n$ with 
$n+1$-dimensional representation space $W^\n \isom \Sym^n(\C^2)$. The space of smooth $\SU(2)$-equivariant maps from $S^7$ to $W^\n$ is defined by
\be
\label{eq:equivariant-S4}
C_{\SU(2)}^\infty(S^7, W^\n):=\{\varphi:  S^7 \to W^\n : \varphi( \psi \cdot w) = w^{-1} \cdot \varphi( \psi)\}.
\ee
It forms the $\Cinf(M)$-module of smooth sections of the associated vector bundle $S^7 \times_{\SU(2)} W^\n \to S^4$.
We will now construct projections $p_\n$ as $N\times N$ matrices taking values in $C^\infty(S^4))$, such that $\Gamma^\infty(S^4, E^\n):=p_\n C^\infty(S^4)^N $ is isomorphic to $C_{\SU(2)}^\infty(S^7, W^\n)$ as right $C^\infty(S^4)$-modules. As the notation suggests, $E^\n$ is the vector bundle over $S^4$ associated with the corresponding representation. Let us first recall the case $n=1$ from \cite{Lnd00} and then use this to generate the vector bundles for any $n$. The $\SU(2)$-equivariant maps from $S^7$ to $W^{(1)} \isom \C^2$ are of the form
\be \label{equivariantmaps}
\varphi_{(1)}(\psi) = \begin{pmatrix} \psi^*_1 \\ -\psi_2\end{pmatrix} f_1 +\begin{pmatrix} \psi_2^* \\ \psi_1\end{pmatrix} f_2 +\begin{pmatrix} \psi^*_3 \\ -\psi_4\end{pmatrix} f_3+ \begin{pmatrix} \psi_4^* \\ \psi_3\end{pmatrix} f_4,
				 \ee
where $f_1, \ldots,f_4$ are smooth functions that are invariant under the action of $\SU(2)$, {i.e.} they are functions on the base space $S^4$.

A description of the equivariant maps is given in terms of ket-valued functions $\ket{\xi}$ on $S^7$, which are then elements in the free module $\E:=\C^N \otimes C^\infty(S^7)$. The $C^\infty(S^7)$-valued hermitian structure on $\E$ given by $\langle \xi, \eta\rangle=\sum_b \xi^*_b \eta_b$ allows one to associate dual elements $\bra{\xi}\in \E^*$ to each $\ket{\xi}\in \E$ by $\bra{\xi}(\eta):= \langle \xi,\eta \rangle, ~\forall \eta\in\E$. 
If we define $\ket{\psi_1},\ket{\psi_2} \in \A(S^7)^4$ by
\be 
\ket{\psi_1}=(\psi_1 , \psi_2, \psi_3, \psi_4 )^\t,\quad 
\ket{\psi_2}=(-\psi_2^* , \psi^*_1, -\psi_4^*, \psi^*_3 )^\t,\nn
\ee
with $\t$ denoting transposition, the equivariant maps in \eqref{equivariantmaps} are given by 
\be\label{defphiuno}
\varphi_{(1)}(\psi) = \begin{pmatrix} \langle\psi_1|f\rangle \\ \langle\psi_2|f\rangle \end{pmatrix}, 
\ee
where $\ket{f} \in (C^\infty(S^4))^4 := \C^4 \otimes C^\infty(S^4)$. 
Since $\langle \psi_k | \psi_l \rangle =\delta_{kl}$ as is easily seen, we get a projection $p_{(1)}^2=p_{(1)}=p_{(1)}^*$ in $M_4(C^\infty(S^4))$ by 
\be
p_{(1)}=\ket{\psi_1} \bra{\psi_1} + \ket{\psi_2} \bra{\psi_2}.\nn
\ee
Indeed, by explicit computation we find a matrix with entries in $C^\infty(S^4)$ which is the limit of the projection \eqref{projection1} at $\theta=0$. Denoting the right $C^\infty(S^4)$-module $p_{(1)} (C^\infty(S^4))^4$ by $\Gamma^\infty(S^4, E^{(1)})$, we have that
\bd 
\Gamma^\infty(S^4, E^{(1)})  \isom  C_{\SU(2)}^\infty(S^7, \C^2), \quad 
\sigma_{(1)} = p_{(1)} \ket{f} \leftrightarrow \varphi_{(1)}, 
\ed
with $\varphi_{(1)}$ the generic equivariant map given in 
\eqref{defphiuno}. 
For a general representation, $\SU(2)$-equivariant maps from $S^7$ to $W^\n$ are of the form
\be
\label{defphi-classical}
\varphi_\n(\psi)  = \begin{pmatrix} \langle\phi_1|f\rangle \\ \vdots \\ \langle\phi_{n+1}|f\rangle \end{pmatrix},
\ee
where $\ket{f}\in C^\infty(S^4)^{4^n}$ and $\ket{\phi_k}$ is the completely symmetrized form of the tensor product $\ket{\psi_1}^{\otimes n-k+1} \otimes \ket{\psi_2}^{\otimes k-1}$ for $k=1,\ldots,n+1$, normalized to have norm 1. For example, for the adjoint representation $n=2$, we have
\bea
\ket{\phi_1}&:=& \ket{\psi_1} \otimes \ket{\psi_1},\nn \\
\ket{\phi_2}&:=& \frac{1}{\sqrt{2}} \big(\ket{\psi_1} \otimes \ket{\psi_2}+\ket{\psi_2} \otimes \ket{\psi_1}\big),\nn \\ 
\ket{\phi_3}&:=& \ket{\psi_2} \otimes \ket{\psi_2}.\nn
\eea
Since in general $\langle \phi_k | \phi_l\rangle=\delta_{kl}$, the matrix-valued function
$$
p_\n=\ket{\phi_1} \bra{\phi_1} + \ket{\phi_2} \bra{\phi_2}+\cdots  +\ket{\phi_{n+1}} \bra{\phi_{n+1}} \in M_{4^n}(C^\infty(S^4))
$$
is a projection with entries in $C^\infty(S^4)$, since each element
$(p_\n)_{ab}=\sum_k \ket{\phi_k}_a \bra{\phi_k}_b$ is $\SU(2)$-invariant as can be easily seen. We conclude that
\bd 
p_\n (C^\infty(S^4)^{4^n})  \isom  C_{\SU(2)}^\infty(S^7, W^\n), \quad 
\sigma_\n= p_\n \ket{f} \leftrightarrow \varphi_\n , 
\ed
with $\varphi_\n$ the generic equivariant map given in \eqref{defphi-classical}.

\subsection{The $\SU(2)$ principal fibration on $\S^4$}
\label{se:hopf}
Firstly, we remind that while there is a $\theta$-deformation of the manifold $S^3\isom \SU(2)$, to a sphere $\S^3$, on the latter there is no compatible group structure so that there is no $\theta$-deformation of the group $\SU(2)$ \cite{CD02}. Therefore, we must choose the matrix $\theta'_{\mu\nu}$ in such a way that the noncommutative $7$-sphere $\Sk^7$ carries a classical $\SU(2)$ action, which in addition is such that the subalgebra of $\A(\Sk^7)$ consisting of $\SU(2)$-invariant polynomials is exactly $\A(\S^4)$. 
The action of $\SU(2)$ on the generators of $\A(\Sk^7)$ is simply defined by
\be \label{def:actionSU2}
\alpha_w: (\psi_1, -\psi_2^*, \psi_3, -\psi_4^*) \mapsto (\psi_1, -\psi_2^*, \psi_3, -\psi_4^*) 
\begin{pmatrix} w & 0 \\ 0 & w
\end{pmatrix} , \qquad 
w = \begin{pmatrix} w^1 & w^2 \\ -\bar{w}^2 & \bar{w}^1 \end{pmatrix}.
\ee
Here $w^1$ and $w^2$, satisfying $w^1 \bar{w}^1+w^2 \bar{w}^2=1$, are the coordinate functions on $\SU(2)$.
By imposing that the map $w \mapsto \alpha_w$ embeds $\SU(2)$ in $\Aut(\A(\Sk^7))$ we find that $\lambda'_{12}=\lambda'_{34} =1$ and $\lambda'_{14}=\lambda'_{23}=\bar{\lambda'_{24}}=\bar{\lambda'_{13}}=:\lambda'$. 

The subalgebra of $\SU(2)$-invariant elements in $\A(\Sk^7)$ can be found in the following way. From the diagonal nature of the action of $\SU(2)$ on $\A(\Sk^7)$ and the above formul{\ae} for $\lambda'_{ab}$, we see that the action of $\SU(2)$ commutes with the action of $\bT^2$ on $\A(\Sk^7)$. Since $\A(\Sk^7)$ coincides with $\A(S^7)$ as vector spaces, we see that the subalgebra of $\SU(2)$-invariant elements in $\A(\Sk^7)$ is completely determined by the classical subalgebra of $\SU(2)$-invariant elements in $\A(S^7)$.
From Sect.~\ref{se:modules} we can conclude  that 
\be
\inv_{\SU(2)} (\A(S^7_\theta)) = \C [~ 1, \psi_1 \psi^*_3 +\psi_2^* \psi_4, -\psi_1^* \psi_4 +\psi_2 \psi_3^*, \psi_1 \psi^*_1+\psi_2^*\psi_2 ~]\nn
\ee
modulo the relations in the algebra $\A(\Sk^7)$. We identify 
\begin{align} \label{subalgebra}
z_0 &= \psi^*_1 \psi_1 + \psi^*_2 \psi_2 - \psi^*_3 \psi_3 - \psi^*_4 \psi_4 = 2(\psi^*_1 \psi_1 + \psi^*_2 \psi_2) -1 = 1 - 2(\psi^*_3 \psi_3 + \psi^*_4 \psi_4), \nn \\
z_1 &= 2 (\mu \psi_3^* \psi_1 + \psi^*_2 \psi_4) = 2 (\psi_1 \psi^*_3 + \psi^*_2 \psi_4), \nn \\
z_2 &= 2(- \mu \psi_4\psi^*_1 + \psi_2 \psi^*_3)= 2(- \psi^*_1 \psi_4 + \psi_2 \psi^*_3).
\end{align}
and compute that $z_1 z_1^* + z_2 z_2^* + z_0^2 = 1$. By imposing commutation rules 
$z_1 z_2 = \lambda  z_2 z_1$ and $z_1 z_2^* = \bar{\lambda} z_2^* z_1$
, we infer that $\lambda'_{14}=\lambda'_{23}=\bar{\lambda'_{24}}=\bar{\lambda'_{13}}=\sqrt{\lambda}=:\mu$ on $\Sk^7$. (Compatibility requires that $\mu^2=\lambda$; we 
drop the case $\mu = -\sqrt{\lambda}$ since its `classical' limit would 
correspond to `anti-commuting' coordinates.) We conclude that $\inv_{\SU(2)} (\A(\Sk^7)) = \A(\S^4)$ for a matrix of deformation parameters $\lambda'_{ab}=e^{2\pi\ii \theta'_{ab}}$ of the following form:
\be \label{def:lambda}
\lambda'_{ab}= 
\begin{pmatrix} 1 & 1 & \bar\mu & \mu \\ 
1 & 1 & \mu & \bar\mu \\
\mu &\bar\mu &1 & 1\\ 
\bar\mu & \mu &1 & 1 
\end{pmatrix}, \quad \mu = \sqrt{\lambda}, \qquad \mathrm{or} \qquad
\theta'_{ab}=\frac{\theta}{2}\begin{pmatrix} 0 & 0 & -1 & 1 \\ 
0 & 0 & 1 & -1 \\
1 & -1 & 0 & 0 \\ 
-1 & 1 & 0 & 0  \end{pmatrix}.
\ee
The relations \eqref{subalgebra} can be also expressed in the form, 
\begin{align}
\label{zg}
z_\mu =\sum_{ab}\psi^*_a(\gamma_\mu)_{ab}\psi_b,\qquad z_\mu^* = \sum_{ab}\psi^*_a(\gamma^*_\mu)_{ab}\psi_b,
\end{align}
with $\gamma_\mu$ the following twisted $4 \times 4$ Dirac matrices:
\begin{align}
\label{eq:dirac}
\gamma_0=\left(\begin{smallmatrix} 1 &&&\\ &1&&\\&&-1&\\&&&-1 \end{smallmatrix}\right),
\qquad
\gamma_1=2\begin{pmatrix} 0 & \begin{smallmatrix} 0 & 0 \\ 0 & 1 \end{smallmatrix} \\  \begin{smallmatrix} \mu & 0 \\ 0 & 0 \end{smallmatrix}& 0 \end{pmatrix}, \qquad
\gamma_2=2\begin{pmatrix} 0 & \begin{smallmatrix} 0 & -1 \\ 0 & 0 \end{smallmatrix} \\  \begin{smallmatrix} 0 & \bar\mu \\ 0 & 0 \end{smallmatrix}& 0 \end{pmatrix}.
\end{align}
Note that $\gamma_0$ is the usual grading given by 
$$
\gamma_0 = -\frac{1}{4}[\gamma_1,\gamma_1^*][\gamma_2,\gamma_2^*].
$$
These matrices satisfy the following relations of the twisted Clifford algebra \cite{CD02}:
\be \label{eq:clifford}
 \gamma_j \gamma_k +\lambda_{j k} \gamma_k\gamma_j=0,  
 \qquad \gamma_j \gamma_k^* +\lambda_{k j} \gamma_k^*\gamma_j=4\delta_{jk}, 
\quad j,k=1,2 .
\ee
There are compatible toric actions on $\S^4$ and $\Sk^7$. The torus $\bT^2$ acts on $\A(\S^4)$ as 
\be\label{eq:act-S4}
\sigma_s(z_0, z_1, z_2) = (z_0, e^{2\pi i s_1} z_1, e^{2\pi i s_2} z_2), \quad s\in\bT^2 .
\ee
This action is lifted to a double cover action on $\A(\Sk^7)$. The double cover map  $p: \tilde\bT^2 \to \bT^2$ is  
given explicitly by $p:(s_1,s_2) \mapsto (s_1+s_2,-s_1+s_2)$.
Then $\tilde\bT^2$ acts on the $\psi_a$'s as:
\be \label{eq:lift-S7}
\tilde\sigma: \left(  \psi_1, \psi_2, \psi_3, \psi_4 \right) 
\mapsto \Big(e^{2\pi i s_1}~\psi_1, ~e^{-2\pi i s_1}~\psi_2, ~e^{-2\pi i s_2}~\psi_3, ~e^{2\pi i s_2}~\psi_4 \Big) .
\ee
Equation \eqref{subalgebra} shows that $\tilde\sigma$ is indeed a lifting to $\Sk^7$ of the action of $\bT^2$ on $\S^4$. Clearly, this compatibility is built in the construction of the Hopf fibration $\Sk^7 \to \S^4$ as a deformation of the classical Hopf fibration $S^7 \to S^4$ with respect to an action of $\bT^2$, a fact that also dictated the form of the deformation parameter $\lambda'$ in \eqref{def:lambda}.
As we shall see, the previous double cover of tori comes from a spin cover $\Spin_\theta(5)$ of $\SO_\theta(5)$ deforming the usual action of $\Spin(5)$ on $S^7$ as a double cover of the action of $\SO(5)$ on $S^4$.

\subsection{The noncommutative instanton bundle}\label{se:ncinstbun}

There is a description of the instanton projection constructed in \cite{CL01} in terms of ket-valued polynomial functions  on $\Sk^7$. The latters are elements in the right $\A(\Sk^7)$-module $\E:=\C^4 \otimes \A(\Sk^7)=: \A(\Sk^7)^4$ with a $\A(\Sk^7)$-valued hermitian structure $\langle \xi, \eta\rangle=\sum_b \xi^*_b \eta_b$. To any $\ket{\xi}\in \E$ one associates its dual $\bra{\xi}\in \E^*$ by setting $\bra{\xi}(\eta):= \langle \xi,\eta \rangle$, $\forall \eta\in\E$. 

Similarly to the classical case (see Sect.~\ref{se:modules}), we define a $2 \times 4$ matrix $\Psi$ in terms of two ket-valued polynomials $\ket{\psi_1}$ and $\ket{\psi_2}$ by
\be 
\label{Psi}
\Psi = (\ket{\psi_1},\ket{\psi_2}) =
\begin{pmatrix} 
\psi_1 & -\psi_2^* \\ 
\psi_2 & \psi^*_1 \\
\psi_3 & -\psi_4^* \\
\psi_4 & \psi^*_3
\end{pmatrix}.
\ee
Then $\Psi^* \Psi = \I_2 $ so that the $4\times 4$-matrix, 
\be
p = \Psi \Psi^*=\ket{\psi_1}\bra{\psi_1}+\ket{\psi_2}\bra{\psi_2},\nn
\ee
is a projection, $p^2=p=p^*$, with entries in $\A(\S^4)$.
The action \eqref{def:actionSU2} becomes 
\be \label{actionSU2}
\alpha_w (\Psi) = \Psi w,
\ee
from which the invariance of the entries of $p$ follows at once. Explicitly one finds 
\be \label{projection1}
p=\frac{1}{2}\begin{pmatrix}
1+z_0 & 0 & z_1 & -\bar\mu z_2^* \\
0 & 1+z_0 & z_2 & \mu z_1^* \\
z_1^* & z_2^* & 1-z_0 & 0\\
-\mu z_2 & \bar\mu z_1 & 0 & 1-z_0 
\end{pmatrix}.
\ee
The projection $p$ is easily seen to be equivalent to the projection describing the instanton on $\S^4$ constructed in \cite{CL01}. Indeed, if one defines 
\be  
\ket{\widetilde{\psi}_1}=(\psi_1 , \psi_2, \psi_3, \mu \psi_4 )^\t , \quad 
\ket{\widetilde{\psi}_2}=(-\psi_2^* , \psi^*_1, -\psi_4^*, \mu\psi^*_3 )^\t , \nn
\ee
one obtains after a substitution $z_2 \mapsto -\bar\lambda z_2^*$ exactly the projection obtained therein: 
\be
\widetilde{p}=\frac{1}{2}\begin{pmatrix}
1+z_0 & 0 & z_1 & -\bar\lambda z_2^* \\
0 & 1+z_0 &z_2 & - z_1^* \\
z_1^*& z_2^* & 1-z_0 & 0\\
-\lambda z_2 &- z_1 & 0 & 1-z_0 
\end{pmatrix} \nn
\ee
One shows by direct calculation that the first component of the Chern Character -- defined in Thm~\ref{thm:cm} -- of $p$ vanishes,
\[
\chern_1(p) = 0. 
\]
It follows that $\chern_2(p)$ is a Hochschild cycle, i.e. $b \chern_2(p) = 0$, and play the role of the round volume form on $\S^4$. Indeed, it was shown in \cite{CL01} that with the isospectral geometry $(\Cinf(\S^4), D, \ch, \gamma_5)$, the image of $\chern_2(p)$ via the map $\pi_D$ in \eqref{pid}, satisfies the following quartic equation in $D$
\be
\pi_D(\chern_2(p)=3 \gamma_5. 
\ee
As we shall see in Sect.~\ref{se:basinst}, this means that the `bundle' $p$ has the correct `topological numbers'. In that section we shall also show that the canonical connection $\nabla=p\circ d$ has selfdual curvature. Hence the name instanton bundle for the module determined by (the class of) $p$. 

\bigskip

The projection \eqref{projection1} can be written in terms of the Dirac matrices defined in \eqref{eq:dirac}.
\begin{lma}
The matrices $\tilde\gamma_0:=\gamma_0$, $\tilde\gamma_1:=\bar\mu\gamma_1^t$ and $\tilde\gamma_2:=\mu\gamma_2^t$ satisfy the relations
\be
\label{eq:clifford-opp}
\tilde\gamma_j \tilde\gamma_k +\lambda_{k j} \tilde\gamma_k \tilde\gamma_j=0, \quad
\tilde\gamma_j \tilde\gamma_k^* +\lambda_{j k} \tilde\gamma_k^* \tilde\gamma_j=4\delta_{j k},
\qquad j,k=1,2 ,
\ee
and the above projection \eqref{projection1} can be expressed as
\be
p= \half \big (1+ \tilde\gamma_0 z_0 + \tilde \gamma_i z_i + \tilde \gamma^*_i z^*_i\big) . \nn
\ee
\end{lma}
\noindent
Note the difference of \eqref{eq:clifford-opp} with \eqref{eq:clifford} in the exchange of $\lambda_{j k}$ by $\lambda_{k j}$. 
\bigskip

We will denote the image of $p$ in $\A(\S^4)^4$ by $\Gamma^\infty(\S^4, E)=p \A(\S^4)^4$ which is clearly a right $\A(\S^4)$-module.
Another description of the module $\Gamma^\infty(\S^4,E)$ comes from considering `equivariant maps' from $\Sk^7$ to $\C^2$, in a way similar to the undeformed case in \eqref{eq:equivariant-S4}. 

Let $\rho: \SU(2) \times \C^2 \to \C^2, (w,v) \mapsto \rho(w) v = w \cdot v$, be the defining left representation of $\SU(2)$. A corresponding equivariant map from $\Sk^7$ to $\C^2$ is an element $\varphi \in \A(\Sk^7)\ot \C^2$, such that
\be 
\label{def:equiv}
(\alpha_w \otimes \id)(\varphi)=(\id \ot \rho(w)^{-1})(\varphi).
\ee
The collection of all such equivariant maps is denoted by $A(\Sk^7)\boxtimes_\rho \C^2$. It is a right $\A(\S^4)$-module (it is in fact a $\A(\S^4)$-bimodule) since multiplication by an element in $\A(\S^4)$ does not affect the equivariance condition \eqref{def:equiv}. 
Since $\SU(2)$ acts classically on $\A(\Sk^7)$, one sees from \eqref{actionSU2} that the equivariant maps are given by elements of the form $\varphi:=\Psi^* f$ for some $f \in \A(\S^4) \ot \C^4$. In terms of the canonical basis $\{e_1,e_2\}$ of $\C^2$, we can write $\varphi=\sum_k \braket{\psi_k}{f} \ot e_k$ for $\ket{f} = \ket{f_1,f_2,f_3,f_4}^\t$, with $f_a \in \A(\S^4)$. We then have the following isomorphism
\be 
\Gamma^\infty(\S^4, E) \isom \A(\Sk^7)\boxtimes_\rho \C^2 , \qquad
\sigma= p \ket{f} \leftrightarrow \varphi =\Psi^* f=\sum \braket{\psi_k}{f} \ot e_k .
\ee

\subsection{Associated modules and their properties}\label{se:associated-modules}

More generally, one can define the right $\A(\S^4)$-module $\Gamma^\infty(\S^4, E^\n)$ associated with any irreducible representation $\rho_\n:\SU(2) \to \GL(W^\n)$, with $W^\n = \Sym^n(\C^2)$ for a positive integer $n$. The module of $\SU(2)$-equivariant maps from $\Sk^7$ to $W^\n$ is defined as
\be
\A(\Sk^7) \boxtimes_{\rho_\n} W:= \big\{ \varphi \in \A(\Sk^7) \otimes W: (\alpha_w\otimes \id) (\varphi) =(\id \otimes \rho_\n(w)^{-1})(\varphi)\big\}.\nn
\ee
It is easy to see that these maps are of the form $\varphi_\n = \sum_k \braket{\phi_k}{f}\ot e_k$ on the basis $\{e_1, \ldots, e_{n+1}\}$ of $W^\n$ where now $\ket{f}\in \A(\S^4)^{4^n}$ and 
\be 
\ket{\phi_k} = \frac{1}{a_k} \ket{\psi_1}^{\otimes(n-k+1)} \otimes_S \ket{\psi_2}^{\otimes(k-1)}  \quad(k=1,\ldots,n+1),\nn
\ee
with $\otimes_S$ denoting symmetrization  and $a_k$ are suitable normalization constants. These vectors $\ket{\phi_k} \in \C^{4^n} \otimes \A(\Sk^7)=: \A(\Sk^7)^{4^n}$ are orthogonal (with the natural hermitian structure), and with $a_k^2=\binom{n}{k-1}$ they are also normalized. Then 
\be \label{def:proj}
p_\n:=\ket{\phi_1} \bra{\phi_1} + \ket{\phi_2} \bra{\phi_2}+\cdots + \ket{\phi_{n+1}} \bra{\phi_{n+1}} \in \Mat_{4^n}(\A(\S^4))
\ee
defines a projection $p^2=p=p^*$. That its entries are in $\A(\S^4)$ and not in $\A(\Sk^7)$ is easily seen. Indeed, much as it happens for the vector $\Psi$ in eq.~\eqref{actionSU2}, for every $i=1,\ldots,4^n$, the vector $\big( \ket{\phi_1}_i ,  \ket{\phi_2}_i, \ldots , \ket{\phi_{n+1}}_i \big)$ transforms under the action of $\SU(2)$ to the vector $\big( \ket{\phi_1}_i , \ldots , \ket{\phi_{n+1}}_i \big) \cdot \rho_{(n)} (w)$ so that each entry $\sum_k \ket{\phi_k}_i \bra{\phi_k}_j$ of $p_\n$ is $\SU(2)$-invariant and hence an element in $\A(\S^4)$. With this we proved the following.
\hyphenation{iso-mor-phic}
\begin{prop} \label{prop:modulesisomorphism}
The right $\A(\S^4)$-module of equivariant maps $\A(\Sk^7) \boxtimes_{\rho_\n} W^\n$ is isomorphic to the right $\A(\S^4)$-module $\Gamma^\infty(\S^4,E^\n):=p_\n (\A(\S^4)^{4^n})$ with the isomorphism given explicitly by:
\begin{eqnarray*}
\Gamma^\infty(\S^4,E^\n) &\isom&  \A(\Sk^7) \boxtimes_{\rho_\n} W^\n \\\nn
\sigma_\n= p_\n \ket{f} &\leftrightarrow& \varphi_\n= \sum_k \braket{\phi_k}{f}\ot e_k .
\end{eqnarray*}
\end{prop}
With the projections $p_\n$ one associates (Grassmann) connections on the right $\Cinf(\S^4)$-modules $\Gamma^\infty(\S^4, E^\n)$ in a canonical way:
\be \label{def:gras}
\nabla = p_\n \circ \dd : \Gamma^\infty(\S^4, E^\n) \to \Gamma^\infty(\S^4, E^\n) \otimes_{\A(\S^4)} \Omega^1(\S^4)
\ee
where $(\Omega^*(\S^4), \dd)$ is the differential calculus defined in the previous section. 
An expression for these connections as acting on coequivariant maps can be obtained using the above isomorphism and results in:
\be \label{def:grasconn}
\nabla (\phi_k) =\dd (\phi_k) + A_{kl} \phi_l
\ee
where $A_{kl}=\langle \phi_k | \dd \phi_l \rangle \in \Omega^1(\Sk^7)$. The corresponding matrix 
$A$ is called the connection one-form; it is clearly anti-hermitian, and it is valued in the derived representation space, $\rho'_n : su(2) \to \End(W^\n)$, of the Lie algebra $su(2)$.

\bigskip
Let us now discuss some properties of the associated modules, like hermitian structures and the structure of the algebra of endomorphisms on them.

We observe that one can lift the construction of the previous section to the smooth level by replacing polynomial algebras by their smooth completions as defined in \ref{se:toric-ncm}. 
Then, with $\rho$ any representation of $\SU(2)$ on an $n$-dimensional vector space $W$, the $\Cinf(\S^4)$ bimodule associated to $W$ is defined by
\be\label{equimaps7}
\E:=\Cinf(\Sk^7) \boxtimes_\rho W:= \big\{ \eta \in \Cinf(\Sk^7) \otimes W: (\alpha_w\otimes \id) (\eta) =(\id \otimes \rho(w)^{-1})(\eta)\big\}. 
\ee
Along the line of the above proof, the module $\E$ is a finite projective $\Cinf(\S^4)$ module. Note that the choice of a projection for a finite projective module requires the choice of one of the two (left or right) module structures. Similarly, the definition of a hermitian structure requires the choice of a left or right module structure. In the following, we will always work with the right structure for the associated modules.
A natural (right) Hermitian structure on 
$\Cinf(\Sk^7) \boxtimes_\rho W$, defined in terms of a fixed inner product of $W$ 
as, \be \label{def:hest} \langle \eta, \eta' \rangle := \sum_i \bar \eta_i 
\eta'_i. \ee where we denoted $\eta=\sum_i\eta_i \otimes e_i$ and $\eta'=\sum_i 
\eta'_i \otimes e_i $, given an orthonormal basis $\{e_i, \, i=1, \cdots, \dim 
W\}$ of $W$. One quickly checks that $\langle \eta, \eta' \rangle$ is an element 
in $\Cinf(\S^4)$, and that $\langle \mb , \mb \rangle$ satisfies all conditions 
of a right Hermitian structure.

The bimodules $\Cinf(\Sk^7) \boxtimes_\rho W$ are of the type described in 
Sect.~\ref{se:nc-vb}. The associated vector bundle $E=S^7 \times_\rho W$ 
on $S^4$ carries an action $V$ of $\tilde\bT^2$ induced from its action on $S^7$, 
which is obviously $\sigma$-equivariant. By the very definition of $\Cinf(\Sk^7)$ 
and of $\Gamma^\infty(\S^4,E)$ in Sect.~\ref{se:nc-vb}, it follows that 
$\Cinf(\Sk^7) \boxtimes_\rho W \isom \Gamma^\infty(\S^4,E)$. Indeed, from the 
undeformed isomorphism, $\Gamma^\infty(S^4,E)\isom \Cinf(S^7)\boxtimes_\rho W$, 
the quantization map $L_{\theta'}$ of $\Cinf(S^7)$, acting only on the first leg 
of the tensor product, establishes this isomorphism, \be 
\label{eq:quantization-vb} L_{\theta'} : \Cinf(S^7) \boxtimes_\rho W \to 
\Cinf(\Sk^7)\boxtimes_\rho W . \ee The above is well defined since the action of 
$\tilde\bT^2$ commutes with the action of $\SU(2)$. Also, it is such that 
$L_{\theta'}(f \lt_{\theta'} \eta)=L_{\theta'}(f) L_{\theta'} 
(\eta)=L_{\theta}(f) L_{\theta'} (\eta)$ for $f \in \Cinf(S^4)$ and $\eta \in 
\Cinf(S^7) \boxtimes_\rho W$, due to the identity $L_{\theta'}=L_{\theta}$ on 
$\Cinf(S^4)\subset\Cinf(S^7)$. A similar result holds for the action 
$\rt_{\theta'}$.

\begin{prop} \label{rem:associated-modules} 
The right $\Cinf(\S^4)$-modules 
$\Cinf(\Sk^7) \boxtimes_\rho W$ admits a homogeneous module basis 
\mbox{$\{e_\alpha, \, \alpha = 1, \cdots, N \}$} -- with a suitable $N$ -- that 
is, under the action $V$ of the torus $\tilde \bT^2$, its elements transform as, 
\be V_{s}(e_\alpha) =e^{2\pi i s\cdot r_{\alpha}}{e_\alpha} , \quad s\in \tilde 
\bT^2 . \ee with $r_\alpha \in \IZ^2$ the degree of $e_\alpha$. \end{prop} 
\begin{proof} The vector space $W$ is a direct sum of irreducible representation 
spaces of $\SU(2)$ and the module $\Cinf(\Sk^7) \boxtimes_\rho W$ decomposes 
accordingly. Thus, we can restrict to irreducible representations. The latter are 
labeled by an integer $n$ with $W\simeq\IC^{n+1}$.

Consider first the case $W=\C^2$. Using \cite[Prop. 2]{LS04}, a basis 
$\{e_1, \cdots, e_4\}$ of the right module $\Cinf(\Sk^7) \boxtimes_\rho \C^2$ is 
given by the columns of $\Psi^\dagger$ where $\Psi$ is the matrix in \eqref{Psi}: 
\be e_1 :=\left(\begin{smallmatrix} ~\psi^*_{1} \\ -\psi_{2} \end{smallmatrix} 
\right), \quad e_2 :=\left(\begin{smallmatrix} \psi^*_{2} \\ \psi_{1} 
\end{smallmatrix} \right), \quad e_3 :=\left(\begin{smallmatrix} ~\psi^*_{3} \\ 
-\psi_{4} \end{smallmatrix} \right), \quad e_4 :=\left(\begin{smallmatrix} 
\psi^*_{4} \\ \psi_{3} \end{smallmatrix} \right) . \ee Using the explicit action 
\eqref{eq:lift-S7} it is immediate to compute the corresponding degrees, \be 
r_1=(-1,0), \quad r_2=(1,0), \quad r_3=(0,1), \quad r_4=(0,-1). \ee More 
generally, with $W=\C^{n+1}$ a homogeneous basis $\{e_\alpha, \alpha = 1, \cdots, 
4^n\}$ for the right module $\Cinf(\Sk^7) \boxtimes_\rho \C^{n+1}$ can be 
constructed from the columns of a similar $(n+1)\times 4^n$ matrix 
$\Psi_\n^\dagger$ given in \cite{LS04}.  \end{proof} The above property allows us 
to prove a useful result for the associated modules. \begin{prop} 
\label{prop:tensor-modules} Let $\rho_1$ and $\rho_2$ be two finite dimensional 
representations of $\SU(2)$ on the vector spaces $W_1$ and $W_2$, respectively. 
There is the following isomorphism of right $\Cinf(\S^4)$-modules, $$ 
\left(\Cinf(\Sk^7) \boxtimes_{\rho_1} W_1 \right) \bar\otimes_{\Cinf(\S^4)} 
\left(\Cinf(\Sk^7) \boxtimes_{\rho_2} W_2 \right) \isom \Cinf(\Sk^7) 
\boxtimes_{\rho_1\otimes \rho_2} (W_1 \otimes W_2). $$ \end{prop} \begin{proof} 
Let $\{ e^1_\alpha, \, \alpha=1,\cdots, N_1 \}$ and $\{ e^2_\beta, \, 
\beta=1,\cdots, N_2\}$ be homogeneous bases for the right modules 
$\Cinf(\Sk^7)\boxtimes_{\rho_1} W_1$ and $\Cinf(\Sk^7) \boxtimes_{\rho_2} W_2$ 
respectively. Then, the right module $\Cinf(\Sk^7) \boxtimes_{\rho_1\otimes 
\rho_2} (W_1 \otimes W_2)$ has a homogeneous basis given by $\{ e^1_\alpha 
\otimes e^2_\beta \}$. We define a map $$ \phi: \left(\Cinf(\Sk^7) 
\boxtimes_{\rho_1} W_1 \right) \bar\otimes_{\Cinf(\S^4)} \left(\Cinf(\Sk^7) 
\boxtimes_{\rho_2} W_2 \right) \to \Cinf(\Sk^7) \boxtimes_{\rho_1\otimes \rho_2} 
(W_1 \otimes W_2) $$ by $$ \phi(e^1 _\alpha \times_\theta f^1_\alpha \otimes 
e^2_\beta \times_\theta f^2_\beta) = (e^1_\alpha \otimes e^2_\beta) \times_\theta 
\sigma_{r_\beta \theta}(f^1_\alpha) \times_\theta f^2_\beta, $$ with summation 
over $\alpha$ and $\beta$ understood. Here $r_\beta \in \IZ^2$ is the degree of 
$e^2_\beta$ under the action of $\tilde\bT^2$, so that $e^2_\beta \times_\theta 
\sigma_{r_\beta \theta} (f) = f \times_\theta e^2_\beta$ for any $f \in 
\Cinf(\S^4)$. Note that this map is well-defined since $$ \phi\left(e^1_\alpha 
\times_\theta f^1_\alpha \times_\theta f \otimes e^2_\beta \times_\theta 
f^2_\beta - e^1_\alpha \times_\theta f^1_\alpha \otimes f \times_\theta e^2_\beta 
\times_\theta f^2_\beta\right)=0. $$ Moreover, it is clearly a map of right 
$\Cinf(\S^4)$-modules. In fact, it is an isomorphism with its inverse given 
explicitly by $$ \phi^{-1} \left( (e^1_\alpha \otimes e^2_\beta) \times_\theta 
f_{\alpha\beta} \right) = e_1^1 \otimes (e^2_\beta \times_\theta f_{1\beta}) + 
\cdots + e_{N_1}^1 \otimes (e^2_\beta \times_\theta f_{N_1\beta}) $$ with 
$f_{\alpha\beta} \in \Cinf(\S^4)$. \end{proof}

\subsection{The adjoint bundle}\label{se:adjbundle}

Given a right $\Cinf(\S^4)$-module $\E$, its {\it dual module} is 
defined by \be \E' := \big\{ \phi: \E \to \Cinf(\S^4) : \phi(\eta f ) = 
\phi(\eta) f, \, \forall ~f \in\Cinf(\S^4) \big\}, \ee and is naturally a left 
$\Cinf(\S^4)$-module. In the case that $\E$ is also a left $\Cinf(\S^4)$-module, 
then $\E'$ is also a right $\Cinf(\S^4)$-module. If $\E:=\Cinf(\Sk^7) 
\boxtimes_\rho W$ comes from the $\SU(2)$-representation $(W,\rho)$, by using the 
induced dual representation $\rho'$ on the dual vector space $W'$ given by \be 
\big( \rho'(w) v'\big) (v) := v'\big(\rho(w)^{-1} v\big); \qquad \forall ~v' \in 
W', v \in W , \ee we have that \begin{align} \E' &\isom \Cinf(\Sk^7) 
\boxtimes_{\rho'} W' \nn \\ &:= \big\{ \phi \in \Cinf(\Sk^7) \otimes W': 
(\alpha_w \otimes \id) (\phi) =(\id \otimes \rho'(w)^{-1})(\phi), \, \forall ~w 
\in \SU(2) \big\} . \end{align} Next, let $L(W)$ denote the space of linear maps 
on $W$, so that $L(W)=W\otimes W'$. The adjoint action of $\SU(2)$ on $L(W)$ is 
the tensor product representation $\ad:=\rho \otimes \rho'$ on $W \otimes W'$. We 
define \begin{align} \Cinf(\Sk^7) \boxtimes_{\ad} L(W) := \big\{ & T \in 
\Cinf(\Sk^7) \otimes L(W) : \nn \\ & \qquad : (\alpha_w \otimes \id) (T) =(\id 
\otimes \ad(w)^{-1})(T), \, \forall ~w \in \SU(2) \big\} , \end{align} and write 
$T=T_{ij} \otimes e_{ij}$ with respect to the basis $\{ e_{ij} \}$ of $L(W)$ 
induced from the basis $\{e_i\}_{i=1}^{\dim W}$ of $W$ and the dual one 
$\{e'_i\}_{i=1}^{\dim W}$ of $W'$.

On the other hand, there is the endomorphism algebra \be \End_{\Cinf(\S^4)}(\E):= 
\big\{ T: \E \to \E : T(\eta f ) = T (\eta) f, \, \forall ~f \in\Cinf(\S^4) 
\big\}.  \ee We will suppress the subscript $\Cinf(\S^4)$ from $\End$ in the 
following. As a corollary to the previous Proposition, we have the following. 
\begin{prop} \label{prop:end} Let $\E:=\Cinf(\Sk^7) \boxtimes_\rho W$ for a 
finite-dimensional representation $\rho$. Then there is an isomorphism of 
algebras \[ \End(\E) \isom \Cinf(\Sk^7) \boxtimes_{\ad} L(W). \] \end{prop} 
\begin{proof} By Prop.~\ref{prop:tensor-modules}, we have that 
$\Cinf(\Sk^7) \boxtimes_{\ad} L(W) \equiv \Cinf(\Sk^7) \boxtimes_{\rho\otimes 
\rho'} (W \otimes W')$ is isomorphic to $\E \bar\otimes_{\Cinf(\S^4)} \E'$ as a 
right $\Cinf(\S^4)$-module. Since $\E$ is a finite projective 
$\Cinf(\S^4)$-module, there is an isomophism $\End(\E) \isom \E 
\bar\otimes_{\Cinf(\S^4)} \E'$, 
whence the result. 
\end{proof} We see that the algebra of endomorphisms of the module $\E$ can be 
understood as the space of sections of the noncommutative vector bundle 
associated to the adjoint representation on $L(W)$ -- exactly as it happens in 
the classical case. This also allows an identification of skew-hermitian 
endomorphisms $\End^s(\E)$ -- which were defined in general in \eqref{def:skew} 
-- for the toric deformations at hand. \begin{corl} There is an identification
 \[
 \End^s(\E) \isom \Cinf_\bR(\Sk^7) \boxtimes_\ad u(n),
 \]
  with $\Cinf_\bR(\Sk^7)$ denoting the subspace of self-adjoint elements in 
$\Cinf(\Sk^7)$ and $u(n)$ consists of skew-adjoint matrices in $M_n(\C) \isom 
L(W)$, with $n=\dim W$. \end{corl} \begin{proof} Note that the involution $T 
\mapsto T^*$ in $\End(\E)$ reads in components $T_{ij} \mapsto \bar{T_{ji}}$ so 
that, with the identification of Prop.~\ref{prop:end}, the space 
$\End^s(\E)$ is made of elements $X \in \Cinf(\Sk^7) \boxtimes_\ad L(W)$ 
satisfying $\bar{X_{ji}} = -X_{ij}$. Since any element in $\Cinf(\Sk^7)$ can be 
written as the sum of two self-adjoint elements, $X_{ij} = X_{ij}^\Re + i 
X_{ij}^\Im$, we can write $$ X=\sum_i X_{ii}^\Im\otimes i e_{ii}+ \sum_{i \neq j} 
X_{ij}^\Re \otimes (e_{ij} - e_{ji})  + X_{ij}^\Im \otimes (i e_{ij}+i e_{ji}) = 
\sum_a X_a \otimes \sigma^a, $$ with $X_a$ generic elements in $\Cinf_\bR(\Sk^7)$ 
and $\{\sigma^a, a=1,\ldots,n^2\} $ the generators of $u(n)$. \end{proof}

\begin{ex} \label{ex:instanton-bundle}
Let us return to the instanton bundle $\E= \Cinf(\Sk^7) \boxtimes_\rho \C^2$. In this case, $\End(\E) \isom \Cinf(\Sk^7) \boxtimes_\ad M_2(\C))$. Since the matrix algebra $M_2(\C)$ decomposes into the adjoint representation $su(2)$ and the trivial representation $\C$ and $\Cinf(\Sk^7 )\boxtimes_{\id} \C \isom \Cinf(\S^4)$, we conclude that 
\be
\End(\E) \isom \Gamma^\infty(\ad(\Sk^7)) \oplus \Cinf(\S^4),
\ee
where we have set $\Gamma^\infty(\ad(\Sk^7)):=\Cinf(\Sk^7) \boxtimes_\ad su(2)$. The latter $\Cinf(\S^4)$-bimodule will be understood as the space of (complex) sections of the adjoint bundle. It is the complexification of the traceless skew-hermitian endomorphism $\Cinf_\bR(\Sk^7) \boxtimes_\ad su(2)$.
\end{ex}

 \subsection{Index of twisted Dirac operators} 
\label{se:index}
In this section, we shall compute explicitly the index of the Dirac operator with coefficients in the bundles $E^\n$, that is the index of the operator 
of $D_{p_\n}:= p_\n (D \otimes \I_{4^n})p_\n$. We compute this index using the special form of the Connes Moscovici local index formula, as given in Thm \ref{prop:lif-theta}.
For the present case, the index of the Dirac operator on $\S^4$ twisted by a 
$p \in K_0(C(\S^4))$, is given by
\bea
\ind(D_{p}) &=& \langle\phi^*,\chern_*(p)\rangle \nn \\ 
&=& \resz z^{-1} \tr \big( \gamma \pi_D (\chern_0(p)) |D|^{-2z} \big) \nn \\
&& + \frac{1}{2!} \resz  \tr \big( \gamma \pi_D (\chern_1(p)) |D|^{-2-2z} \big) 
+ \frac{1}{4!} \resz  \tr \big( \gamma \pi_D (\chern_2(p)) |D|^{-4-2z} \big). \nn
\eea
The Chern character classes and their realization as operators are described in Sect.~\ref{se:liftoric}. In particular, 
$\pi_D$ represents the universal differential calculus  as operators by the map
\be
\pi_D : \Omega^p_\un(\A(\S^4)) \to \cB(\cH) , \quad 
a^0 \delta a^1 \cdots \delta a^p \mapsto a^0 [D,a^1] \cdots [D,a^p].\nn
\ee
We know from Sect.~\ref{se:diff-calc} that these operators are noncommutative forms (see eq.~\eqref{formdirac}) provided we quotient them by junk forms, that is operators of the kind in eq.~\eqref{formjunk}.
We will avoid a discussion on junk-forms and introduce instead a different quotient of $\Omega_\un(\A(\S^4))$. We define $\Omega_D(\S^4)$ to be $\Omega_\un(\A(\S^4))$ modulo the relations 
\begin{align}
&\alpha \delta \beta - \lambda (\delta \beta) \alpha = 0, \quad
(\delta \alpha) \beta - \lambda \beta \delta \alpha = 0, \nn \\
&\alpha \delta \beta^* - \bar \lambda (\delta \beta^*) \alpha = 0, \quad 
(\delta \alpha^*) \beta - \bar \lambda \beta \delta \alpha^* = 0, \nn \\
& a \delta x - (\delta x) a = 0 , \qquad \forall a \in \A(\S^4),\nn
\end{align}
avoiding the second order relations that define $\Omega(\S^4)$. One proves that the above relations are in the kernel of $\pi_D$: for instance, 
$\alpha [D,\beta]-\lambda [D,\beta] \alpha = 0$, so that $\pi_D$ is well-defined on $\Omega_D(\S^4)$. The differential calculus $\Omega_D(\Sk^7)$ is the quotient of $\Omega_\un(\A(\Sk^7))$ by only the relations in \eqref{rel:diff-Rm} of order one, that is by the relations:
\[
z_\mu \delta z_\nu= \lambda_{\mu\nu} (\delta z_\nu)z_\mu, \quad z_\mu \delta z_\nu^* = 
\lambda_{\nu\mu} \delta z_\nu^* z_\mu .  
\]
For a proof of the following result we refer to \cite{LS04}.
\begin{lma}
The images under $\pi_D$ of the Chern characters of the projections $p_\n$ are given by
\begin{eqnarray*}
\pi_D(\chern_0(p_\n))&=& n+1,\\
\pi_D(\chern_1(p_\n))&=&0,\\
\pi_D(\chern_2(p_\n))&=&\frac{1}{6} n(n+1)(n+2)\pi_D( \chern_2(p_{(1)})), \\
\end{eqnarray*}
up to the coefficients $\mu_k=(-1)^k \frac{(2k)!}{k!}$.
\end{lma}
\noindent
Combining this with the simple form of the index formula given above,
while taking the proper coefficients, we find that
\be
\ind(D_{p_\n}) = \frac{1}{4!} \frac{4!}{2!} \frac{1}{6} n(n+1)(n+2) \resz  \tr\big(\gamma_5 \pi_D(\chern_2(p_{(1)})) |D|^{-4-2z} \big)\nn
\ee
where the vanishing of the first term, -- the one involving $\pi_D(\chern_0(p_\n))$ -- follows from  the fact that $\ind{D}=0$ on $S^4$. Thm I.2 in \cite{CM95} allows one to express the residue as a Dixmier trace. Combining this with $\pi_D(\chern_2(p_{(1)}))=3 \gamma_5$ (as computed in \cite{CL01}), we obtain
\bd
3 \cdot \resz  \tr (|D|^{-4-2z} ) = 6 \cdot \dix( |D|^{-4} )= 2
\ed
since the Dixmier trace of $|D|^{-m}$ on the $m$-sphere equals $8/m!$ (cf. for instance \cite{GVF01,Lnd97}). This combines to give:
\begin{prop}
The index of the Dirac operator on $\S^4$ with coefficients in $E^\n$ is 
\bd
\ind(D_{p_\n}) =  \frac{1}{6} n(n+1)(n+2).
\ed
\rightbox
\end{prop}
\noindent Note that this coincides with the classical result.

\subsection{The noncommutative principal bundle structure} 
\label{se:hopf-galois}

In this section, we apply the general theory of Hopf-Galois extensions 
to the inclusion $\A(\S^4) \into \A(\Sk^7)$. As explained in Sect.~\ref{se:hg-general}, such extensions can be understood as noncommutative principal bundles. We will first dualize the construction of the previous section, {i.e.} replace the action of $\SU(2)$ on $\A(\Sk^7)$ by a coaction of $\A(\SU(2))$. Then, we show that $\A(\S^4) \into \A(\Sk^7)$ is a not-cleft (i.e. not-trivial) Hopf-Galois extension and compare the connections on the associated bundles, induced from the strong connection, with the Grassmann connection defined in Sect.~\ref{se:associated-modules}.

The action of $\SU(2)$ on $\A(\Sk^7)$ by automorphisms can be easily dualized to a coaction $\Delta_R: \A(\Sk^7) \to \A(\Sk^7) \otimes \A(\SU(2))$, where now $\A(\SU(2))$ is the unital complex $*$-algebra generated by $w^1, \bar{w}^1,w^2, \bar{w}^2$ with relation $w^1 \bar{w}^1+w^2 \bar{w}^2=1$. Clearly, $\A(\SU(2))$ is a Hopf algebra with comultiplication
\be
\Delta : \begin{pmatrix} w^1 &w^2\\-\bar{w}^2 & \bar{w}^1 \end{pmatrix} \mapsto  \begin{pmatrix} w^1 &w^2\\-\bar{w}^2 & \bar{w}^1 \end{pmatrix} \otimes  \begin{pmatrix} w^1 &w^2\\-\bar{w}^2 & \bar{w}^1 \end{pmatrix},\nn
\ee
antipode $S(w^1) = \bar{w}^1, S(w^2)= -w^2$ and counit $\epsilon(w^1)=\epsilon(\bar{w}^1)=1, \epsilon(w^2)=\epsilon(\bar{w}^2)=0$. The coaction of $\A(\SU(2))$ on $\A(\Sk^7)$ is given by
\be 
\Delta_R: (\psi_1, -\psi_2^*, \psi_3, -\psi_4^*) \mapsto (\psi_1, -\psi_2^*, \psi_3, -\psi_4^*) \otimes \begin{pmatrix} w^1 & w^2 & 0 & 0 \\ -\bar{w}^2 & \bar{w}^1  & 0 & 0 \\ 0 & 0 & w^1 & w^2 \\ 0 & 0 & -\bar{w}^2 & \bar{w}^1 \end{pmatrix}.\nn
\ee
The algebra of coinvariants in $\A(\Sk^7)$, which consists of elements $z\in \A(\Sk^7)$ satisfying $\Delta_R(z)= z \otimes 1$, can be identified with $\A(\S^4)$ for the particular values of $\theta'_{ij}$ found before, in the same way as in Sect.~\ref{se:hopf}.

\begin{thm}
\label{thm:hopfgalois}
The inclusion $\A(\S^4) \into \A(\Sk^7)$ is a not-cleft faithfully flat $\A(\SU(2))$-Hopf-Galois extension. 
\end{thm}
\begin{proof}
Since $\A(\SU(2))$ is cosemisimple,
from the general considerations of Sect.~\ref{se:hg-general}, 
we simply need to prove that $ 1\ot  h$ is in the image of $\chi$, for $h$ a generator of $\A(\SU(2))$. But it is straightforward to check that using the ket-valued polynomials in \eqref{Psi} we have
\bean
&\chi(\sum_a \bra{\psi_1}_a \ot_{\A(\S^4)} \ket{\psi_1}_a)=1\otimes w^1, 
& \quad \chi(\sum_a \bra{\psi_1}_a \ot_{\A(\S^4)} \ket{\psi_2}_a)=1\otimes w^2, \nn \\ 
&\chi(\sum_a \bra{\psi_2}_a \ot_{\A(\S^4)} \ket{\psi_1}_a)=-1\otimes \bar{w}^2,  & \quad \chi(\sum_a \bra{\psi_2}_a \ot_{\A(\S^4)} \ket{\psi_2}_a)=1\otimes \bar{w}^1.\nn
\eean
Non-cleftness is a simple consequence of the nontriviality of the Chern characters of the projection $p_\n$ as seen in Sect.~\ref{se:index}. Indeed, this implies that the associated modules are nontrivial, which would not be true were the extension cleft. 
\end{proof}
The existence of a strong connection follows from general properties. However, one can easily write an explicit expression in terms of the inverse of the canonical map. If we denote the latter when lifted to $P \otimes P$ by $\tau$ it follows that the map $\ell:H \to P\otimes P$ defined by $\ell(h) = \tau(1\otimes h)$ 
enjoys the properties in \eqref{ell}, hence determines a strong connection. Furthermore, this map 
satisfies recursive relations similar to the one for $\chi^{-1}$: if $\ell(h)=h_l \otimes h_l'$ and $\ell(g)=g_k \otimes g_k'$, then 
\be \label{recursive}
\ell(hg)=g_k h_l \otimes h_l' g_k' .
\ee
As said in the Sect.~\ref{se:hg-general}, such a map $\ell:H \to P\otimes P$ yields a strong connection. 
\begin{prop}
On the Hopf-Galois extension $\A(\S^4) \into \A(\Sk^7)$, the following formul{\ae} on the generators of 
$\A(\SU(2))$, 
\bea \label{def:lGenerators}
&\ell(w^1) = \sum_a \bra{\psi_1}_a \otimes \ket{\psi_1}_a, \quad \ell(w^2)= \sum_a \bra{\psi_1}_a \otimes \ket{\psi_2}_a,& \\ \nn
&\ell(\bar{w}^2)=-\sum_a \bra{\psi_2}_a \otimes \ket{\psi_1}_a, \quad \ell(\bar{w}^1)=\sum_a \bra{\psi_2}_a \otimes \ket{\psi_2}_a.&  \nn
\eea
define a strong connection.
\end{prop}
\noindent
\begin{proof}
The expressions \eqref{def:lGenerators} are extended to the all of $\A(\SU(2))$ by the recursive formula \eqref{recursive}. Recall the usual vector basis $\{ r^{klm} : k \in \Z, m,n \geq 0\}$ in $\A(\SU(2))$ given by
\be
r^{klm}:=\left\{ \begin{array}{ll} (-1)^n (w^1)^k(w^2)^m(\bar{w}^2)^n & k \geq 0,\\
(-1)^n (w^2)^m(\bar{w}^2)^n (\bar{w}^1)^{-k} & k<0. \end{array} \right.
\ee
The recursive expressions on this basis are explicitly given by
\bea \nn \label{recursiveEll}
\ell(r^{k+1,mn}) &=& \psi^*_1 \ell(r^{kmn}) \psi_1 +\psi_2^* \ell(r^{kmn})\psi_2+\psi^*_3 \ell(r^{kmn}) \psi_3+\psi_4^* \ell(r^{kmn}) \psi_4,  \quad k\geq 0, \\ \nonumber
\ell(w^{k-1,mn}) &=&  \psi_2 \ell(r^{kmn}) \psi_2^*+\psi_1 \ell(r^{kmn}) \psi^*_1 +\psi_4 \ell(r^{kmn}) \psi_4^*+ \psi_3 \ell(r^{kmn}) \psi^*_3,  \quad k <0, \\ \nonumber
\ell(w^{k,m+1,n}) &=&  -\psi^*_1 \ell(r^{kmn}) \psi_2^*+\psi_2^* \ell(r^{kmn}) \psi^*_1 -\psi^*_3 \ell(r^{kmn}) \psi_4^* +\psi_4^* \ell(r^{kmn}) \psi^*_3, \\
\ell(w^{km,n+1}) &=& \, -\psi_2 \ell(r^{kmn}) \psi_1 +\psi_1 \ell(r^{kmn}) \psi_2-\psi_4 \ell(r^{kmn}) \psi_3+\psi_3 \ell(r^{kmn}) \psi_4, 
\eea
while setting $\ell(1)=1 \otimes 1$. One systematically prove that the map $\ell$ defined by the above recursive relations indeed satisfies all conditions of a strong connection. 
\end{proof}

The associated modules $\Gamma^\infty(\S^4,E^\n)$ are described in the following way. Given an irreducible corepresentation of $\A(\SU(2))$, $\rho_\n : W^\n \to \A(\SU(2))\otimes W^\n$, we  denote $\rho_\n(v)=v_{(0)} \otimes v_{(1)}$. Then, the the associated (right $\A(\S^4)$-)modules of coequivariant maps $\Hom^{\rho_\n} (W^\n, \A(\Sk^7))$ consists of maps $\varphi: W^\n \to \A(\Sk^7)$ satisfying
\[ 
\varphi( v_{(1)}) \otimes S v_{(0)}=\Delta_R \varphi( v),  \quad v \in W^\n. 
\]
Again, such maps are $\C$-linear maps written -- on the basis $\{e_i\}_{i = 1, \cdots, n+1}$ of $W^\n$ -- as $\varphi_\n(e_k) = \langle\phi_k|f\rangle$, in the notation of the previous section. With the projections defined in 
eq.~\eqref{def:proj}, proposition \ref{prop:modulesisomorphism} translates straightforwardly into an isomorphism 
\[
\Hom^{\rho_\n} (W^\n, \A(\Sk^7)) \isom p_\n (\A(\S^4))^{4^n} \isom \A(\Sk^7) \boxtimes_{\rho_\n} W^\n  
\]
with a slight abuse of notation -- $\rho_\n$ being the corepresentation of $\A(\SU(2))$ in the last expression and the representation of $\SU(2)$ on the first expression.

As explained in Sect.~\ref{se:hg-general}, the strong connection on the extension $\A(\S^4) \into \A(\Sk^7)$ induces connections on $\Hom^{\rho_\n} (W^\n, \A(\Sk^7))$, with respect to the universal calculus. 
Recall that for $\varphi \in \Hom^{\rho_\n} (W^\n, \A(\Sk^7))$, we have
\be
\nabla_\omega (\varphi)(v) \mapsto \delta \varphi(v) + \omega(v_{(0)}) 
\varphi(v_{(1)}).\nn
\ee
which is in fact a map
$$\nabla_\omega: \Hom^{\rho_\n} (W^\n, \A(\Sk^7)) \to \Hom^{\rho_\n} (W^\n, \Omega_\un^1(\A(\Sk^7))).
$$ 
It turns out that the connection one-form $\omega$ coincides with the connection one-form $A$ of eq.~\eqref{def:grasconn}, on the quotient $\Omega^1(\Sk^7)$ of $\Omega^1_\un(\A(\Sk^7))$. More precisely, let $\{e^\n_k\}$ be a basis of $W^\n$, and $e^\n_{kl}$ the corresponding matrix coefficients of $\A(\SU(2))$ in the representation $\rho_\n$. An explicit expression for $\omega(e^{(n)}_{kl})$ can be obtained from eqs.~\eqref{recursiveEll}; for example $\omega(e^{(1)}_{kl})=\braket{\psi_k}{\delta \psi_l}, \, k,l=1,2$.
Writing $\braket{\phi_k}{\dd \phi_l}$ in terms of $\braket{\psi_k}{\dd \psi_l}$, this leads indeed to 
$$\pi(\omega(e^\n_{kl}))=A^\n_{kl}= \braket{\phi_k}{\dd \phi_l},$$ 
where $\pi: \Omega_\un (\A(\Sk^7)) \to \Omega(\Sk^7)$ is the quotient map.

\section{Yang-Mills theory on toric manifolds}\label{se:ym} 

We now introduce the Yang-Mills action functional together with the corresponding equations of motion. We will show how instantons naturally arise. We first work out in detail the case of $\S^4$ and then sketch the theory on a generic four dimensional manifold.

\subsection{Yang-Mills theory on $\S^4$} \label{se:gt4sphere}

Before we proceed we recall the noncommutative spin structure $(\Cinf(\S^4),\ch,D,\gamma_5$) of $\S^4$ with $\ch=L^2(S^4,\cs)$ the Hilbert space of spinors, $D$ the undeformed Dirac operator, and $\gamma_5$ -- the even structure -- the fifth Dirac matrix.

Let $\E=\Gamma^\infty(\S^4,E)$ for some $\sigma$-equivariant vector bundle $E$ on $S^4$, so that there exists a projection $p \in M_N(\Cinf(\S^4))$ such that $\E\isom p (\Cinf(\S^4)^N$. Recall from Sect.~\ref{se:connections} that a connection $\nabla$ on $\E=\Gamma^\infty(\S^4,E)$ for some vector bundle $E$ on $S^4$, is a map from $\E$ to $\E\otimes \Omega(\S^4)$. The Yang-Mills action functional is defined in terms of the curvature of a connection on $\E$, which is an element in $\Hom_{\Cinf(\S^4)} (\E, \E \otimes \Omega^2(\S^4) )$; equivalently, it is an element in $\End_{\Omega(\S^4)}(\E\ot \Omega(\S^4))$ of degree 2 (see Sect.~\ref{se:connections}). We define an inner product on the latter algebra as follows \cite[III.3]{C94}. An element $T \in \End_{\Omega(\S^4)}(\E\ot \Omega(\S^4))$ of degree $k$ can be understood as an element in $p M_N(\Omega^k(\S^4))p$, since $\E\ot \Omega(\S^4)$ is a finite projective module over $\Omega(\S^4)$. A trace over internal indices together with the inner product defined in \eqref{eq:inner-product-forms}, defines the inner product $(\cdot,\cdot)_2$ on $\End_{\Omega(\S^4)}(\E\ot \Omega(\S^4))$. In particular, we can give the following definition.
\begin{defn}
\label{def:YM}
The {\rm Yang-Mills action functional} on the collection $C(\ce)$ of compatible connections $\nabla$ on the module $\E$ is defined by
$$
\YM(\nabla)=\big( F,F\big)_2 = \ncint *_\theta \tr (F *_\theta F),
$$
for any connection $\nabla$ with curvature $F$.
\end{defn}
\noindent
Recall from Sect.~\ref{se:connections} that gauge transformations are given by unitary endomorphisms $\cU(\E)$ of the module $\E$. 
\begin{lma} 
The Yang-Mills action functional is gauge invariant, positive and quartic.
\end{lma}
\begin{proof} 
From eq.~\eqref{ugcur}, under a gauge transformation $u \in \cU(\E)$ the curvature $F$ transforms as $F \mapsto u^* F u$. Since $\cU(\E)$ can be identified with the unitary elements in $p M_N(\A) p$, it follows that 
\begin{align*}
\YM(\nabla^u)=\ncint \sum_{i,j,k,l} *_\theta (\bar{u_{ji}} F_{jk} *_\theta F_{kl} ,u_{li} )=\YM(\nabla)
\end{align*}
using the tracial property of the Dixmier trace and the fact that $u_{li} \bar{u_{ji}}=\delta_{lj}$. \\
Positiveness of the Yang-Mills action functional follows from Lemma~\ref{lma:innerprod-forms} giving
$$
(F,F)_2=(F_D,F_D)_D=\ncint F_D^* F_D,
$$
which is clearly positive.
\end{proof}

The Yang-Mills equations (equations for critical points) are obtained from the Yang-Mills action functional by a variational principle. Let us describe how this principle works in our case. We consider a linear perturbation $\nabla_t=\nabla + t \alpha$ of a connection $\nabla$ on $\E$ by an element $\alpha \in \Hom(\E,\E\ot_{\Cinf(\S^4)} \Omega^1(\S^4)) $. The curvature $F_t$ of $\nabla_t$ is readily computed as $F_t = F + t [\nabla, \alpha] + \cO (t^2)$. If we suppose that $\nabla$ is an extremum of the Yang-Mills action functional, this linear perturbation should not affect the action. In other words, we should have
\be 
\frac{\partial}{\partial t}\bigg|_{t=0} \YM(\nabla_t)=0.
\ee
If we substitute the explicit formula for $F_t$, we obtain
\be
\big( [\nabla, \alpha],F \big)_2 + \bar{\big( [\nabla, \alpha] , F\big)_2} = 0.
\ee
using properties of  the complex scalar product $(\cdot,\cdot)_2$ on $\Hom(\E,\E\ot \Omega(\S^4))$. Then, its positive definiteness implies that $(F_t,F_t)=\bar{(F_t,F_t)}$, which when differentiated with respect to $t$, at $t=0$, yields $\big( [\nabla, \alpha],F \big)_2 = \bar{\big( [\nabla, \alpha] , F\big)_2}$; hence, $( [\nabla, \alpha],F )_2=0$. Since $\alpha$ was arbitrary, we derive the following equations of motion
\be
[\nabla^*, F\big] = 0.
\ee
where the adjoint of $[\nabla,\cdot~ ]$ is defined with respect to the scalar product by
\be
\big( [\nabla^*,\alpha], \beta \big)_2 = \big( \alpha, [\nabla, \beta] \big)_2 
\ee
for $\alpha \in \Hom(\E,\E \ot \Omega^3(\S^4))$ and $\beta \in \Hom(\E,\E\ot \Omega^1(\S^4))$.
From Lemma~\ref{lma:d-dstar}, it follows that $[\nabla^* ,F]= \ast_\theta [\nabla, \ast_\theta F]$, and the equations of motion can also be written as the more familiar \textit{Yang-Mills equations}:
\be \label{eq:ym}
[\nabla, \ast_\theta F]=0.
\ee
Note that connections with a self-dual or antiself-dual curvature $\ast_\theta F= \pm F$ are special solutions of the Yang-Mills equation. Indeed, in this case the latter equation follows directly from the Bianchi identity $[\nabla,F]=0$, given in Prop. \ref{ubianchi}.

\bigskip
We will next establish a relation between the Yang-Mills action functional and the so-called topological action \cite[VI.3]{C94} on $\S^4$. Suppose $\E$ is a finite projective module over $\Cinf(\S^4)$ defined by a projection $p \in M_N(\Cinf(\S^4))$. The topological action for $\E$ is given by teh pairing between the class of $p$ in K-theory and the cyclic cohomology of $\Cinf(\S^4)$. For computational purposes, we give the following definition in terms of the curvature of a connection on the module $\E$
\begin{defn}\label{topact}
Let $\nabla$ be a connection on the module $\E$ with curvature $F$. The {\rm topological action} is given by 
$$
\Top(\E)= (F,\ast_\theta F)_2= \ncint \ast_\theta \tr(F^2)
$$
where the trace is taken over internal indices, and in the second equality we have used the identity $\ast_\theta\circ\ast_\theta=\id$ on $\S^4$.
\end{defn}
Let us show that $\Top(\E)$ does not depend on the choice of a connection on $\E$. Since two connections differ by an element $\alpha$ in $\Hom_{\Cinf(\S^4)}(\E,\E\otimes \Omega^1(\S^4))$, we have to establish that $(F',\ast_\theta F')_2=(F,\ast_\theta F)_2$ where $F'=F+t [\nabla,\alpha] + \cO(t^2)$ is the curvature of $\nabla':=\nabla + t \alpha$, $t \in \bR$ . By definition of the inner product $(\cdot,\cdot)_2$ we then have
\begin{align*}
(F', \ast_\theta F')_2 - (F, \ast_\theta F)_2 &= t (F, \ast_\theta [\nabla, \alpha])_2+ t([\nabla,\alpha],\ast_\theta F)_2 +\cO(t^2) \nn \\
&= t(F, [\nabla^*,*_\theta \alpha])_2+ t([\nabla^*,*_\theta\alpha],F)_2 +\cO(t^2),
\end{align*}
which vanishes due to the Bianchi identity $[\nabla,F]=0$.

The Hodge star operator $\ast_\theta$ splits $\Omega^2(\S^4)$ into a self-dual and antiself-dual space,
\be
\Omega^2(\S^4)=\Omega_+^2(\S^4) \op \Omega_-^2(\S^4).
\ee
In fact, $\Omega_\pm^2(\S^4)=L_\theta \left(\Omega_\pm^2(S^4) \right)$. This decomposition is orthogonal with respect to the inner product $( \cdot,\cdot)_2$ which follows from the property $(\alpha,\beta)_2=\bar{(\beta,\alpha)_2}$, so that we can write the Yang-Mills action functional as
\be
\YM(\nabla) =  \big( F_+,F_+\big)_2 + \big( F_-,F_-\big)_2.
\ee
Comparing this with the topological action,
\be 
\Top(\E)= \big( F_+,F_+\big)_2 - \big( F_-,F_-\big)_2,
\ee
we see that $\YM(\nabla) \geq |\Top(\E)|$, with equality holding iff 
\be
\ast_\theta F = \pm F. 
\ee
Solutions of these equations are called instantons. We conclude that instantons are absolute minima of the Yang-Mills action functional. 

\subsection{On a generic four dimensional $\M$} \label{se:general}

We shall briefly describe how the just constructed Yang-Mills theory on $\S^4$ can be generalized to any four-dimensional toric noncommutative manifold $\M$. 

With $G$ a compact semisimple Lie group, let $P \to M$ be a principal $G$ bundle on $M$. We take $M$ to be a compact four-dimensional Riemannian manifold equipped with an isometrical action $\sigma$ of the torus $\bT^2$. For the construction to work, we assume that this action can be lifted to an action $\tilde\sigma$ of a cover $\tilde\bT^2$ on $P$, while it commutes with the action of $G$. As in Sect.~\ref{se:toric-ncm}, we define the noncommutative algebras $\Cinf(\P)$ and $\Cinf(\M)$ as the vector spaces $\Cinf(P)$ and $\Cinf(M)$ with star products defined like in \eqref{eq:star-product} with respect to the action of $\tilde \bT^2$ and $\bT^2$ respectively; or, equivalently as the images of $\Cinf(P)$ and $\Cinf(M)$ under the corresponding quantization map $L_\theta$. 
Since the action of $\tilde \bT^2$ commutes with the action of $G$ on $P$, the corresponding  action $\alpha$ of $G$ on the algebra $\Cinf(P)$ by 
\be
\alpha_g(f)(p)=f(g^{-1}\cdot p) ,
\ee
induces an action of $G$ by automorphisms on the algebra $\Cinf(\P)$. This means that also the inclusion $\Cinf(M) \subset \Cinf(P)$ as $G$-invariant elements in $\Cinf(P)$ extends to an inclusion $\Cinf(\M) \subset \Cinf(\P)$ of $G$-invariant element in $\Cinf(\P)$. Clearly, the action of $G$ translates trivially into a coaction of the Hopf algebra $\Cinf(G)$ on $\Cinf(\P)$.
\begin{prop}
The inclusion $\Cinf(\M) \into \Cinf(\P)$ is a (principal) Hopf-Galois $\Cinf(G)$ extension.
\end{prop}
\begin{proof}
As in \cite{LS04}, it is enough to establish surjectivity of the canonical map
\begin{align*}
\chi : \Cinf(\P) \ot_{\Cinf(\M)} \Cinf(\P) &\to \Cinf(\P) \otimes \Cinf(G), \\
f' \otimes_{\Cinf(\M)} f &\mapsto f' \Delta_R(f) = f' f_{(0)} \otimes f_{(1)};
\end{align*}
all additional nice properties would then follow from the cosemisimplicity of the Hopf algebra $\Cinf(G)$. Now, for the undeformed case, the bijectivity of the canonical map $\chi^\class:\Cinf(P)\ot_{\Cinf(M)} \Cinf(P) \to \Cinf(P) \otimes \Cinf(G)$ follows by the very definition of a principal bundle. Furthermore, there is an isomorphism of vector spaces:
\begin{align*}
T:\Cinf(\P) \ot_{\Cinf(\M)} \Cinf(\P) &\to \Cinf(P)\ot_{\Cinf(\M)} \Cinf(\P) \\
f' \ot_{\Cinf(\M)} f &\mapsto \sum f'_r \ot_{\Cinf(M)} \tilde\sigma_{r\theta} (f)
\end{align*}
where $f'=\sum_r f'_r$ is the homogeneous decomposition of $f'$ under the action of $\tilde \bT^2$. We claim that the canonical map is given as the composition $\chi = \chi^\class \circ T$; hence, it is bijective. Indeed,
\begin{align*}
\chi^\class \circ T \big( f' \ot_{\Cinf(\M)} f \big) &=  \sum_r f'_r \tilde\sigma_{r\theta}(f_{(0)}) \ot f_{(1)}\\
&= f' \times_\theta f_{(0)} \ot f_{(1)}=\chi(f' \ot_{\Cinf(\M)} f) ,
\end{align*}
since the action of $\tilde \bT^2$ on $\Cinf(\P)$ commutes with the coaction of $\Cinf(G)$.
\end{proof}
Noncommutative associated bundles are defined as in \eqref{equimaps7}  
by setting 
$$
\E = \Cinf(\P) \boxtimes_\rho W:= \big\{ f \in \Cinf(\P)\ot W | (\alpha_g \ot \id)(f)=(\id \ot \rho(g)^{-1})(f) \big\}
$$ 
for a representation $\rho$ of $G$ on $W$. These $\Cinf(\M)$ bimodules are finite projective since they are of the form of the modules defined in Sect.~\ref{se:nc-vb} (cf. Prop.~\ref{rem:associated-modules}). 
Moreover, Prop. \ref{prop:end} generalizes and reads $\End(\E) \isom \Cinf(\P) \boxtimes_\ad L(W)$, where $\ad$ is the adjoint representation of $G$ on $L(W)$. Also, one identifies the adjoint bundle as the module coming from the adjoint representation of $G$ on $\g \subset L(W)$, namely $\Gamma^\infty(\ad(\P)) := \Cinf(\P)\boxtimes_{ad} \g$.

For a (right) finite projective $\Cinf(\M)$-module $\E$ we define an inner product $(\cdot, \cdot)_2$ on $\Hom_{\Cinf(\M)} (\E,\E \ot_{\Cinf(\M)} \Omega(\M))$ as in Sect.~\ref{se:ym}. The Yang-Mills action functional on the space $C(\ce)$ of compatible connections $\nabla$ on $\E$ is then given in terms of the corresponding curvatures $F$ as before  by
\be
\YM(\nabla)=(F,F)_2 ,
\ee
and is a gauge invariant, positive and quartic functional. The derivation of the Yang-Mills eqs. \eqref{eq:ym} on $\S^4$ does not rely on the specific properties of $\S^4$ and continues to hold on $\M$. The same is true for the topological action, and $\YM(\nabla)\geq |\Top(\E)|$ with equality iff $\ast_\theta F=\pm F$. In other words, instanton connections are absolute minima of the Yang-Mills action.

\section{Let us twist symmetries} \label{se:twist}

The noncommutative sphere $\S^4$ can be realized as a quantum homogeneous space of the quantum orthogonal group $\SO_\theta(5)$ \cite{Var01,CD02}. In other words, $\A(\S^4)$ can be 
obtained as the subalgebra of $\A(\SO_\theta(5))$ made of elements that are 
coinvariant under the natural coaction of $\SO_\theta(4)$ on $\SO_\theta(5)$. For 
our purposes, it turns out to be more convenient to take a dual point of view and 
consider an {\it action} instead of a coaction. One obtains a twisted 
action of the Lie algebra $so(5)$ on $\S^4$ and elements of $so(5)$ act as twisted derivations on the algebra $\A(\S^4)$.  Similar considerations hold for any noncommutative sphere $S_{\Theta}^N$.

When lifted to $\Sk^7$, the twisted rotational symmetry  leaves invariant  
the basic instanton $\nabla_0$ described above.  In {\cite{LS06} we used a twisted conformal symmetry to construct instantons on $\S^4$, a construction that we shall review later on.

In fact, what we are really describing are Hopf algebras $\U_\theta(so(5))$ and 
$\U_\theta(so(5,1))$ which are obtained from the undeformed Hopf algebras 
$\U(so(5))$ and $\U(so(5,1))$ via a twist of a Drinfel'd type. Twisting of 
algebras and coalgebras has been known for some time \cite{Dr83,Dr90,GZ94}. The twists relevant for toric noncommutative manifolds are associated to the Cartan subalgebra of a Lie algebra and were already introduced in \cite{Re90}. Their use to implement symmetries of toric noncommutative manifolds was made explicit in \cite{Sit01}. 

The geometry of multi-parametric 
quantum groups and quantum enveloping algebras coming from twists has been 
studied in \cite{AC1,AC2}. Interesting consequences, e.g. for the nonassociativity of differential calculi were studied in \cite{BM05}. 
For symmetries of the usual noncommutative planes and their use for quantum field theories on it, one has the approach of \cite{Wa99,Oe00,Wa00}. More recently \cite{CKNT04,We04}, a twist was used to implement Poincar\`e symmetry on the Moyal plane while conformal transformations are twisted in \cite{Ma05}.
Twisting of infinitesimal diffeomorphisms and their use for gravity theories are in \cite{ABDMSW05,ADMW05} and infinite dimensional (infinitesimal) conformal symmetries on a two dimensional Moyal plane are twisted in \cite{LVV06}.
There are also studies of spin and statistics and their relations in the context of these twisted symmetries of the Moyal plane \cite{BMPV05}. 
Finally, a deformation of nonrelativistic
Schr\"odinger symmetry is in \cite{Ba05} while extensions to superspace, including super Poincar\`e and superconformal simmetries,  were treated in \cite{BLS05}.

\subsection{Twisting Hopf algebras and their actions}

We review the known algebraic construction of twisting a Hopf algebra and its actions, for which we refer, for instance, to \cite{Ma92} or \cite{CP94}  for details. The Hopf algebra that is relevant in the present paper is just the universal enveloping algera 
$H=\U(\g)$ with $\g$ a Lie algebra. On elements $X\in\g$, we have  
coproduct: 
\be\label{primcp}
\Delta  : \U(\g) \to \U(\g) \ot \U(\g), 
\qquad X \mapsto \Delta(X)=X \ot \II + \II \ot X ,
\ee
counit:
\be\label{primcu}
\varepsilon : \U(\g) \to \IC, \qquad X \mapsto \varepsilon(X)=0 ,
\ee
and antipode: 
\be\label{priman}
S : \U(\g) \to \U(\g), \qquad X \mapsto S(X)=-X.
\ee
When $\g$ is realized as a Lie algebra of vector fields acting on an algebra of functions $A=\Cinf(M)$, the coproduct \eqref{primcp} is just the implementation of the Leibniz rule for any $X\in\g$,
\be
X(a b) := \Delta X(a\ot b)= X(a) b + a X(b) ,
\ee
saying that $X$ is a derivation of $A$.
Then, suitably twisting the Hopf algebra $H$ goes together with twisting the product in $A$ to a noncommutative algebra that carries an action of the twisted Hopf algebra.

Let us start with a Hopf algebra $H = (H, \mu, \II, \Delta, \varepsilon, S)$  over $\IC$ (say), with multiplication $\mu: H\ot H \to H$; comultiplication $\Delta: H \to H\ot H$ (for which we use Sweedler notation, 
$\Delta(h) = h_{(1)} \ot h_{(2)}$); unit $\II : \IC \to H$ and counit $\varepsilon : H  \to \IC$; antipode $S: H \to H$. This structure is twisted by an invertible element $\cf\in H\ot H$ with properties,
\begin{align}
& (\cf \ot \II) (\Delta \ot \id) \cf = (\II \ot \cf) (\id \ot \Delta) \cf , \label{cocyf} \\
& (\varepsilon \ot \id) \cf = \II = (\id \ot \varepsilon ) \cf .   
\end{align}
Then, the element of $A$
\[
v=\mu (\id \ot S) \cf
\] 
is invertible with inverse given by
\[
v^{-1}=\mu (S \ot \id) \cf^{-1}.
\]
The twisted Hopf algebra $H_{\cf} = (H, \mu, \II, \Delta_{\cf}, \varepsilon, S_{\cf})$ has the same algebra and counit as $H$ but twisted coproduct $\Delta_{\cf} : H \to H\ot H$ and antipode $S_{\cf}: H \to H$,
\be 
\Delta_{\cf}(h) = \cf \Delta (h) \cf^{-1} , \qquad S_{\cf}(h) = v S(h) v^{-1}.
\ee
The twist $\cf\in H\ot H$ is called a $2$-cochain in general and the condition \eqref{cocyf} is a cocycle condition that assures coassociativity of the twisted coproduct $\Delta_{\cf}$. By dropping condition \eqref{cocyf}, $\Delta_{\cf}$ is however `almost coassociative' and this leads to the notion of a {\em quasi-Hopf algebra} \cite{Dr90}.  

The cocycle $\cf$ can be used to twist the multiplication of any left $H$-module algebra $A$. This twisting yields a new algebra $A_{\cf}$  which is naturally a left $H_{\cf}$-module algebra.

Let us recall that a left $H$-module algebra $A$ is first of all a left $H$-module, and this means that there is a map
$\lambda : H \ot A \to A$ such that the association $H \ni h \mapsto \lambda(h \ot \cdot)$ is a homomorphism of algebras from $H$ into the endomorphisms of $A$. In addition there is compatibility with respect to the algebra structure of $A$, 
\be
h \lt ( a b ) := \Delta h (a \ot b) =  (h_{(1)} \lt a)  (h_{(2)} \lt b), \qquad h \lt 1 = \varepsilon (h) 1 ,
\ee
for all $h \in H$, and $a,b \in A$; and we have used the notation $\lambda(h \ot a) = h \lt a$. 

Now, if $m : A\ot A \to A, m(a \ot b) = ab$ denotes the multiplication in $A$, the new algebra $A_{\cf}$ is defined to be $A$ with multiplication given by 
\[
m_{\cf} = m \circ \cf^{-1} , 
\]
and associativity of this product is guaranteed by the cocycle condition \eqref{cocyf}. Suggestively, the new product can be indicated as 
\be
a \times_\cf b = m \left( (\cf^{-1} \lt a \ot b) \right). 
\ee
As mentioned, the algebra $A_{\cf}$  is  a left $H_{\cf}$-module algebra; in fact the action of any $h\in H_{\cf}$ on any $a\in A_{\cf}$ is just the old one $h\lt a$ but extended on products via the twisted comultiplication:
\be
h \lt (a \times_\cf b) = \Delta_{\cf}(a \ot b).
\ee 
Dually, the cocycle $\cf$ could be used to twist the comultiplication of any right $H$-module coalgebra $B$ to get a coalgebra $B_\cf$ carrying a natural action of $H_\cf$.

\begin{rem}
To be precise, the twists that we use in the present paper are rather formal power series, that is 
$\cf\in H \ot H [[\lambda]]$, with $\lambda=e^{2\pi \ii \theta}$ the deformation parameter. We shall avoid the use of formal power series and of $\lambda$-adic topology by working in explicit representation spaces and with explicit operators (see also \cite{LS06}).
\end{rem}

\subsection{The rotational symmetry of $\S^4$}\label{subsect:infconf}

Let us start with the construction of the twisted symmetry $\U_\theta(so(5))$. The eight roots of the Lie algebra $so(5)$ are two-component vectors $r=(r_1,r_2)$ of the form $r=(\pm1,\pm1)$ and $r=(0,\pm1), r=(\pm1,0)$.  There are corresponding generators $E_r$ together 
with two mutually commuting generators $H_1,H_2$ of the Cartan subalgebra. The Lie brackets are
\bea\label{lie-so5}
&& [H_1,H_2] = 0, \quad [H_j,E_r] = r_j E_r , \nn \\ &&
[E_{-r},E_{r}] = r_1 H_1 + r_2 H_2, \quad  [E_{r},E_{r'}] = N_{r,r'} E_{r+r'}, 
\eea
 with $N_{r,r'}=0$ if $r+r'$ is not a root. The universal enveloping algebra $\U(so(5))$ is the algebra 
 generated by elements $\{H_j, E_r\}$ modulo relations given by the previous Lie brackets -- The
 additional Serre relations; they generate an ideal that needs to be quotiented out. This is not 
 problematic and we shall not dwell upon this point here. It is a Hopf algebra with the undeformed structure as in \eqref{primcp}, \eqref{primcu} and \eqref{priman}.
 
The twisted universal enveloping algebra $\U_\theta(so(5))$ is 
generated as above (i.e. one does not change the algebra structure) but is endowed with a twisted 
coproduct, 
\bd 
\Delta_\theta: \U_\theta(so(5)) \to \U_\theta(so(5)) \ot \U_\theta(so(5)), \quad
\quad X \mapsto \Delta_\theta(X) = \cf \Delta_0(X) \cf^{-1}. 
\ed
For the symmetries studied in the present paper the twist $\cf$ is given explicitly by 
\be
\cf = \lambda^{\half (-H_1 \ot H_2 + H_2 \ot H_1) } .
\ee
On the generators $E_r$, $H_j$, the twisted coproduct reads
\bea\label{twdel} 
&& \Delta_\theta(E_r) = E_r \ot \lambda^{\half (-r_1 H_2 + r_2 H_1) } + \lambda^{\half (r_1 H_2 - r_2 H_1) } \ot E_r , \nn\\
&& \Delta_\theta(H_j) = H_j  \ot \II + \II \ot H_j .
\eea
This coproduct allows one to represent $\U_\theta(so(5))$ as an algebra of twisted derivations on both $\S^4$ and 
$\Sk^7$ as we shall see  below. With counit and antipode given by 
\bea\label{twhopf} 
&& \varepsilon(E_r) = \varepsilon(H_j) = 0, \nn\\
&& S(E_r) = - \lambda^{\half (r_2 H_1-r_1 H_2) }  
E_ r\lambda^{\half (r_1 H_2 - r_2 H_1) }, \quad S(H_j) = -H_j ,
\eea
the algebra $\U_\theta(so(5))$ becomes a Hopf algebra \cite{CP94}. At the classical value of the deformation 
parameter, $\theta=0$, one recovers the Hopf algebra structure of $\U(so(5))$.

\begin{rem} 
The operators $\lambda^{\pm \half r_i H_j}$ in \eqref{tder} are understood as exponentials of diagonal matrices due to the fact that on generators the operators $H_1$ and $H_2$ can be written as finite dimensional matrices. This will be  clear presently when acting on both $\S^4$ and $\Sk^7$ \end{rem}

We are ready for the representation of $\U_\theta(so(5))$ on $\S^4$. For convenience, we introduce `partial derivatives', 
$\partial_\mu$ and 
 $\partial_\mu^*$ with the usual action on the generators of the algebra $\A(\S^4)$ {i.e},  
 $\partial_\mu(z_\nu)=\delta_{\mu\nu}$, $\partial_\mu(z_\nu^*)=0$, and 
 $\partial_\mu^*(z_\nu^*)=\delta_{\mu\nu}$, $\partial_\mu^*(z_\nu)=0$. Then, the action of $\U_\theta(so(5))$ on 
$\A(\S^4)$ is given by the following operators,
\be\label{act4}
\begin{aligned}
H_1 &= z_1 \partial_1 - z_1^* \partial_1^* , \\
E_{+1,+1}&= z_2 \partial_1^*   - z_1 \partial_2^*,\\
E_{+1,0} &= \tfrac{1}{\sqrt{2}} (2 z_0 \partial_1^* - z_1 \partial_0), 
\end{aligned}
\qquad
\begin{aligned}
H_2 &= z_2 \partial_2 - z_2^* \partial_2^* \\
E_{+1,-1} &= z_2^* \partial_1^* - z_1 \partial_2  \, , \\
E_{0,+1} &=  \tfrac{1}{\sqrt{2}} (2 z_0 \partial_2^* - z_2 \partial_0) \, , 
\end{aligned}
\ee
and $E_{-r}=(E_{r})^*$, with the obvious meaning of the adjoint.  
A comparison with 
eq.~\eqref{eq:act-S4} shows that $H_1$ and $H_2$ in~\eqref{act4} are the infinitesimal generators of the action of $\bT^2$ on $\S^4$.  
These operators (not the partial 
derivatives!) are extended to the whole of $\A(\S^4)$ as twisted derivations via the coproduct \eqref{twdel},
\bea\label{tder} 
&& E_r( a b ) = \Delta_\theta(E_r) (a \ot b) = E_r(a) 
\lambda^{\half (-r_1 H_2 + r_2 H_1) }(b) + 
\lambda^{\half (r_1 H_2 - r_2 H_1) }(a) E_r(b) , 
\nn\\
&& H_j( a b ) = \Delta_\theta(H_j) (a \ot b) = H_j(a) b + a H_j(b) , 
\eea
for any two elements $a,b\in \A(\S^4)$. With these twisted rules, one readily 
checks compatibility with the commutation relations \eqref{s4t} of $\A(\S^4)$. 

\bigskip
We can write the twisted action of $\U_\theta(so(5))$ on $\A(\S^4)$ by using the quantization map $L_\theta$ of Sect.~\ref{se:toric-ncm}. For 
$L_\theta(a) \in \A(\S^4)$ and $t \in \U(so(5))$ a twisted action is defined by 
\be 
\label{eq:def-twisted} T \cdot L_\theta (a) = L_\theta(t\cdot a) \ee where $T$ is 
the `quantization' of $t$ 
and $t \cdot a$ is the classical action of $\U(so(5))$ on 
$\A(S^4)$ (a better but heavier notation for the action $T\cdot~$ would be 
$t\cdot_\theta~$). One checks that both of these definitions of the twisted action coincide. The latter definition allows one to define an action of $\U(so(5))$ on $\Cinf(\S^4)$ by allowing $a$ to be in $\Cinf(S^4)$ in 
eq.~\eqref{eq:def-twisted}. Furthermore, as operators on the Hilbert space 
$\ch$ of spinors, one could identify $\lambda^{{\frac{1}{2}}({r_1 H_2-r_2 
H_1})}=U(\frac{1}{2} r\cdot \theta)$, with $r=(r_1,r_2)$, $\theta$ the 
antisymmetric two by two matrix with $\theta_{12}=-\theta_{21}=\theta$ and $U(s)$ the representation of $\IT^2$ on $\ch$ as in Sect~\ref{subse:def-torus}.

The twisted action of the Hopf algebra $\U(so(5))$ on $\A(\S^4)$ is extended to the differential calculus $(\Omega(\S^4),\dd)$ by requiring it to commute with the exterior derivative, 
$$ T \cdot \dd \omega := \dd (T \cdot \omega). 
$$ for $T\in \U(so(5)), ~ \omega \in \Omega(\S^4)$. Then, we need to use the rule \eqref{tder} on a generic form. For instance, on 1-forms we have, \begin{align}\label{onforms} & 
E_r( \sum_k a_k \dd b_k ) = \sum_k \Big( E_r(a_k) \dd \big( 
\lambda^{{\frac{1}{2}}({-r_1 H_2+r_2 H_1})}(b_k) \big) + 
\lambda^{{\frac{1}{2}}({r_1 H_2-r_2 H_1})}(a_k) \dd \big( E_r(b_k) \big) \Big), 
\nn \\ & H_j( \sum_k a_k \dd b_k ) = \sum_k \Big(H_j(a_k) \dd b_k + a_k \dd 
\big(H_j (b_k) \big) \Big)\, . 
\end{align}

\bigskip

The representation of $\U_\theta(so(5))$ on $\S^4$ given in \eqref{act4} is the fundamental vector representation.  
 When lifted to $\Sk^7$ one gets the fundamental spinor representation: as we see from the 
 quadratic relations among corresponding generators, as given in \eqref{subalgebra}, the lifting amounts to 
take the `square root' representation.
 The action on $\U_\theta(so(5))$ on $\A(\Sk^7)$ is constructed by requiring twisted derivation properties via the 
 coproduct \eqref{tder} so as to reduce to the action \eqref{act4} on $\A(\S^4)$ when using the defining 
 quadratic relations \eqref{subalgebra}. The action on  $\A(\Sk^7)$ can be given as the action of matrices 
$\Gamma$'s on the $\psi$'s,
\begin{align}\label{act7}
\psi_a \mapsto \sum_b \Gamma_{ab} \psi_b, \qquad \psi^*_a \mapsto  \sum_b \tilde\Gamma_{ab} \psi^*_b
\end{align}
with the matrices $\Gamma = \{H_j, E_r\}$ given explicitly by
\be\label{tgamma}
\begin{aligned}
&H_1 = \half\left( \begin{smallmatrix}
1  & & & \\
 & -1 &  &   \\
 &   & -1 &  \\
 &  & & 1 
\end{smallmatrix} \right), \\
&E_{+1,+1} =\begin{pmatrix} 
0 & 0 \\ 
0 & \begin{smallmatrix} 0 & -1 \\ 0 & 0 \end{smallmatrix} \\
\end{pmatrix},  \\
&E_{+1,0} =  \tfrac{1}{\sqrt{2}} \begin{pmatrix} 
0 & \begin{smallmatrix} 0 & 0\\ 0 & -1 \end{smallmatrix} \\ 
\begin{smallmatrix} \mu  & 0 \\ 0 & 0 \end{smallmatrix} & 0 \\
\end{pmatrix}, 
\end{aligned}
\qquad
\begin{aligned}
&H_2 = \half\left( \begin{smallmatrix}
-1  & & & \\
 & 1 &  &   \\
 &   & -1 &  \\
 &  & & 1 
\end{smallmatrix} \right), \\
&E_{+1,-1} =\begin{pmatrix} 
\begin{smallmatrix} 0 & 0 \\ -\mu & 0 \end{smallmatrix} & 0 \\ 
0 & 0 \\
\end{pmatrix},  \\
&E_{0,+1} =  \tfrac{1}{\sqrt{2}} \begin{pmatrix} 
0 & \begin{smallmatrix} 0 & \bar{\mu} \\ 0 & 0 \end{smallmatrix} \\ 
\begin{smallmatrix} 0  & 1 \\ 0 & 0 \end{smallmatrix} & 0 \\
\end{pmatrix}, 
\end{aligned}
\ee
and 
\be\label{tildelta}
\tilde\Gamma:= \sigma \Gamma \sigma^{-1}, \qquad 
\sigma := \begin{pmatrix} \begin{smallmatrix} 0 & -1 \\ 1 & 0 \end{smallmatrix} & 0 \\  0 & \begin{smallmatrix} 0 & -1 \\ 1 & 0 \end{smallmatrix} \end{pmatrix}.
\ee
Furthermore, $E_{-r}=(E_{r})^*$. 
With the twisted rules \eqref{tder} for the action on products, one checks compatibility of the above 
action with the commutation relations \eqref{s7t} of $\A(\Sk^7)$. 
Again, the operators $\lambda^{\pm \half r_i H_j}$ in \eqref{tder} 
are exponentials of diagonal matrices $H_1$ and $H_2$ given in the representation 
\eqref{tgamma} and as above, one could think of $\lambda^{{\frac{1}{2}}({r_1 
H_2-r_2 H_1})}$ as the operator $U(\frac{1}{2} r\cdot \theta)$. \begin{rem} 
Compare the form of the matrices $H_1$ and $H_2$ in the representation 
\eqref{tgamma} above with the lifted action $\tilde \sigma$ of $\tilde \bT^2$ on 
$\Sk^7$ as defined in \eqref{eq:lift-S7}. One checks that \[ \tilde\sigma_s = e^{ 
\pi i \left((s_1+s_2) H_1 + (-s_1+s_2) H_2\right)} , \] when acting on the spinor 
$(\psi_1, \cdots, \psi_4)$. \end{rem}

Notice that $\tilde\Gamma=-\Gamma^t$ at $\theta=0$. 
There is a beautiful correspondence between the matrices in the representation \eqref{tgamma} and the 
twisted Dirac matrices introduced in \eqref{eq:dirac},\be
\begin{aligned}
&\tfrac{1}{4}[\gamma_1^*, \gamma_1]=2 H_1\\
&\tfrac{1}{4}[\gamma_1, \gamma_2]=(\mu+\bar\mu) E_{+1,+1}\\
&\tfrac{1}{4}[\gamma_1, \gamma_0]= \sqrt{2}  E_{+1,0}
\end{aligned}
\qquad
\begin{aligned}
&\tfrac{1}{4}[\gamma_2^*, \gamma_2]=2 H_2\\
&\tfrac{1}{4}[\gamma_1, \gamma_2^*]=(\mu+\bar\mu) E_{+1,-1}\\
&\tfrac{1}{4}[\gamma_2, \gamma_0]=\sqrt{2}\bar\mu E_{0,+1}
\end{aligned}
\ee
Also, the twisted Dirac matrices satisfy the following relations under conjugation by $\sigma$:
\be
\label{eq:gamma-gammat}
(\sigma\gamma_0\sigma^{-1})^t = \gamma_0,\qquad
(\sigma\gamma_1\sigma^{-1})^t = \gamma_1 \lambda^{H_2}, \qquad (\sigma\gamma_2\sigma^{-1})^t = \gamma_2 \lambda^{H_1}.
\ee

As for $\S^4$, the twisted action of $so(5)$ on $\A(\S^7)$ is straightforwardly 
extended to the differential calculus $(\Omega(\S^7),\dd)$ by requiring that the action commutes with the exterior differential $\dd$ and using the twisted rule when action on products.

\section{Instantons from twisted conformal symmetries}
\label{se:constrinst}

Different instantons are obtained by a twisted symmetry action of $so(5,1)$. 
Classically, $so(5,1)$ is the conformal Lie algebra consisting of the 
infinitesimal diffeomorphisms leaving the conformal structure invariant. The Lie 
algebra $so(5,1)$ is given by adding 5 generators to $so(5)$. We explicitly 
describe its action on $\S^4$ together with its lift to $\Sk^7$ as an algebra of 
twisted derivations. The induced action on forms leaves the conformal structure 
invariant and when acting on $\nabla_0$ eventually results in a five-parameter 
family of instantons.

The Hopf algebra $\U_\theta(so(5))$ described in the previous section is made of twisted infinitesimal symmetries under which a basic instanton -- associated canonically with the noncommutative instanton bundle constructed previously -- is invariant. We construct a collection of (infinitesimal) gauge-nonequivalent instantons, by acting with a twisted conformal symmetry $\U_\theta(so(5,1))$ on the basic one. A completeness argument on this collection is provided using an index theoretical argument, similar to \cite{AHS78}. The dimension of the `tangent' of the moduli space can be computed as the index of a twisted Dirac operator and it turns out to be equal to its classical value which is five. 

Here, one has to be careful with the notion of tangent space to the moduli space. As will be discussed elsewhere \cite{LPRS05}
one can construct a noncommutative family of instantons, that is instantons parametrized by the quantum quotient space of the deformed conformal group $\SL_\theta(2,\bH)$ by the deformed gauge group $\Sp_\theta(2)$. 
It turns out that the basic instanton of \cite{CL01} is a `classical point' in this moduli space of instantons. We perturb this connection $\nabla_0$ linearly by sending $\nabla_0 \mapsto \nabla_0+ t \alpha$ where $t\in \bR$ and $\alpha \in \Hom(\E,\E \ot \Omega^1(\S^4))$. In order for this new connection still to be an instanton, we have to impose the self-duality equation on its curvature. After deriving this equation with respect to $t$, at $t=0$, we obtain the linearized self-duality equation to be fulfilled by $\alpha$. It is in this sense that we are considering the tangent space to the moduli space of instantons at the origin $\nabla_0$. 

\subsection{The basic instanton}\label{se:basinst}

We start with a technical lemma that simplifies 
the discussion. Let $\E = \Cinf(\Sk^7)\boxtimes_\rho W$ be a module of sections 
associated to a finite dimensional representation of $\SU(2)$, as defined in 
eq.~\eqref{equimaps7}.
 
\begin{lma} 
There is the following isomorphism of right $\Cinf(\S^4)$-modules, $$ 
\E \ot_{\Cinf(\S^4)} \Omega(\S^4) \isom \Omega(\S^4)\ot_{\Cinf(\S^4)}\E. $$ 
Consequently, $\Hom(\E,\E\ot_{\Cinf(\S^4)} \Omega(\S^4)) \isom 
\Omega(\S^4)\ot_{\Cinf(\S^4)} \End(\E)$. \end{lma} \begin{proof} Recall from 
Prop.~\ref{rem:associated-modules} that the right $\Cinf(\S^4)$-module $\E$ 
has a homogeneous module-basis $\{e_\alpha,\alpha=1,\cdots, N\}$ for some $N$ and 
each $e_\alpha$ of degree $r_\alpha$. A generic element in $\E \ot_{\Cinf(\S^4)} 
\Omega(\S^4)$ can be written as a sum $\sum_\alpha e_\alpha \otimes_{\Cinf(\S^4)} 
\omega^\alpha$, with $\omega^\alpha$ an element in $\Omega(\S^4)$. Now, for every 
$\omega \in \Omega(\S^4)$ there is an element $\tilde \omega \in \Omega(\S^4)$ -- 
given explicitly by $\tilde\omega = \sigma_{r_\alpha \cdot \theta} (\omega)$ -- 
such that $e_\alpha \omega = \tilde \omega e_\alpha$ where the latter equality 
holds inside the algebra $\Omega(\Sk^7)$ (recall that $\Omega(\S^4) \subset 
\Omega(\Sk^7)$). We can thus define a map $$ T: \E \ot_{\Cinf(\S^4)} \Omega(\S^4) 
\isom \Omega(\S^4)\ot_{\Cinf(\S^4)}\E, $$ by $T(e_\alpha \otimes_{\Cinf(\S^4)} 
\omega^\alpha) = \tilde \omega^\alpha \otimes e_\alpha$; it is a right 
$\Cinf(S^4)$-module map: $$ T\left(e_\alpha \otimes_{\Cinf(\S^4)} (\omega^\alpha 
\times_\theta f) \right) = (\tilde \omega^\alpha \times_\theta \tilde f) 
\otimes_{\Cinf(\S^4)} e_\alpha = T(e_\alpha \otimes_{\Cinf(\S^4)} \omega^\alpha) 
\times_\theta f. $$ Since an inverse map $T^{-1}$ is easily constructed, $T$ 
gives the desired isomorphism. \end{proof} \noindent Thus, we can unambiguously 
use the notation $\Omega(\S^4,\E)$ for the above right $\Cinf(\S^4)$-module $\E 
\ot_{\Cinf(\S^4)} \Omega(\S^4)$.

We let $\nabla_0=p\circ \dd$ be the canonical (Grassmann) connection on the projective module 
$\E=\Cinf(\Sk^7) \boxtimes_\rho \C^2 \isom p (\Cinf(\S^4) )^4$, with the projection $p= \Psi^\dagger \Psi$ of \eqref{projection1} and $\Psi$ is the matrix \eqref{Psi}. When acting on equivariant maps, 
we can write $\nabla_0$  as
\be\label{cancon}
\nabla_0: \E \to \E \ot_{\Cinf(\S^4)} \Omega^1(\S^4),  \qquad
(\nabla_0 f)_i = \dd f_i + \omega_{ij} \times_\theta f_j,
\ee
where $\omega$ -- called the gauge potential -- is given in terms of the matrix $\Psi$ by
\be\label{cangp}
\omega=\Psi^\dagger \dd \Psi , 
\ee
The above, is a $2\times 2$-matrix with entries in $\Omega^1(\Sk^7)$ satisfying $\bar{\omega_{ij} }=\omega_{ji}$ and $\sum_i \omega_{ii}=0$. 
Note here that the entries $\omega_{ij}$ commute with all elements in $\Cinf(\Sk^7)$. Indeed, from \eqref{Psi} we see that the elements in $\omega_{ij}$ are $\bT^2$-invariant and hence central (as one forms) in $\Omega(\Sk^7)$. 
In other words $L_\theta(\omega)=\omega$, which shows that for an element $f \in \E$ as above, we have $\nabla_0(f)_i=\dd f_i + \omega_{ij} \times_\theta f_j= \dd f_i + \omega_{ij} f_j$ which coincides with the action of the classical connection $\dd+\omega$ on $f$.
The curvature $F_0=\nabla_0^2=\dd \omega + \omega^2$ of $\nabla_0$ is an element of $\End(\E) \ot_{\Cinf(\S^4)} \Omega^2(\S^4)$ that satisfies \cite{AB02,CD02} the self-duality equation,
\be
\ast_\theta F_0 = F_0;
\ee 
hence this connection is an instanton. At the classical value of the deformation parameter, $\theta=0$, the connection \eqref{cangp} is nothing but the $\SU(2)$ instanton of \cite{BPST75}. 

Its `topological charge', i.e. the values of \eqref{topact}, was already computed in \cite{CL01}. Clearly it depends only the class $[p]$ of the bundle and can be evaluated as the index 
\be\label{topbasic}
\Top([p]) = \ind(D_p) =  \ncint \gamma_5 \pi_D(\chern_2(p)) 
\ee
 of the twisted Dirac operator 
\[
D_p = p (D \ot \II_4) p .
\]
The last equality in \eqref{topbasic} follows from the vanishing of the class $\chern_1(p)$ of the bundle. The Chern character classes and their realization as operators are in the Sect.~\ref{se:liftoric}. On the other hand, one finds 
\[
\pi_D \big(\chern_2(p))\big) = 3 \gamma_5, 
\]
which, together with the fact that 
\[
\ncint 1 = \dix (|D|^{-4}) = \frac{1}{3},
\] 
on $S^4$ (see for instance \cite{GVF01,Lnd97}),
gives  the value $\Top([p])=1$. 

We aim at constructing all connections $\nabla$ on $\E$ whose curvature satisfies this self-dual equation and of topological charge equal to $1$. 
We can write any such connection in terms of the canonical connection as in eq.~\eqref{uconn}, {i.e.} $\nabla=\nabla_0 + \alpha$ with $\alpha$ a one-form valued endomorphism of $\E$. Clearly, this will not change the value of the topological charge. We are interested in $\SU(2)$-instantons, so we impose that $\alpha$ is traceless and skew-hermitian, with the trace  taken in the second leg of $\End (\E) \isom P \boxtimes_\ad M_2(\C)$. When complexified, this yields an element $\alpha \in \Omega^1(\S^4)\ot_{\Cinf(\S^4)}\Gamma^\infty(\ad(\Sk^7)) =: \Omega^1(\ad(\S^4))$ 
(cf. Example \ref{ex:instanton-bundle}).

As usual, we impose an irreducibility condition on the instanton connections, a connection on $\E$ being 
{\it irreducible} if it cannot be written as the sum of two other connections on $\E$. We are interested only in the irreducible instanton connections on the module $\E$. 

\begin{rem}\label{rem:conn-form-central-general}
In Sect.~\ref{se:associated-modules}, we constructed projections $p_{(n)}$ for all modules $\Cinf(\Sk^7) \boxtimes_{\rho} \C^n$ over $\Cinf(\S^4)$ associated to the irreducible representations $\C^n$ of $\SU(2)$. The induced Grassmann connections $\nabla_0^{(n)}:= p_{(n)} \dd$, when acting on $\Cinf(\Sk^7) \boxtimes_{\rho} \C^n$, were written as $\dd + \omega_{(n)}$, with $\omega_{(n)}$ an $n\times n$ matrix with entries in $\Omega^1(\Sk^7)$. A similar argument as above then shows that all $\omega_{(n)}$ have entries that are central (as one forms) in $\Omega(\Sk^7)$; again, this means that $L_\theta(\omega_\n)=\omega_\n$. In particular, this holds for the adjoint bundle associated to the adjoint representation on $su(2)_\C \isom \C^3$ (as complex representation spaces), from which we conclude that $\nabla_0^{(2)}$ coincides with $[\nabla_0,\cdot]$ (since this is the case if $\theta=0$).
\end{rem}

\bigskip
The instanton potential in eq.~\eqref{cangp} is invariant under the action of $\U_\theta(so(5))$. Recall the latter's action on $\Sk^7$ given in eqs. \eqref{act7}-\eqref{tildelta} 9and extended to canonically to forms). 
Due to the form of $\tilde\Gamma$ in \eqref{tildelta} and the property $\Psi_{a2} = \sigma_{ab}  \psi^*_b$ for the second column of the matrix $\Psi$ in \eqref{Psi}, the algebra $\U_\theta(so(5))$ acts on $\Psi$ by left matrix multiplication by $\Gamma$, and by right matrix multiplication on $\Psi^*$ by the matrix transpose $\tilde \Gamma^t$ as follows
\be
\Psi_{ai} \mapsto \sum_b \Gamma_{ab} \Psi_{bi}, \qquad
\Psi^*_{ia} \mapsto \sum_a \Psi^*_{ib} \tilde\Gamma_{ab}.
\ee
These are used in the following
\begin{prop}
The instanton gauge potential $\omega$ is invariant under the twisted action of $\U_\theta(so(5))$. 
\end{prop}
\begin{proof}
From the above observations, the gauge potential transforms as:
$$
\omega = \Psi^* \dd \Psi \mapsto \Psi^* \big( \tilde\Gamma^t \lambda^{-r_1 H_2} + \lambda^{r_2 H_1} 
\Gamma \big) \dd \Psi.
$$
where  
$\lambda^{-r_i H_j}$ is understood in its representation \eqref{tgamma} on $\A(\Sk^7)$. Direct computation for $\Gamma=\{H_j, E_r\}$ shows that $\tilde\Gamma^t \lambda^{-r_1 H_2} + \lambda^{r_2 H_1} \Gamma=0$, which finishes the proof.
\end{proof}

\subsection{The infinitesimal conformal symmetry}

Different instantons are obtained by a twisted symmetry action of $\U_\theta(so(5,1))$. Classically, $so(5,1)$ is the conformal Lie algebra consisting of the infinitesimal diffeomorphisms leaving the conformal structure invariant. We construct the Hopf algebra $\U_\theta(so(5,1))$ by adding 5 generators to $\U_\theta(so(5))$ and describe its action on $\S^4$ together with its lift to $\Sk^7$. The induced action of $\U_\theta(so(5,1))$ on forms leaves the conformal structure invariant. The action of  $\U_\theta(so(5,1))$ on $\nabla_0$  eventually results in a five-parameter family of (infinitesimal) instantons. 

The conformal Lie algebra $so(5,1)$ consists of the generators of $so(5)$ together with the dilation and the so-called special conformal transformations. On $\bR^4$ with coordinates $\{x_\mu, \mu=1,\ldots,4 \}$ they are given  by the operators $H_0=\sum_\mu x_\mu \partial/\partial x_\mu$ and $G_\mu = 2 x_\mu \sum_\nu x_\nu \partial/\partial x_\nu -  \sum_\nu x_\nu^2 (\partial/\partial x_\nu)$, respectively \cite{Lie70}. 
In the definition of the enveloping algebra $\U_\theta(so(5,1))$ we do not change the algebra structure, {i.e.} we take the relations of $\U(so(5,1))$, as we did for $\U_\theta(so(5))$. We thus define $\U_\theta(so(5,1))$ as the algebra $\U_\theta(so(5))$ with five extra generators adjoined,  $H_0, G_r$, 
$r=(\pm 1,0),(0,\pm 1)$, subject to the relations of $\U_\theta(so(5))$ of eq.~\eqref{lie-so5} together with the (undeformed) relations,
\begin{align}\label{lie-confa}
\begin{aligned}
&[H_0,H_i]=0, \\ & [H_0,G_r]=\sqrt{2} E_r, 
\end{aligned} \qquad 
\begin{aligned}
&[H_j, G_r] = r_j G_r, \\ & [H_0,E_r]=(\sqrt{2})^{-1} G_r,
\end{aligned}
\end{align}
whenever $r=(\pm 1,0),(0,\pm 1)$, and 
\begin{align}\label{lie-confb}
\begin{aligned}
&[G_{-r},G_r]=2 r_1 H_1 + 2 r_2 H_2, \\ & [E_r,G_{r'}]=\tilde N_{r,r'} G_{r+r'}, 
\end{aligned}
\qquad
\begin{aligned}
&[G_r,G_{r'}] =N_{r,r'} E_{r+r'}, \\ &[E_{-r},G_r]=\sqrt{2} H_0,
\end{aligned}
\end{align}
with  as before, the constant $N_{r,r'}=0$ if $r+r'$ is not a root of $so(5)$ and the constant $\tilde N_{r,r'}=0$ if $r+r' \notin \{(\pm 1,0),(0,\pm 1) \}$.
Although the algebra structure is unchanged, again the Hopf algebra structure of $\U_\theta(so(5,1))$ gets twisted. The twisted structures are given by eqs. \eqref{twdel} and \eqref{twhopf} together with
\be
\begin{aligned}
\Delta_\theta(G_r) &= G_r \ot \lambda^{\half (-r_1 H_2 + r_2 H_1) } + \lambda^{\half (r_1 H_2 - r_2 H_1) } \ot G_r , \\ 
S(G_r) &= - \lambda^{\half (r_2 H_1-r_1 H_2) }  G_ r
\lambda^{\half (r_1 H_2 - r_2 H_1) }, \\
\varepsilon(G_r) &= 0 ,
\end{aligned}
\qquad
\begin{aligned}
\Delta_\theta(H_0) &= H_0 \ot 1+ 1 \ot H_0 ,\\
S(H_0) &=-H_0, \\
\varepsilon(H_0) &= 0
\end{aligned}
\ee
making  $\U_\theta(so(5,1))$ an Hopf algebra. 

The action of $\U_\theta(so(5,1))$ on $\A(\S^4)$ is given by the operators \eqref{act4} together with
\begin{align}\label{confact4}
&H_0 = \partial_0 - z_0 (z_0 \partial_0 + z_1 \partial_1 + z_1^* \partial_1^* +
 z_2 \partial_2 + z_2^* \partial_2^*), \nn \\
&G_{1,0}=  2 \partial_1^* - z_1 (z_0 \partial_0 + z_1 \partial_1 + z_1^* \partial_1^* + \bar\lambda z_2 \partial_2 + \lambda z_2^* \partial_2^*),\\
&G_{0,1}=  2 \partial_2^* - z_2 (z_0 \partial_0 + z_1 \partial_1 + z_1^* \partial_1^* + z_2 \partial_2 + z_2^* \partial_2^*), \nn
\end{align}
and $G_{-r} = (G_r)^*$. The introduction of the extra $\lambda$'s in $G_{1,0}$ (and $G_{-1,0}$) are necessary for the algebra structure of $\U_\theta(so(5,1))$, as dictated by the Lie brackets in \eqref{lie-confa}, 
\eqref{lie-confb} to be preserved. Since the operators $H_0$ and $G_r$ are quadratic in the $z$'s, one has to be careful when checking the Lie brackets and use the twisted rules \eqref{tder}. For instance, on the generator $z_2$, we have
\begin{align*}
[E_{-1,-1},G_{1,0}](z_2)&=E_{-1,-1}(-\bar\lambda z_1 z_2)+G_{1,0}(z_1^*)\\
&=-\bar\lambda (E_{-1,-1}(z_1) \lambda^{H_2} (z_2)+ \lambda^{H_1}(z_1) E_{-1,-1}(z_2))+ G_{1,0}(z_1^*) \\
&=-z_2^* z_2 + z_1 z_1^* +2 - z_1 z_1^* = G_{0,-1}(z_2)
\end{align*}
The operators in \eqref{act4} 
and \eqref{confact4} give a well defined action of $\U_\theta(so(5,1))$ on the algebra $\A(\S^4)$ provided one extends them to the whole of $\A(\S^4)$ as twisted derivations via the rules \eqref{tder} together with 
\bea 
&& G_r( a b ) = G_r(a) \lambda^{{\frac{1}{2}}({-r_1 H_2+r_2 H_1})}(b) + \lambda^{{\frac{1}{2}}({r_1 
H_2-r_2 H_1})}(a) G_r(b) , \nn\\ 
&& H_0( a b ) = H_0(a) b + a H_0(b) , 
\eea for 
any two elements $a,b\in \A(\S^4)$. 

Equivalently, the Hopf algebra $\U_\theta(so(5,1))$ could be  defined to act on $\A(\S^4)$ by
\be\label{action:hopf-so51}
T \cdot L_\theta(a) = L_\theta(t\cdot a) ,
\ee
for $T \in \U_\theta(so(5,1))$ deforming $t \in \U(so(5,1))$ and $L_\theta(a) \in \A(\S^4)$ 
deforming $a \in \A(S^4)$. Again, equation~\eqref{action:hopf-so51} makes sense for $a \in \Cinf(S^4)$, which defines an action of $so(5,1)$ on $\Cinf(\S^4)$. As before, the action on the differential calculus $(\Omega(\S^4),\dd)$ is obtained by requiring commutation with the exterior derivative: $ T \cdot \dd \omega = \dd (T \cdot \omega)$, for $T\in so(5,1)$ and $\omega \in \Omega(\S^4)$. On products 
we shall have formul{\ae} like the one in \eqref{onforms}, \begin{align} & G_r( 
\sum_k a_k \dd b_k ) = \sum_k \Big( G_r(a_k) \dd \big( 
\lambda^{{\frac{1}{2}}({-r_1 H_2+r_2 H_1})}(b_k) \big) + 
\lambda^{{\frac{1}{2}}({r_1 H_2-r_2 H_1})}(a_k) \dd \big( G_r(b_k) \big) \Big), 
\nn \\ & H_0( \sum_k a_k \dd b_k ) = \sum_k \Big(H_0(a_k) \dd b_k + a_k \dd 
\big(H_0 (b_k) \big) \Big)\, . \end{align} What we are dealing with are 
`infinitesimal' twisted conformal transformations: 
\begin{lma} \label{lma:so51-hodge} 
The Hodge $\ast_\theta$-structure of $\Omega(\S^4)$ is 
invariant for the twisted action of $\U_\theta(so(5,1))$, 
$$ T\cdot (\ast_\theta\omega)= 
\ast_\theta (T \cdot \omega) , 
$$ 
\end{lma} 
\begin{proof} Recall that 
$T(L_\theta(a))=L_\theta(t\cdot a)$ for $a \in \A(S^4)$ and $T$ is the 
`quantization' of $t \in \U(so(5,1))$. Then, since $so(5,1)$ leaves the Hodge 
$\ast$-structure of $\Omega(S^4)$ invariant and the differential $\dd$ commutes 
with the action of $\U_\theta(so(5,1))$, if follows that the latter algebra leaves the Hodge 
$\ast_\theta$-structure of $\Omega(\S^4)$ invariant as well. 
\end{proof}

\subsection{The construction of instantons}
Again, 
the action of $so(5,1)$ on $\S^4$ can be lifted to an action on $\Sk^7$. And the 
latter action can be written as in \eqref{act7} in terms of matrices $\Gamma$'s 
acting on the $\psi$'s, \begin{align} \psi_a \mapsto \sum_b \Gamma_{ab} \psi_b, 
\qquad \psi^*_a \mapsto \sum_b \tilde\Gamma_{ab} \psi^*_b , \end{align} where in 
addition to \eqref{tgamma} we have also the matrices $\Gamma=\{H_0,G_r\}$, given 
explicitly by \begin{align} H_0 &= \half (-z_0 \I_4 + \gamma_0), \nn\\ 
G_{1,0}&=\half (-z_1 \lambda^{-H_2} + \gamma_1), \qquad G_{0,1}=\half (-z_2 + 
\lambda^{-H_1} \gamma_2), \end{align} with $G_{-r}= (G_r)^*$ and $\tilde\Gamma = 
\sigma \Gamma \sigma^{-1}$. Notice the reappearance of the twisted Dirac matrices 
$\gamma_\mu, \gamma_\mu^*$ of \eqref{eq:dirac} in the above expressions. In the 
above expressions, the operators $\lambda^{-H_j}$ are $4 \times 4$ matrices 
obtained from the spin representation \eqref{tgamma} of $H_1$ and $H_2$ and 
given explicitily by 
\begin{align}\label{expspin} 
\lambda^{-H_1} = \left( \begin{smallmatrix} \bar\mu & & & \\
 & \mu & & \\
 & & \mu & \\
 & & & \bar\mu \end{smallmatrix} \right), \qquad \lambda^{-H_2} = \left( 
\begin{smallmatrix} \mu & & & \\
 & \bar\mu & & \\
 & & \mu & \\
 & & & \bar\mu \end{smallmatrix} \right), \qquad \mu=\sqrt{\lambda}. 
\end{align} 
As for $so(5)$, the action of $so(5,1)$ on the matrix $\Psi$ is found to be by 
left matrix multiplication by $\Gamma$ and on $\Psi^*$ by $\tilde \Gamma$, \be 
\Psi_{ai} \mapsto \sum_b \Gamma_{ab} \Psi_{bi}, \qquad \Psi^*_{ia} \mapsto \sum_a 
\tilde\Gamma_{ab} \Psi^*_{ib}. \ee Here we have to be careful with the ordering 
between $\tilde\Gamma$ and $\Psi^*$ in the second term since the $\tilde\Gamma$'s 
involve the (not-central) $z$'s. 
There are the following useful commutation 
relations between the $z_\mu$'s and $\Psi$: \be \label{eq:z-Psi} \begin{aligned} 
&z_1 \Psi_{ai} = (\lambda^{-H_2})_{ab} \Psi_{bi} z_1, \qquad z_2 \Psi_{ai} = 
(\lambda^{-H_1})_{ab} \Psi_{bi} z_2, \\ &z_1 \Psi^*_{ia} = \Psi^*_{ib} 
(\lambda^{-H_2})_{ba} z_1, \qquad z_2 \Psi^*_{ia} = \Psi^*_{ib} 
(\lambda^{-H_1})_{ba} z_2, 
\end{aligned} \ee
with $\lambda^{-H_j}$ understood as the explicit matrices \eqref{expspin}.

\begin{prop}
\label{prop:instantons}
The instanton gauge potential $\omega=\Psi^* \dd \Psi$ transforms under the action of the Hopf algebra $\U_\theta(so(5,1))$ as $\omega \mapsto \omega+\delta \omega_i$, where
\begin{align*}
&\delta \omega_0 := H_0(\omega)=-z_0 \omega - \half d z_0 \I_2 + \Psi^* ~\gamma_0 ~\dd\Psi,\\
&\delta \omega_1 :=G_{+1,0}(\omega)= -z_1 \omega - \half d z_1 ~\I_2 + \Psi^* ~\gamma_1~ \dd\Psi,\\
&\delta \omega_2 :=G_{0,+1}(\omega)= -z_2 \omega - \half d z_2 ~ \I_2 + \Psi^* ~\gamma_2~ \dd\Psi,\\
&\delta \omega_3 :=G_{-1,0}(\omega)= - \omega \bar z_1-\half d \bar z_1 ~\I_2+\Psi^*~\gamma_1^* ~\dd\Psi,\\
&\delta \omega_4 :=G_{0,-1}(\omega)= -\omega \bar z_2- \half d \bar z_2 ~ \I_2+\Psi^*~\gamma_2^*~ \dd\Psi,
\end{align*}
with $\gamma_\mu, \gamma_\mu^*$ the twisted $4\times 4$ Dirac matrices defined in \eqref{eq:dirac}.
\end{prop}
\begin{proof}
The action of $H_0$ on the instanton gauge potential $\omega=\Psi^* \dd \Psi$ takes the form
\begin{align*}
H_0(\omega) &=H_0(\Psi^*) \dd \Psi + \Psi^* \dd(H_0(\Psi))= \Psi^* 
(- z_0 \I_4 + \gamma_0 )\dd \Psi - \half d z_0 \Psi^* \Psi,
\end{align*}
since $z_0$ is central. Direct computation results in the above expression for $\delta \omega_0$. Instead, the twisted action of $G_r$ on $\omega$ takes the form,
$$
G_r: \omega_{ij}
\mapsto \sum_{a,b,c} \tilde\Gamma_{ab} \Psi^*_{ib} (\lambda^{-r_1 H_2})_{ac} \dd \Psi_{cj} +(\lambda^{r_2 H_1})_{ab} \Psi^*_{ib} \Gamma_{ac} \dd \Psi_{cj} + (\lambda^{r_2 H_1})_{ab} \Psi^*_{ib} (\dd \Gamma_{ac}) \Psi_{cj} ,
$$
where we used the fact that $\tilde H_j= \sigma H_j \sigma^{-1} = - H_j$.
Let us consider the case $r=(+1,0)$. Firstly, note that the complex numbers $(\lambda^{-H_2})_{ac}$ commute with $\Psi^*_{ib}$ so that from the definition of $\Gamma$ and $\tilde\Gamma$, and using \eqref{eq:z-Psi}, we obtain for the first two terms,
\begin{align*}
-z_1 (\Psi^* \dd \Psi)_{ij} + 
\half \Psi^*_{ib} 
(\sigma\gamma_1\sigma^{-1})_{cb} (\lambda^{-H_2})_{cd} \dd \Psi_{dj} +\half \Psi^*_{ib} (\gamma_1)_{bc} \dd \Psi_{cj}.
\end{align*}
The first term forms the matrix $-z_1 \omega$ whereas the second two terms combine to give $\Psi^* \gamma_1 \dd \Psi$, due to relation \eqref{eq:gamma-gammat}. 
Finally, using eq.~\eqref{eq:z-Psi} the term $ \Psi^*_{ib} (\dd \Gamma_{ac}) \Psi_{cj}$ reduces to $-\half\dd z_1 \Psi^*_{ib} \Psi_{bj}=-\half \dd z_1 \I_2$. The formul{\ae} for $r=(-1,0)$ and $r=(0, \pm 1)$ are established in likewise manner.
\end{proof}

The transformations in Prop.~\ref{prop:instantons} of the gauge potential $\omega$ under the twisted symmetry $\U_\theta(so(5,1))$ induce natural transformations of the canonical connection $\nabla_0$ in \eqref{cancon} 
to $\nabla_{t,i}:=\nabla_0 + t \delta \omega_i + \cO(t^2)$. We shall presently see explicitly that these new connections are (infinitesimal) instantons, {i.e.} their curvatures are self-dual. 
In fact, this also follows from Lemma~\ref{lma:so51-hodge} which statesthat $\U_\theta(so(5,1))$ acts by conformal transformation therefore leaving the self-duality equation $\ast_\theta F_0 = F_0$ for the basic instanton $\nabla_0$ invariant.

We start by writing $\nabla_{t,i}$ in terms of the canonical connection on $\E\isom p \big(\A(\S^4)\big)^4$. Using the explicit isomorphism of Prop.~\ref{prop:modulesisomorphism}, between this module and the module of equivariant maps $\A(\Sk^7)\boxtimes_\rho \C^2$, we find that $\nabla_{t,i}=p \dd + t \delta \alpha_i + \cO(t^2)$ with explicit expressions 
\be
\begin{aligned}
\delta \alpha_0 = p \gamma_0 (\dd p) p - \half \Psi \dd z_0 \Psi^*, \\
\delta \alpha_1 = p \gamma_1 (\dd p) p - \half \Psi \dd z_1 \Psi^*,\\
\delta \alpha_2 = p \gamma_2 (\dd p) p - \half \Psi \dd z_2 \Psi^*,
\end{aligned}
\qquad
\begin{aligned}
~ \\
\delta \alpha_3 = p \gamma_1^* (\dd p) p - \half \Psi \dd z_1^* \Psi^*,\\
\delta \alpha_4 = p \gamma_2^* (\dd p) p - \half \Psi \dd z_2^* \Psi^*,
\end{aligned}
\ee
The  $\delta \alpha_i's$ are $4 \times 4$ matrices with entries in the one-forms $\Omega^1(\S^4)$ and  satisfying the relations $p \delta \alpha_i = \delta \alpha_i p = p \delta \alpha_i p = \delta \alpha_i$, as expected from the general theory on connections on modules in Sect.~\ref{se:connections}). Indeed, using relations \eqref{eq:z-Psi} one can move the $dz$'s to the left of $\Psi$ at the cost of some $\mu$'s, so getting expression like $dz_i ~p \in M_4(\Omega^1(\S^4))$.

From eq.~\eqref{ucurv}, the curvature $F_{t,i}$ of the connection $\nabla_{t,i}$ is given by 
\be
F_{t,i}=F_0 + t p \dd (\delta \alpha_i) + \cO(t^2).
\ee
In order to check self-duality  (modulo $t^2$) of this curvature, we express it in terms of the projection $p$ and consider $F_{t,i}$ as a two-form valued endomorphism on $\E\isom p \big(\A(\S^4)\big)^4$. 
\begin{prop}
\label{prop:curv-transf}
The curvatures $F_{t,i}$ of the connections $\nabla_{t,i}$, $i=0,\ldots,4$, are given by $F_{t,i} =F_0 + t \delta F_i+ \cO(t^2)$, where $F_0=p \dd p \dd p$ and 
\be
\begin{aligned}
\delta F_0&=-2 z_0 F_0, \\
\delta F_1&=-2 z_1 \lambda^{H_2} F_0, \\
\delta F_2&=-2 z_2 \lambda^{H_1} F_0,
\end{aligned}
\qquad
\begin{aligned}
~ \\
\delta F_3=-2 z_1^*\lambda^{-H_2} F_0, \\
\delta F_4=-2 z_2^*\lambda^{-H_1} F_0.
\end{aligned}
\ee
\end{prop}
\begin{proof}
A small computation yields for $\delta F_i=p \dd (\delta \alpha_i)$, thought of as an $\Omega^2(\S^4)$-valued  endomorphism on $\E$ the expression, 
$
\delta F_i = p (\dd p) \gamma_i (\dd p) p - p \gamma_i (\dd p) (\dd p) p,
$
with the notation $\gamma_3=\gamma_1^*$ and $\gamma_4=\gamma_2^*$, and using $p (\dd p) p=0$. 
Then, the crucial property $p (\dd p\gamma_i +\gamma_i \dd p) (\dd p) p = 0$ all $i=0,\ldots,4$ yields  $\delta F_i = -2 p \gamma_i \dd p \dd p p$. This is expressed as $\delta F_i = -2 p \gamma_i p dp dp$ by using 
$\dd p = (\dd p) p + p \dd p $. Finally, $p \gamma_i p=\Psi (\Psi^* \gamma_i \Psi) \Psi^*$, so that the result follows from the definition of the $z$'s in terms of the Dirac matrices given in eq.~\eqref{zg}, together with the commutation relations between them and the matrix $\Psi$ in eq.~\eqref{eq:z-Psi}
\end{proof}
\begin{prop} The connections $\nabla_{t,i}$ are (infinitesimal) 
instantons, {i.e.} \be \ast_\theta F_{t,i}=F_{t,i} \qquad \mathrm{mod} \, t^2 . 
\ee Moreover, the connections $\nabla_{t,i}$ are not gauge equivalent to 
$\nabla_0$.\end{prop} \begin{proof} The first point follows directly from the 
above expressions for $\delta_i F$ and the self-duality of $F_0$.  To establish 
the gauge inequivalence of the connections $\nabla_{t,i}$ with $\nabla_0$, we 
recall that an infinitesimal gauge transformation is given by $\nabla_0 \mapsto 
\nabla_0 + t [\nabla_0,X]$ for $X \in \Gamma^\infty(\ad(\Sk^7))$. We need to show 
that $\delta_i \omega$ is orthogonal to $[\nabla_0,X]$ for any such $X$, {i.e.} 
$$ ([\nabla_0,X],\delta_i \omega )_2= 0, $$ with the natural inner product on 
$\Omega^1(\ad(\Sk^7)):=\Omega^1(\S^4)\ot_{\Cinf(\S^4)} 
\Gamma^\infty(\ad(\Sk^7))$. From Remark~\ref{rem:conn-form-central-general}, it 
follows that $$(\nabla_0^{(2)}(X),\delta_i \omega)_2 = (X, 
\left(\nabla_0^{(2)}\right)^* (\delta_i \omega) )_2,$$ which then should vanish 
for all $X$. From equation~\eqref{action:hopf-so51}, we see that $\delta_i \omega 
= T_i(\omega)$ coincides with $L_\theta (t_i \cdot \omega^\class)$ with $t_i$ and 
$\omega^\class$ the classical counterparts of $T_i$ and $\omega$, respectively. 
In the undeformed case, the infinitesimal gauge potentials generated by acting 
with elements in $so(5,1)-so(5)$ on the basic instanton gauge potential 
$\omega^\class$ satisfy $(\nabla_0^{(2)} )^* (\delta_i \omega^\class)=0$ as shown 
in \cite{AHS78}. The result then follows from the observation that 
$\nabla_0^{(2)}$ commutes with the quantization map $L_\theta$ (cf. 
Remark~\ref{rem:conn-form-central-general}). \end{proof}

\subsection{Instantons on $\R^4$}
\label{subsect:localexpr}
In this section, we obtain `local expressions' for the instantons on $\S^4$ constructed in the previous section; that is we map then to a noncommutative $\R^4$ obtained by `removing a point' from $\S^4$.
With $\lambda=e^{2\pi \ii \theta}$ as above, the algebra $\A(\R^4)$ of polynomial functions on the 4-plane $\R^4$ is defined to be the $*$-algebra generated by $\zeta_1, \zeta_2$ satisfying
\be
\zeta_1 \zeta_2 = \lambda \zeta_2 \zeta_1, \qquad \zeta_1  \zeta_2^* = \bar\lambda \zeta_2^* \zeta_1.
\ee
At  $\theta=0$ one recovers the $*$-algebra of polynomial functions 
on the usual 4-plane $\bR^4$. 

The algebra $\A(\R^4)$ can also be defined as the vector space $\A(\bR^4)$
 equipped with a deformed product $\times_\theta$ as in
 equation~\eqref{eq:star-product}. Indeed, the torus $\bT^2$ acts naturally on
 the two complex coordinates of $\bR^4 \isom \C^2$. This also allows us to define
 the smooth algebra $C_b^\infty(\R^4)$ as the vector space $C_b^\infty(\bR^4)$ of
 bounded smooth functions on $\bR^4$ equipped with a deformed product
 $\times_\theta$. However, for our purposes it suffices to consider the
 polynomial algebra $\A(\R^4)$ with one self-adjoint central generator $\rho$
 added together with relations $\rho^2 (1+|\zeta|^2)=(1+|\zeta|^2) \rho^2 = 1 $
 where $|\zeta|^2:=\zeta_1^* \zeta_1 + \zeta_2^* \zeta_2$ (this enlargement was
 already done in \cite{CD02}). In the following, we will denote the enlarged
 algebra by $\tilde\A(\R^4)$ and will also use the notation 
\be
\rho^2=(1+|\zeta|^2)^{-1}= \frac{1}{1+|\zeta|^2}.
\ee 
Note that $\rho^2$ is an element in $C^\infty_b(\R^4)$.
One defines elements $\tilde z_\mu, \mu=0,1,2$ in $\tilde\A(\R^4)$ by 
\be \label{chart}
\tilde z_0 = (1-|\zeta|^2)(1+|\zeta|^2)^{-1}, \qquad \tilde z_j = 2 \zeta_j (1+|\zeta|^2)^{-1} \quad j=1,2,  
\ee
and checks that they satisfy the same relations as in \eqref{s4t} of the generators $z_\mu$ of $\A(\S^4)$. The difference is that the classical point $z_0=-1, z_j=z_j^*=0$ of $\S^4$ is not in the spectrum of $\tilde z_\mu$. We interpret the noncommutative plane $\R^4$ as a `chart' of the noncommutative 4-sphere $\S^4$ and the eq.~\eqref{chart} as the (inverse) stereographical projection from $\S^4$ to $\R^4$. In fact, one can cover $\S^4$ by two such charts with domain $\R^4$, and transition functions on $\R^4 \backslash \{0\}$, where $\{0\}$ is the classical point $\zeta_j=\zeta_j^* = 0$ of $\R^4$ (cf. 
\cite{CD02} for more details). 

A differential calculus $(\Omega(\R^4),\dd)$ on $\R^4$ is obtained from the general procedure described 
in Sect.~\ref{se:toric-ncm}. Explicitly, $\Omega(\R^4)$ is the graded $*$-algebra generated by the elements $\zeta_\mu$ of degree $0$ and $d\zeta_\mu$ of degree $1$ with relations,
\begin{align}  \label{rel:diff-R4}
&d\zeta_\mu d\zeta_\nu+ \lambda_{\mu\nu} d\zeta_\nu d\zeta_\mu =0 , \qquad d\zeta_\mu^* d\zeta_\nu+ \lambda_{\nu\mu} d\zeta_\nu d\zeta_\mu^* =0, \nn \\
&\zeta_\mu d\zeta_\nu = \lambda_{\mu\nu} d\zeta_\nu \zeta_\mu, \qquad \zeta_\mu^* d\zeta_\nu = \lambda_{\nu\mu} d\zeta_\nu \zeta_\mu^* ,
\end{align}
and $\lambda_{1 2} = \bar{\lambda}_{2 1} =: \lambda$. There is a unique differential $\dd$ on $\Omega(\R^{4})$ such that $\dd: \zeta_\mu\mapsto d\zeta_\mu$ and a Hodge star operator $\ast_\theta : \Omega^p(\R^4) \to \Omega^{4-p}(\R^4)$, obtained from the classical Hodge star operator. In terms of the standard Riemannian metric on $\bR^4$, on two-forms  we have,
\begin{align}
\label{eq:hodge-2forms}
\ast_\theta d\zeta_1 d\zeta_2 = -d\zeta_1 d\zeta_2 ,
\qquad \ast_\theta d\zeta_1 d \zeta_1^* = -d\zeta_2 d \zeta_2^* ,
\qquad\ast_\theta d\zeta_1 d \zeta_2^* =  d\zeta_1 d \zeta_2^* ,
\end{align}
and $\ast_\theta^2 = \id$. These are the same formul{\ae} as the ones for the undeformed Hodge $\ast$ on $\bR^4$ -- since the metric is not changed in an isospectral deformation. 

Again, we slightly enlarge the differential calculus $\Omega(\R^4)$ by adding the 
self-adjoint central generator $\rho$. The differential $\dd$ on $\rho$ is 
derived from the Leibniz rule for $\dd$ applied to its defining relation, 
\begin{align*} (\dd \rho^2) (1+|\zeta|^2) + \rho^2 \dd(1+|\zeta|^2)= \dd 
\left(\rho^2 (1+|\zeta|^2) \right) = 0, \end{align*} so that $\rho \dd \rho = 
\half \dd \rho^2 = - \half \rho^4 \dd(1+|\zeta|^2) = - \half \rho^4 \sum_\mu (\zeta_\mu 
d\zeta_\mu^* + \zeta_\mu^* d\zeta_\mu)$. The enlarged differential calculus will 
be denoted by $\tilde\Omega(\R^4)$.

\bigskip

The stereographical projection from $S^4$ onto $\bR^4$ is a conformal map commuting with the action of $\bT^2$; thus it makes sense to investigate the form of the instanton connections on $\S^4$ obtained in Prop. \ref{prop:instantons} on the local chart $\R^4$. 
As in \cite{Lnd05}, we first introduce a `local section' of the principal bundle  $\Sk^7 \to \S^4$ on the local 
chart of $\S^4$ defined in \eqref{chart}. Let $u=(u_1,u_2)$ be a complex spinor of modulus one, $u_1^* 
u_1 + u_2^*u_2 = 1$, and define
\be \label{lose}
\begin{pmatrix}
\psi_1 \\ \psi_2  
\end{pmatrix} = \rho 
\begin{pmatrix}
u_1 \\ u_2  
\end{pmatrix}
, \qquad 
\begin{pmatrix}
\psi_3 \\ \psi_4  
\end{pmatrix} = \rho \begin{pmatrix}
\zeta_1^*& \zeta_2^* \\ -\mu \zeta_2 & \bar{\mu}   \zeta_1 
\end{pmatrix} 
\begin{pmatrix}
u_1 \\ u_2  
\end{pmatrix}.
\ee
Here $\rho$ is a central element in $\Cinf(\R^4)$ such that $\rho^2 = (1+| \zeta |^2)^{-1}$ and the commutations rules of  the $u_j$'s with the $\zeta_k$'s are dictated by those of the $\psi_j$, 
\be
u_1 \zeta_j = \mu \zeta_j u_1 \, , \quad u_2 \zeta_j = \bar{\mu} \zeta_j u_2 \, , \quad j=1,2 . 
\ee
The right action of $\SU(2)$ rotates the vector $u$ while mapping to  the `same point' of $\S^4$, 
which, using the definition~\eqref{subalgebra}, from the choice in \eqref{lose} is found to be
\be 2 (\psi_1 \psi^*_3+ \psi^*_2 \psi_4) = \wt z_1, \,  \quad 2(- \psi^*_1 \psi_4 + \psi_2\psi^*_3) =\wt{z}_2, \, \quad 2(\psi^*_1 \psi_1 + \psi^*_2 \psi_2) -1 = \wt{z}_0,
\ee 
and which is in the local chart \eqref{chart}, as expected. 
\begin{rem} 
Strictly speaking, the symbols $\psi_a$ here denotes elements in the 
algebra $\A(\Sk^7)$ enlarged by an extra generator which is the inverse of $1+z_0 = 2(1+\psi_1^* \psi_1 + \psi_2^* \psi_2)$. Intuitively, this corresponds to `remove' the fiber $S^3$ in the sphere $\Sk^7$ above the classical point $z_0=-1, 
z_j=z_j^*=0$ of the base space $\S^4$. 
\end{rem}
By writing the unit vector $u$ as an $\SU(2)$ matrix, $u=\left(\begin{smallmatrix} u_1 & -u_2^* \\ u_2 
&   u_1^*\end{smallmatrix}\right)$, we have 
\be
\Psi= \rho \begin{pmatrix} \I_2 & 0 \\ 0 & \cZ \end{pmatrix} \begin{pmatrix} u \\ u \end{pmatrix}, \qquad \text{with } \cZ = \left( \begin{smallmatrix} \zeta_1^*& \zeta_2^* \\ -\mu \zeta_2 & \bar{\mu}   \zeta_1 \end{smallmatrix} \right) .
\ee
Then, by direct computation the gauge potential $\omega= \Psi^* d \Psi$ takes the form
\begin{align}
u \omega u^* &= \rho^{-1} \dd \rho  + \rho^2  \cZ^* \dd \cZ  +  (\dd u) u^* \nn \\
&= \frac{1}{(1+|\zeta|^2)} \begin{pmatrix} \sum_i \zeta_i d \zeta_i^* - d \zeta_i \zeta_i^*  
& 2( \zeta_1 d\zeta_2^* - d \zeta_1 \zeta_2^*) \\ 2(\zeta_2 d \zeta_1^* - d \zeta_2 \zeta_1^*)  
& \sum_i \zeta_i^* d \zeta_i - d \zeta_i^* \zeta_i \end{pmatrix}  +  (\dd u) u^* ,
\end{align}
while its curvature $F=\dd \omega + \omega^2$ is,
\be\label{gcur}
u F u^* = \rho^4 \dd\cZ^* \dd\cZ = \frac{1}{(1+|\zeta|^2)^2} \begin{pmatrix} d \zeta_1 d \zeta_1^* - d \zeta_2 d \zeta_2^*  & 2 d \zeta_1 d \zeta_2^* \\ 2 d \zeta_2 d \zeta_1^* & -d \zeta_1 d \zeta_1^* - d \zeta_2 d \zeta_2^*  \end{pmatrix} .
\ee
From \eqref{eq:hodge-2forms} one checks that this curvature is self-dual:
$\ast_\theta(u F u^*) = u F u^*$, as expected.  

\bigskip
The explicit local expressions for the transformed -- under 
infinitesimal conformal transformations -- gauge potentials and their curvature 
can be obtained in a similar manner. As an example, let us work out the local 
expression for $\delta_0 \omega$ which is the most transparent one. Given the 
expression for $\delta_0 \omega$ in Prop.~\ref{prop:instantons}, a direct 
computation shows that its local counterpart is \be u \delta_0\omega u^* = - 2 
\rho \dd \rho \, -2 \rho^4 \cZ^* \dd \cZ , \ee giving for the transformed 
curvature, \be u F_{t,0} u^* =F_0 + 2t (1-2\rho^2) F_0 + \cO(t^2). \ee It is 
clear that this rescaled curvature still satisfies the self-duality equation; 
this is also in concordance with Prop.~\ref{prop:curv-transf}, being 
$\tilde z_0 =2\rho^2-1$.

\subsection{Moduli space of instantons}\label{subsect:modulispace}
We will closely follow the infinitesimal construction in \cite{AHS78} of instantons for the undeformed case. This will eventually result in the computation of the dimension of the `tangent space' to the moduli space of instantons on $\S^4$ by index methods. It will turn out that the five-parameter family of instantons constructed in the previous section is indeed the complete set of infinitesimal instantons on $\S^4$. 

Let us start by considering the following family of connections on $\S^4$,
\be
\nabla_t=\nabla_0 + t \alpha
\ee 
where $\alpha \in \Omega^1(\ad(\Sk^7)) \equiv 
\Omega^1(\S^4) \otimes_{\Cinf(\S^4)} \Gamma^\infty(\ad(\Sk^7))$ and 
after Example~\ref{ex:instanton-bundle} we have denoted 
$\Gamma^\infty(\ad(\Sk^7))=\Cinf(\Sk^7)\otimes_\ad su(2)$. For $\nabla_t$ to be an instanton, we have to impose the self-duality equation $\ast_\theta F_t = F_t$ on the curvature $F_t = F_0^2 + t [\nabla,\alpha] + \cO(t^2)$ of $\nabla_t$. This leads, when differentiated with respect to $t$, at  $t=0$, to 
{\it linearized self-duality equations}
\be
P_- [\nabla_0, \alpha]=0,
\ee
with $P_-:=\half(1-\ast_\theta)$ the projection onto the antiself-dual 2-forms. Here $[\nabla_0, \alpha]$ is an element in $\Omega^2(\Gamma^\infty(\ad(\Sk^7)))$,   
\be
[\nabla_0,\alpha]_{ij} = \dd \alpha_{ij} + \omega_{ik} \alpha_{kj} -\alpha_{ik} \omega_{kj},
\ee
and has vanishing trace, due to the fact that $\omega_{ik} \alpha_{kj} =\alpha_{kj} \omega_{ik}$ (cf. eqs.~\eqref{cancon}, \eqref{cangp} and the related discussion). 

If the family were obtained from an infinitesimal gauge transformation, we would have had $\alpha=[\nabla_0,X]$, for some $X \in \Gamma^\infty(\ad(\Sk^7))$. Indeed, $[\nabla_0,X]$ is an element in $\Omega^1(\ad(\Sk^7))$ and 
$P_-[\nabla_0,[\nabla_0,X]]=[P_- F_0,X]=0$, since $F_0$ is self-dual. 
Hence, we have defined an element in the first cohomology group $H^1$ of the so-called {\it self-dual complex}:
\be\label{complex}
0 \to \Omega^0(\ad(\Sk^7)) \overset{\dd_0}{\longrightarrow} \Omega^1(\ad(\Sk^7)) \overset{\dd_1}{\longrightarrow} \Omega^2_-(\ad(\Sk^7)) \to 0 ,
\ee
with $\Omega^0(\ad(\Sk^7))=\Gamma^\infty(\ad(\Sk^7))$ and $\dd_0=[\nabla_0,\cdot~]$, 
$\dd_1:= P_-[\nabla_0,\cdot~]$. Note that these operators are Fredholm operators, so that the cohomology groups of the complex are finite dimensional. 
As usual, the complex can be replaced by a single Fredholm operator
\be
\label{eq:op-fredholm}
\dd_0^* + \dd_1 : \Omega^1(\ad(\Sk^7)) \to \Omega^0(\ad(\Sk^7)) \op \Omega^2_-(\ad(\Sk^7)) ,
\ee
where $\dd_0^*$ is the adjoint of $\dd_0$ with respect to the inner product \eqref{eq:inner-product-forms}. 

Our goal is to compute $h^1=\dim H^1$ -- the number of `true' instantons. This is achieved by calculating the alternating sum $-h^0+h^1-h^2$ of Betti numbers from the index of the above Fredholm operator, 
\be
\ind(d_0^* + d_1) = -h^0 + h^1 - h^2,
\ee
while showing that $h^0=h^2=0$. 
By definition, $H^0$ consists of the covariant constant elements in $\Gamma^\infty(\ad(\Sk^7))$. 
Since $[\nabla_0,\cdot]$ commutes with the action of $\bT^2$ and coincides with $\nabla_0^{(2)}$ on $\Gamma^\infty(\ad(\Sk^7))$ (cf. Remark~\ref{rem:conn-form-central-general}), being covariantly constant means
\be
[\nabla_0,X]=\nabla_0^{(2)}(X)=0.
\ee
If we write once more $X=L_\theta(X^\class)$ in terms of its classical counterpart, we find that this condition entails
\be
\nabla_0^{(2)}(L_\theta(X^\class)) = L_\theta( \nabla_0^{(2)}(X^\class))=0
\ee
since $\nabla_0^{(2)}$ commutes with $L_\theta$ (cf. Remark~\ref{rem:conn-form-central-general}). Since for the undeformed case, there are no covariant constant elements in $\Gamma^\infty(\ad(S^7))$ for an irreducible self-dual connection on $\E$, we conclude that $h^0=0$. A completely analogous argument for the kernel of the operator $d_1^*$ shows that also $h^2=0$.

\subsection{Dirac operator associated to the complex}
The next step consists in 
computing the index of the Fredholm operator $\dd_0^* + \dd_1$ defined in 
\eqref{eq:op-fredholm}. Firstly, this operator can be replaced by a Dirac 
operator on the spinor bundle $\cS$ with coefficients in the adjoint bundle. For 
this, we need the following lemma, which is a straightforward modification of its 
classical analogue \cite{AHS78}. Recall that the $\IZ^2$-grading $\gamma_5$ 
induces a decomposition of the spinor bundle $\cS= \cS^+ \oplus \cS^-$. Note also 
that $\cS^-$ coincides classically with the charge $-1$ anti-instanton bundle. 
Indeed, the Levi-Civita connection -- when lifted to the spinor bundle and 
restricted to negative chirality spinors -- has antiself-dual curvature. 
Similarly, $\cS^+$ coincides with the charge $+1$ instanton bundle. Then 
Remark~\ref{rem:associated-modules} implies that the $\Cinf(S^4)$-modules 
$\Gamma^\infty(S^4,\cS^\pm)$ have a module-basis that is homogeneous under the 
action of $\tilde\bT^2$. We conclude from $\tilde\bT^2$-equivariance that 
$\Gamma^\infty(\S^4, \cS^-)$ is isomorphic to the charge $-1$ anti-instanton 
bundle $\Gamma^\infty(\Sk^7\times_{\SU(2)} \C^2$) on $\S^4$. Similarly 
$\Gamma^\infty(\S^4, \cS^+)$ is isomorphic to the charge $+1$ instanton bundle.

\begin{lma}\label{spbundles} 
There are the following isomorphisms of right 
$\Cinf(\S^4)$-modules, \begin{align*} \Omega^1(\S^4) &\isom 
\Gamma^\infty(\S^4,\cS^+ \ot \cS^-)\isom 
\Gamma^\infty(\S^4,\cS^+)\ot_{\Cinf(\S^4)} \Gamma^\infty(\S^4,\cS^-) , \\ 
\Omega^0(\S^4) \oplus \Omega^2_-(\S^4) &\isom \Gamma^\infty(\S^4,\cS^- \ot \cS^-) 
\isom \Gamma^\infty(\S^4,\cS^-)\ot_{\Cinf(\S^4)} \Gamma^\infty(\S^4,\cS^-) . 
\end{align*} 
\end{lma} 
\begin{proof} Since classically $\Omega^1(S^4) \isom 
\Gamma^\infty(S^4, \cS^+ \otimes \cS^-)$ as $\sigma$-equivariant 
$\Cinf(S^4)$-bimodules, Lemma~\ref{lma:modulesA} shows that $\Omega^1(\S^4) \isom 
\Gamma^\infty(\S^4, \cS^+ \otimes \cS^-)$ as $\Cinf(\S^4)$-bimodules. The 
observations above the Lemma indicate that $\cS^\pm \isom S^7 \times_{\rho^\pm} 
\C^2$ for the spinor representation $\rho^+ \oplus \rho^-$ of $\Spin(4) \isom 
\SU(2) \times \SU(2)$ on $\C^4$, so that, using 
Prop.~\ref{prop:tensor-modules}, 
\begin{align*} \Gamma^\infty(\S^4, \cS^+ 
\otimes \cS^-) &\isom \Cinf(\Sk^7) \boxtimes_{\rho^+ \otimes \rho^-} (\C^2 
\otimes \C^2)\\ &\isom \left(\Cinf(\Sk^7) \boxtimes_{\rho^+} \C^2\right) 
\otimes_{\Cinf(\S^4)} \left( \Cinf(\Sk^7) \boxtimes_{\rho^-} \C^2 \right) , 
\end{align*}
This 
proves our claim. An analogous statement holds for the second isomorphism. 
\end{proof} 
Let us forget for the moment the adjoint bundle $\ad(\Sk^7)$. Since 
$\Omega(\S^4) \isom \Omega(S^4)$ as vector spaces and both $\dd$ and the Hodge 
$\ast$ commute with the action of $\bT^2$, the operator $\dd^* + P_- \dd$ can be 
understood as a map from $\Omega^1(S^4) \to \Omega^0(S^4) \op \Omega^2_-(S^4)$ (see Sect.~\ref{se:diff-calc}). Under the isomorphisms of the above Lemma, 
this operator is replaced \cite{AHS78} by a Dirac operator with coefficients in 
$\cS^-$, 
\be \label{eq:dirac-spinors} 
D': \Gamma^\infty(\S^4,\cS^+ \ot \cS^-) \to 
\Gamma^\infty(\S^4,\cS^- \ot \cS^-). 
\ee 
Twisting by the adjoint bundle 
$\ad(\Sk^7)$, merely results into a composition with the projection $p_{(2)}$ 
defining the bundle $\ad(\Sk^7)$. Hence, eventually the operator $\dd_0^* + 
\dd_1$ is replaced by the Dirac operator with coefficients in 
the vector bundle $\cS^-\otimes \ad(\Sk^7)$ on $\S^4$: 
\be \label{eq:dirac-twisted} 
\cD: 
\Gamma^\infty(\S^4, \cS^+ \otimes \cS^-\ot \ad(\Sk^7)) \to 
\Gamma^\infty(\S^4,\cS^- \otimes\cS^- \ot \ad(\Sk^7)). 
\ee 

We have finally arrived to the computation of the index of this Dirac operator 
which we do by means of the Connes-Moscovoci local index formula. It is given by the pairing, 
\be \ind(\cD) = \langle \phi, \chern(\cS^- \ot \ad(\Sk^7)) \rangle = 
\langle \phi, \chern(\cS^-) \cdot \chern(\ad(\Sk^7)) \rangle. 
\ee 
Both the cyclic cocycle $\phi^*$ and the Chern characters, as well as their realization 
as operators $\pi_D(\chern (\ce)$ on the the Hilbert space $\cH$ are recalled in Sect.~\ref{se:liftoric}. In \cite{LS04} we computed these 
operators for all modules associated to the principal bundle $\Sk^7 \to \S^4$. In particular, for the adjoint bundle we found that 
\begin{align*} \pi_D 
\big(\chern_0(\ad(\Sk^7))\big) = 3, \qquad \pi_D \big(\chern_1(\ad(\Sk^7))\big) = 0, \qquad \pi_D \big(\chern_2(\ad(\Sk^7))\big) = 4 (3 \gamma_5). 
\end{align*} 
To compute the Chern character of the spinor bundle $\cS^-$ we use its mentioned identification with the charge $-1$ instanton bundle 
$\Gamma^\infty(\Sk^7\times_{\SU(2)} \C^2$) on $\S^4$. It then follows from 
\cite{CL01} (cf. also \cite{LS04}) that 
\begin{align*} 
\pi_D 
\big(\chern_0(\cS^-)\big) = 2, \qquad\pi_D \big(\chern_1(\cS^-)\big) = 0, 
\qquad\pi_D \big(\chern_2(\cS^-)\big) = - 3 \gamma_5. 
\end{align*} 
Combining both 
Chern characters and using the local index formula on $\S^4$, we have 
\be 
\ind(\cD) = 6 ~\resz ~z^{-1} \tr (\gamma_5 |D|^{-2z}) + 0 + \half (2\cdot 4 - 3 
\cdot 1) \resz \tr ( 3 \gamma_5^2 |D|^{-4-2z}), 
\ee 
with $D$ identified with the 
classical Dirac operator on $S^4$ (recall that we do not change it in the 
isospectral deformation). Now, the first term vanishes due to $\ind(D)=0$ for 
this classical operator. On the other hand $\gamma_5^2 = \I_4$, and \[ 3 \resz 
\tr ( |D|^{-4-2z}) = 6 \Tr_{\omega}(|D|^{-4}) = 2, \] since the Dixmier trace of 
$|D|^{-m}$ on the $m$-sphere equals $8/m!$ (cf. for instance \cite{GVF01,Lnd97}). 
We conclude that $\ind(\cD) = 5$ and for the moduli space of instantons on 
$\S^4$, we have:
\begin{thm} The tangent space at the base point 
$\nabla_0$ to the moduli space of (irreducible) $\SU(2)$-instantons on $\S^4$ is five-dimensional. 
\end{thm}

For the global geometry of the moduli space, it appears that 
one can construct a noncommutative family of instantons, that is instantons parametrized by the quantum quotient space of the deformed conformal group $\SL_\theta(2,\bH)$ by the deformed gauge group $\Sp_\theta(2)$; this will be reported elsewhere \cite{LPRS05}.

\section{Final remarks}\label{se:fire}
A different quantum version of  the $\SU(2)$ Hopf bundle $S^7 \rightarrow S^4$ was constructed in \cite{LPR06}. The quantum sphere $S^7_q$  arises from the symplectic group $Sp_q(2)$ and a quantum $4$-sphere $S^4_q$ is obtained via a suitable self-adjoint idempotent $p$ whose entries generate the algebra
$A(S^4_q)$ of polynomial functions over it. This projection determines a
deformation of an instanton bundle over the classical sphere $S^4$.

One starts with the symplectic quantum groups $A(Sp_q(2))$, 
i.e. the Hopf algebras generated by matrix elements
$T_i^j$'s with commutation rules coming  from the
$R$ matrix of the $C$-series \cite{frt}. The symplectic quantum
$7$-sphere $A(S_q^7)$ is generated by the matrix elements of
the first and
the last  column of $T$. The
algebra $A(S_q^7)$ is the  quantum version of the homogeneous space
$Sp(2)/Sp(1)$
and the injection
$A(S^7_q)\hookrightarrow A(Sp_q(2))$ is a quantum principal bundle with `structure Hopf algebra'  $A(Sp_q(1))$,
an example of the general construction of \cite{BM93}.
In turn, the sphere $S_q^7$ is the total space of a
quantum $\SU_q(2)$
principal bundle over a quantum $4$-sphere $S_q^4$. 
As mentioned, the algebra $A(S_q^4)$  is
constructed as the subalgebra of $A(S_q^7)$ generated by the matrix 
elements of a
self-adjoint projection $p$ which generalizes (at $q\not=1$) the instanton of 
charge $-1$ (at $q=1$).
Unlike the construction for $A(S_q^7)$ out of $A(Sp_q(2))$, one does not have a quantum homogeneous structure. Still, there is a natural coaction of
$\SU_q(2)$ on $A(S^7_q)$ with coinvariant algebra $A(S^4_q)$ and the injection $A(S^4_q)\hookrightarrow A(S^7_q)$ turns out to be an faithfully flat $A(\SU_q(2))$ noncommutative principal bundle (a Hopf-Galois extension).

To compute the charge of the projection and to prove the non triviality of this principal bundle, one follows a general strategy of noncommutative index theorem \cite{C94}. One constructs the 
representations of the algebra $A(S_q^4)$ and the corresponding 
$K$-homology. 
The
analogue of the fundamental class of $S^4$ is given by a non trivial Fredholm
module $\mu$. The natural coupling between $\mu$ and the projection 
$p$ is computed
via the pairing of the corresponding Chern characters $\chern ^*(\mu)\in
HC^*[A(S_q^4)]$ and
$\chern_*(p)\in HC_*[A(S_q^4)]$ in cyclic cohomology and homology respectively. 
As expected the result of this pairing, which is an integer by
principle being the index of a Fredholm operator, is actually $-1$ and therefore the bundle is non trivial.

Clearly the next step would be to repeat the analysis of the toric four sphere and define a Yang-Mills action functional  and self-duality equations. To this end one needs a `metric structure' on the bundle; for this, the recently found \cite{DLSSV04} isospectral noncommutative geometry for  $\SU_q(2)$ promises to be useful. 

\newpage

\end{document}